\documentclass[journal,article,submit,moreauthors,pdftex]{Definitions/mdpi}

\firstpage{1} 
\pubvolume{1}
\issuenum{1}
\articlenumber{0}
\pubyear{2021}
\copyrightyear{2020}
\datereceived{} 
\dateaccepted{} 
\datepublished{} 
\hreflink{https://doi.org/} % If needed use \linebreak
\Title{Causal inference in time series in terms of R\'{e}nyi transfer entropy }
\TitleCitation{Title}
\Author{Petr Jizba $^{1,\ddagger,*}$\orcidA{}, Hynek Lavi\v{c}ka $^{1,\dagger,\ddagger}$\orcidB{} and Zlata Tabachov\'{a} $^{1,2 \ddagger}$\orcidC{}}

\AuthorNames{Petr Jizba, Hynek Lavi\v{c}ka and Zlata Tabachov\'{a}}

\AuthorCitation{Jizba, P.; Lavi\v{c}ka, H.; Tabachov\'{a}, Z.}
\address{%
$^{1}$
\quad Faculty of Nuclear Sciences and Physical Engineering, Czech Technical University in Prague, Břehová 7, 115 19 Praha 1, Czech Republic\\
$^{2}$  
\quad Complexity Science Hub Vienna, Josefst\"{a}adter Strasse 39, 1080 Vienna, Austria;\\ \quad \hspace{1.5mm}  p.jizba@fjfi.cvut.cz; hynek.lavicka@fjfi.cvut.cz; Zlata.Tabachova@fjfi.cvut.cz}

\corres{Correspondence: p.jizba@fjfi.cvut.cz.; Tel.: +420-775-317-309}

\firstnote{Current address: Blocksize Capital GmbH, Taunusanlage 8, D-60329 Frankfurt am Main, Germany %Former address: Thetaris GmbH, Leopoldstraße 244, D-80807 Munich, Germany
} 
\secondnote{These authors contributed equally to this work.}
\abstract{
Uncovering causal interdependencies from observational data is one of the great challenges of nonlinear time series analysis. In this paper, we discuss this topic with the help of information-theoretic concept known as R\'{e}nyi's information measure. In particular, we tackle the directional information flow between bivariate time series in terms of R\'{e}nyi’s transfer entropy. 
We show that by choosing R\'{e}nyi’s parameter $\alpha$ appropriately we can control information that is transferred only between selected parts of underlying distributions. This, in turn, provides particularly potent tool for quantifying causal interdependencies in time series, where the knowledge of ``black swan'' events such as spikes or sudden jumps are of a key importance.  In this connection, we first prove that for Gaussian variables, Granger causality and
R\'{e}nyi transfer entropy are entirely equivalent. Moreover, we also partially extend this results to heavy-tailed $\alpha$-Gaussian variables.
These results allow to establish connection between autoregressive and 
R\'{e}nyi entropy based information-theoretic approaches
to data-driven causal inference. 
To aid our intuition 
we employ Leonenko {\em et al.} entropy estimator and analyze R\'{e}nyi's information flow between bivariate time series generated from two unidirectionally coupled R\"{o}ssler systems. 
Notably, we find that R\'{e}nyi’s transfer entropy
not only allowed us to detect a threshold of synchronization but it also
provided a non-trivial insight into the structure 
of a transient regime that exists between region of chaotic correlations and
synchronization threshold.
In addition,  from R\'{e}nyi’s transfer entropy we could reliably infer the direction of coupling -- and hence causality, only for coupling strengths smaller that the onset value of transient regime , i.e.
when two R\"{o}ssler systems were coupled, but have not yet entered  a synchronization. 
 }

\newcommand{\figuresizeRoessler}{0.36\textwidth}

\begin{document}

%%%%%%%%%%%%%%%%%%%%%%%%%%%%%%%%%%%%%%%%%%
%\setcounter{section}{-1} %% Remove this when starting to work on the template.
\section{Introduction}

%Complex dynamical systems are characterized by non-trivial interactions among internal
%constituents so that the entire system evolves differently than just the sum of its parts.

The time evolution of complex systems is usually recorded in the form of time series. Time series analysis is a traditional field of mathematical statistics, however, development in nonlinear dynamical systems and theory of deterministic chaos has opened up new vistas in analysis of nonlinear time series~\cite{SchreiberII,Kanz-book}. The discovery of synchronization of chaotic systems~\cite{Pecora} has changed the study of interactions and cooperative behaviour of complex systems and also brought new approaches to study relations between nonlinear time series~\cite{Boccaletti}. During the process of synchronization
two systems can either mutually interact or only one can influence the other. In order to distinguish these two ways, and also to find which system is the driver (``master'') and which the response (``slave'') system, a number of approaches from the dynamical system theory 
have been proposed~\cite{Quiroga,Nawrath,Sugihara,Feldhof}.
The aforementioned problem of synchronization can be
seen as part of a broader framework 
known \textit{as causality or causal relations between systems},
processes or phenomena. The mathematical formulation of causality in terms of predictability was first proposed by Wiener~\cite{Wiener} and formulated for time series by Granger~\cite{Granger}. In particular, Granger introduced what is now known as \textit{Granger causality}, which is a statistical concept of causality that is based 
on the evaluation of predictability in bivariate autoregressive models.

%introduced in his influential 1969 paper a specific notion of causality
%that is particularly useful in time series analysis  
%into time series analysis 
%by evaluating the predictability in bivariate autoregressive models.

Extracting causal interdependencies from observational data is presently one of the key tasks in nonlinear time series analysis. Apart from the linear Granger causality and various nonlinear extensions thereof~\cite{Ancona,Chen,Wismller2021}, existing methods for this purpose include, for instance, state-space based approaches such as conditional probabilities of recurrence~\cite{Zou,Donner,Romano}, or  information-theoretic quantities such as conditional mutual information~\cite{Vejmelka,palusRossler} and \textit{transfer 
entropies}~\cite{Schreiber,Kantz,Kanz-book,JizbaRE}. 
Especially, the latter information-theoretic quantities
represent powerful instruments in quantifying causality between time-evolving
systems. This is because ensuing information-theoretic functionals (typically based on Shannon entropy)  quantify in a non-parametric and explicitly non-symmetric way the flow of information between two (or more) time series.
Particularly transfer entropies (TEs) have enjoyed recently a considerable attention. The catalyst was infusion of new ideas both from numerical and conceptional side. For instance, the performance of Shannon-entropy based conditional entropies and conditional mutual entropies has been in recent years extensively tested using numerically generated time series~\cite{Vejmelka,Palus:2007a}. 
Sophisticated algorithms have been developed to uncover direct causal relations
in multivariate time series~\cite{Runge:2012a,Feas:2015a,Sun:2015a}. In parallel, increasing attention has been devoted to the development of reliable estimators of entropic functionals with the aim to detect causality from nonlinear 
time series~\cite{Leonenko_Prozanto_Savani,Leonenko_Prozanto}. 
At the same time, it has been recognized that information-theoretic approaches play important role in dealing with complex dynamical systems that are multiscale and/or non-Gaussian~\cite{Lungarella:2007a,Faes:2017a,Palus:2014a,JizbaRE}. The latter class includes complex systems with heavy-tailed probability distributions epitomized, e.g., in financial and climatological time series~\cite{Tsallis:book,Thurner:book}.

In this paper we extend the  popular Shannon-entropy based TE (STE), which represents a prominent tool for assessing directed information flow between joint processes, and instead quantify information transfer in terms of \textit{R\'{e}nyi's TE} (RTE). RTE was introduced by one of us (PJ) in Ref.~\cite{JizbaRE} in the context of bivariate financial time series. The original idea was to use the RTE in order to exploit the theoretical formulation that could identify and quantify peculiar features in multiscale bivariate processes (e.g., multiscale patterns, generalized fractal dimensions or multifractal cross-correlations) that are often seen in finance. In contrast to~\cite{JizbaRE} where the focus was mostly on qualitative aspects of  R\'{e}nyian information flow between selected stock-market time series, in the present work we wish to be more quantitative by analyzing coupled time series that are numerically generated from known dynamics. Specifically, we demonstrate \textit{how the \textit{RTE
method performs in the detection of the coupling direction and onset of synchronization between
two R\"{o}ssler oscillators}}~\cite{Rossler:1976} \textit{that are unidirectionally coupled in the first variable $x$}. R\"{o}ssler system (RS) is paradigmatic and well studied low-dimensional chaotic dynamical system. When coupled, RSs allow for \textit{synchronization} and also for a subtle phenomenon known as ``phase synchronization'', i.e., situation when the amplitudes of both systems
are not correlated while the phases are approximately equal. 
%In the context of time series analysis, the latter implies that estimating only the cross-correlation (or even mutual information) of the measured signal (in this case, the $x$-components of both RSs) is not enough to capture all possible forms of interrelations~\cite{Rosenblum:1996,Cheng:2017}. 
In this respect the synthetic bivariate time series generated from coupled RSs serve as an excellent test-bed allowing to numerically analyze, e.g., drive-response relationships or identify ensuing onset (or threshold) of synchronization.
In doing so, we identify factors and influences that can lead to either decreased in the RTE  sensitivity or false detections and propose some ways to cope with them.
%
%and demonstrate with numerical simulations the much larger estimation accuracy of the RTE approach compared to the STE. 
%The improved computational reliability is exploited to disclose meaningful multiscale patterns of information transfer between ***.
%
Aforementioned issues  have not been so far explicitly studied in the framework of the RTE
and this work presents a first attempt in this direction.

To set the stage, we shall first, in Section~\ref{sec.2.a}, provide some information-theoretic background on R\'{e}nyi entropy (RE) that will be needed in the main body of the text. For a self-consistency of our exposition we briefly review the 
Shannon's transfer entropy of Schreiber and motivate and derive the core quantity of this work --- the R\'{e}nyi transfer entropy.
Issue of causality and its connection to RTE is examined in
Section~\ref{Sec.4.aa}. In particular, we prove
that the Granger causality is entirely equivalent to the RTE for Gaussian processes and show how the Granger causality and the RTE are related in the case of heavy-tailed (namely $\alpha$-Gaussian) processes. 
Section~\ref{Sec.3.cc} is dedicated to the discussion of the estimator of RE introduced by Leonenko, {\em et al.} that will be employed in our numerical analysis.  
The proposed framework is then
illustrated for two unidirectionally coupled R\"{o}ssler systems as a paradigmatic
example. To cultivate our intuition about the latter RSs, we discuss in Section~\ref{Sec.5.cc}
the inner workings of such RSs in terms of simple numerical experiments.  
%that serve as a paradigmatic
%example in this case. 
Ensuing numerical analysis is presented in Section~\ref{Sec.6.cc}.
There we discuss how the RTE 
can be used to detect causality and onset of synchronization in the two coupled RSs.
We also demonstrate how the RTE provides 
a non-trivial insight into the structure of a transient regime that exists between region of chaotic
correlations and onset of synchronization.
%and the
%lengths of the available time series on the detectability of the
%true coupling direction are discussed in some detail.
Finally, 
Section~\ref{Sec.8.cc} summarizes our theoretical and numerical findings and discusses possible extensions of the present work. 

%Inferring the internal interaction patterns of a complex dynamical system is a challenging task. Traditional methods often rely on examining the correlations among time series characterizing the temporal evolution of dynamical units (e.g., long-term records of air temperature from different locations on the Earth, or between two stock-index series in an economic system). The correlation functions, however, have at least two limitations. First, they measure only linear relations, although it is clear that linear models do not faithfully reflect real interactions within (or among) complex systems. Second, all they determine is whether two (or more) observed units have correlated dynamics. Yet they do not provide any directional information about cause and effect. In nonlinear complex systems in general, and in climate or finance, in particular, neither the time lag derived from the maximum lagged cross-correlation  can generally reveal the correct cause-effect temporal order~\cite{Nes}

%%%%%%%%%%%%%%%%%%%%%%%%%%%%%%%%%%%%%%%%%%%
\section{R\'{e}nyi entropy \label{sec.2.a}}
%%%%%%%%%%%%%%%%%%%%%%%%%%%%%%%%%%%%%%%%%%%%

Information-theoretic approaches based on Shannon-entropy currently belong in portfolio of techniques and tools that are indispensable in addressing causality issues in complex dynamical systems. At the same time, Shannon's information theory is limited in its scope. In fact, it has been known already since Shannon's seminal papers~\cite{Shannon:1948a,Shannon:1948b} that Shannon's
information measure (or entropy) represents mere idealized information appearing only
in situations when the buffer memory (or storage capacity) of a transmitting channel
is infinite. In particular, 
Shannon's source coding theorem (or noiseless coding theorem), which establishes the limits to possible data compression, and thus provides operational meaning to the Shannon entropy assumes that the {\em cost} of a codeword is a linear function of its length (so the optimal code has a minimal cost out of all codes).
Linear cost of codewords is, however not always  desirable. For instance, when the storage capacity is finite one would like to penalize excessively lengthy codewords with price that is, e.g. exponential  rather then linear function of the length.

For these reasons information theorists have devised various remedies to deal with such cases. This usually consists of substituting Shannon's information measure with information measures of other types.
Consequently, numerous generalizations of Shannon's entropy have started to proliferate in the information-theory literature ranging from additive
entropies~\cite{JA,Burg:1972a}  through a rich class of non-additive entropies~\cite{Tsallis:1988a,Havrda:1967a,Frank:2000a,Sharma:1978a,Jizba:2016a} to more exotic types of entropies~\cite{Vos:2015a}.
Particularly prominent among such generalizations is a one-parametric class of information measures known as \textit{R\'{e}nyi entropies} that were introduced by Hungarian mathematician and
information theorist Alfred R\'{e}nyi in early 1960's~\cite{Renyi:1970a,Renyi:1976a} . 
Applications
of RE in information theory, namely its generalization to
coding theorems, were carried over by Campbel~\cite{Campbell:1965a},
Csisz\'{a}r~\cite{Csiszar:1995,Czisar:2004a}, Acz\'{e}l~\cite{Aczel:1975} and
others. 
In a physical setting  RE was popularized in the context of chaotic dynamical systems by
Kadanoff, {\em et al}.~\cite{Kadanoff:1986a} and  in connection with multifractals by
Mandelbrot~\cite{Mandelbrot:1977a}. RE is also indispensable in quantum information theory where it quantifies multipartite entanglement~\cite{Bengtsson:2006a}.

%It should be stressed that, apart from the CCT, the RE has yet further operational definitions, e.g., in the theory of
%guessing [33], in the buffer overflow problem [34], or in the theory of error block coding [35]. The RE is also underpinned
%with various axiomatics [25,26,36]. In particular, it satisfies identical Khinchin axioms [18] as Shannon’s entropy save for
%the additivity axiom (chain rule) [11,37,38]:

In its essence, REs constitute a one-parametric family of information measures labeled by parameter $\alpha$ that fulfill the additivity with respect to the composition of
statistically independent systems. The special case with $\alpha = 1$ corresponds to ordinary Shannon’s entropy. REs  belong to a broader class of so-called Uffink entropic functionals~\cite{Jizba:2019c,Jizba:2020c}, i.e., the most general class of solutions that satisfy Shorem--Johnson axioms for the maximum entropy principle in statistical estimation theory. Moreover, it might be shown that R\'{e}nyi entropies belong to the class of the so-called mixing homomorphic functions~\cite{Lesche:1982a} and that they are analytic for
$\alpha \in {\mathbb{C}}_{I \cup IV}$, cf.~\cite{JA}.

\subsection{Definition}

RE is defined as an exponentially weighted mean of {\em Hartley  information measure} $-\log p$ (i.e., elementary measure of information) ~\cite{MM}. 
In fact, it was shown by R\'{e}nyi that, except for a linearly weighted average (which leads to Shannon entropy), exponential weighting is the only possible averaging that is both compatible with the Kolmogorov-- Nagumo average prescription and  leads to entropies that are additive with respect to independent systems~\cite{Renyi:1970a,Renyi:1976a}.  RE associated with a system described with a probability distribution $\mathcal{P}$ reads
%
%of quantifying average value in terms of Kolmogorov--Nagumo.
%
\begin{equation}
    H_{\alpha}[\mathcal{P}]\ = \ \frac{1}{1-\alpha}\log_2\sum_{i=1}^n p_i^{\alpha}\, .
    \label{RE}
\end{equation}
RE has the following properties~\cite{Renyi:1976a,JA}:
\begin{itemize}
    \item RE is symmetric, i.e. $H_{\alpha}[\{p_1,..,p_n\}]=H_{\alpha}[\{p_{\pi(1)},...,p_{\pi(n)}\}]$;
    \item RE is non-negative, i.e. $H_{\alpha}\geq 0$;
    \item $\lim_{\alpha \to 1}H_{\alpha}=H_1$, where $H_1=H$ is the Shannon entropy;
    \item $H_0=\log_2 n$ is the \textit{Hartley} entropy and $H_2 =- \log_2 \sum_{i=1}^n p_i^2$ is the \textit{Collision entropy};
    \item $0\leq H_\alpha[\mathcal{P}]\leq \log_2 n$;
    \item $H_\alpha$ is a positive, decreasing function of $\alpha \geq 0$. 
\end{itemize}

\subsection{Multifractals, chaotic systems and R\'{e}nyi entropy}

Another appealing property of the R\'{e}nyi entropy is its close connection to {\em multifractals}, i.e. mathematical paradigm that is often encountered in complex dynamical systems with examples ranging from turbulence and strange attractors to meteorology and finance, see e.g.~\cite{Harte:2019a}. Aforementioned connection is established through the so-called \textit{generalized dimensions}, which are defined as~\cite{Kanz-book,Kadanoff:1986a}
\begin{equation}
    D_{\alpha} \ = \ -\lim_{\delta \rightarrow 0}\frac{H_{\alpha}(\delta)}{\log \delta}\, 
    \label{gendim}
\end{equation}
where $\delta$ is a size of a $\delta-$mesh covering of a configuration space of a system.
Generalized dimensions $D_{\alpha}$ are conjugete to {\em multifractal spectrum} $f(\beta)$ through the Legendre transform~\cite{Kadanoff:1986a}
\begin{equation}
    (\alpha-1)D_{\alpha} \ = \ \alpha \beta \ - \ f(\beta).
\end{equation}
The function $f(\beta)$ is called multifractal spectrum because $\beta$ plays the role of scaling exponent 
in the local probability distribution, e.g. distribution with support on the $i$-th hypercube of a mesh size $\delta$ scales as $p_i(\delta) \sim \delta^{\beta_i}$. The key assumption in the multifractal analysis is that in the small $\delta-$ limit the local probability distribution depends smoothly on $\beta$. It can be argued that $f(\beta)$ corresponds to the (box-counting) fractal dimension of the portion of configuration space where local probability distributions have scaling exponent $\beta$, cf. e.g. Ref.~\cite{JA}.  In this way, 
multifractal can be viewed as an ensemble of intertwined (uni)fractals each with its own fractal dimension $f(\beta)$.

Multifractal paradigm is particularly pertinent in \textit{theory of chaotic systems}. For instance,  chaotic dynamics and strange attractors, in particular, are uniquely characterised by the infinite sequence of generalized dimensions $D_{\alpha}$, cf.~Ref.~\cite{Hentschel:1983c}. 
%
%In fact, a complete knowledge of the collection of generalized dimensions $D_{\alpha}$ is
%equivalent to a complete physical characterization of the fractal [39]
In particular, the generalized dimensions can help to recognize in a quantitative way main geometric features of chaotic systems. For instance, they may help to distinguish chaotic behaviour from noisy behaviour, determine a number of variables that are needed to model the dynamics of the system or classify systems into universality classes. On the other hand, dynamical features of chaotic systems are often analyzed through such quantifiers as \textit{Lyapunov exponent}, that is a measure of the divergence of nearby trajectories, or ensuing \textit{Kolmogorov-Sinai entropy rate} (KSE), which quantifies the change of entropy as system evolves
and is given by the sum of all positive Lyapunov exponents. 
Connection between KSE and the time evolution
of the information-theoretic or statistical entropy is quite delicate, see e.g. discussion in Ref.~\cite{Latora:1999c}, though the upshot is clear; in order to describe the dynamics of a (complex) system, the temporal change or the difference in entropy is more relevant than the entropy itself. 
Consequently, while RE (alongside with $D_{\alpha}$) is suitable quantifier of geometric properties of chaotic systems, its temporal differences or temporal rates are useful for the description of the dynamics of such systems. R\'{e}nyi's transfer entropy follows the latter route.

%%%%%%%%%%%%%%%%%%%%%%%%%%%%%%%%%%%%%%%%%
\subsection{Shannon transfer entropy}
%%%%%%%%%%%%%%%%%%%%%%%%%%%%%%%%%%%%%%%%%

In order to understand the concept of R\'{e}nyi transfer entropy we recall first its Shannon's counterpart.

Let ${X}= \{x_i\}_{i=1}^N$ be a discrete random variable with ensuing probability distribution $\mathcal{P}_{X}$, then the Shannon entropy of this process is
\begin{equation}
    H({X}) \ \equiv \ H(\mathcal{P}_{X}) \ = \ -\sum_{x\in \mathcal{X}} p(x)\log_2 p(x)\, .
\end{equation}
Let ${Y}=\{y_i\}_{i=1}^N$ be another random variable, then  {\em mutual information} between ${X}$ and ${Y}$ is 
\begin{eqnarray}
        I({X}\!:\!{Y}) \ &=& \ \sum_{x\in {X},\ \! y\in {Y}} p(x,y)\log_2 \frac{p(x,y)}{p(x)p(y)}\nonumber \\[2mm]
        &=& \ H({X})\ - \ H({X}|{Y}) \ = \ H({Y})\ - \ H({Y}|{X}) \, ,
\end{eqnarray}
where quantity $H(X|Y)$ is the \emph{conditional entropy}, defined as
\begin{equation}
    H(X|Y)\ = \ -\sum_{x\in X, \ \! y\in Y} p(x,y)\log_2 p(x|y)\, .
\end{equation}
Mutual information thus quantifies  an average reduction in uncertainty (i.e., gain in information) about $X$ resulting from observation of $Y$, or vice versa. Since 
$I({X}\!:\!{Y})=I({Y}\!:\!{X})$, it can not be used as a measure of directional information flow. Note also that the amount of information contained in $X$ about itself is
just the Shannon entropy, i.e. $I({X}\!:\!{X})=H({X})$.  

The mutual information between two processes $X$ and $Y$ conditioned on the third process $Z$ is called \emph{conditional mutual information} and is defined as
\begin{equation}
\label{CMI}
    I(X:Y|Z) \ = \ H(X|Z)\ - \ H(X|Y,Z)\  = \ I(X:(Y,Z)) \ - \ I(X:Y)\, .
\end{equation}
Let us now consider two time sequences (e.g., two stock market time
series) described by stochastic random variables $X_t$ and $Y_t$. Let us assume
further that the time steps (e.g., data ticks) are discrete with the time step $\tau$ and with $t_n = t_0 + n\tau$ where $t_0$ is some reference time. For a practical purposes it is also useful to assume that $X_t$ and $Y_t$ represent  discrete-time stochastic Markov processes of order $k$ and $l$, respectively.

We wish now to know, what information will be gained  on $X_{t_{n+1}}$  by observing $Y_t$ up to time $t_n$.
To this end we introduce the joint process $X_{t_n},X_{t_{n-1}}, \ldots, X_{t_{n-k+1}}$, which we denote as $X_n^{(k)}$ and similarly we define the joint process  $Y_n^{(l)} \equiv 
Y_{t_n},Y_{t_{n-1}}, \ldots, Y_{t_{n-l+1}}$.
By replacing $X$ in (\ref{CMI}) by $X_{t_{n+1}}$, $Y$ by $Y_n^{(l)}$ and    $Z$ 
by $X_n^{(k)}$, we obtain the desired conditional mutual information 
    \begin{eqnarray}
    \mbox{\hspace{-6mm}}I(X_{t_{n+1}}:Y_n^{(l)}|X_n^{(k)}) \!\! &=& \!\!
    H(X_{t_{n+1}}\!:\!X_n^{(k)}) \ - \ H(X_{t_{n+1}}\!:\!(Y_n^{(l)}, X_n^{(k)}) ) \nonumber \\[2mm]
    &=& \!\!\sum_{x_n^{(k)}\in X_{n+1}^{(k)}, \ \! y_n^{(l)}\in Y_n^{(l)}}p(x_{n+1},x_n^{(k)},y_n^{(l)})\log_2\left(\frac{p(x_{n+1}|x_n^{(k)},y_n^{(l)})}{p(x_{n+1}|x_n^{(k)})}\right)\, .
    \label{8.cc}
 \end{eqnarray}
 %
 %The quantity inside $(\ldots)$ has the meaning of the deviation from the Markov property. 

The conditional mutual information (\ref{8.cc}) is also known as \emph{Shannon Transfer Entropy} from $Y_t$ to $X_t$ (or simply from $Y$ to $X$) and 
%or, in our case, \emph{gain in information about $x_{n+1}$ caused by the history of $Y$ up to $y_n^{(l)}$ under the assumption that the history of $X$ up to $x^{(k)}$ is known}.  
as a measure  of directed (time asymmetric) information transfer between joint processes,  
it was introduced by Schreiber in Ref.~\cite{Schreiber}. The latter is typically denoted as
 \begin{equation}
     T_{Y \rightarrow X}(k,l) \ 
\equiv \ I(X_{t_{n+1}}:Y_n^{(l)}|X_n^{(k)})\, .
 \end{equation}
 As already mentioned, for independent processes TE is equal to zero. For a non-zero cases transfer entropy measures the deviation from the independence of the two processes. An important property of the transfer entropy is that it is directional, i.e. in general $T_{Y\rightarrow X}\neq T_{X\rightarrow Y}$.

\subsection{R\'{e}nyi transfer entropy}
%%%%%%%%%%%%%%%%%%%%%%%%%%%%%%%%%%%%%%%%%%%

%Following the previous ideas one can now obtain generalizations using R\'{e}nyi's information measures with parameter $\alpha>0$ defined for a discrete random variable $X$ as
%
%\begin{equation}
 %   H_{\alpha}(X) \ = \ \frac{1}{1-\alpha}\log_2\sum_{x\in X} p^{\alpha}(x)\, .
%\end{equation}
%
In the same manner as in (\ref{CMI}) we can introduce  {\em R\'{e}nyi transfer entropy of order $\alpha$} from $Y$ to $X$ (see also Ref.~\cite{JizbaRE}) as
  \begin{eqnarray}
    T^R_{\alpha,Y\rightarrow X}(k,l) \ &=& \ H_{\alpha}(X_{t_{n+1}}|X_n^{(k)}) \ - \ H_{\alpha}(X_{t_{n+1}}|X_n^{(k)},Y_n^{(l)})\nonumber \\[2mm]
    &=& \ I_{\alpha}(X_{t_{n+1}}:Y_n^{(l)}|X_n^{(k)})\, ,
      \label{RTE}
\end{eqnarray}
where $H_{\alpha}(X|Y)$ is the \emph{conditional entropy of order $\alpha$} and $I_{\alpha}(X:Y)$ is the \emph{mutual information of order $\alpha$}. 
These can be explicitly written as~\cite{Renyi:1976a,JizbaRE}
\begin{eqnarray}
    H_{\alpha}(X|Y)\ &=& \ \frac{1}{1-\alpha}\log_2 \frac{\sum_{x\in X, y\in Y}p^{\alpha}(x,y)}{\sum_{y\in Y}p^{\alpha}(y)}\, , \nonumber \\[2mm]
    I_{\alpha}(X:Y) \ &=& \ \frac{1}{1-\alpha}\log_2\frac{\sum_{x\in X, y\in Y}p^{\alpha}(x)p^{\alpha}(y)}{\sum_{x\in X, y\in Y}p^{\alpha}(x,y)}\, .
    \label{12.gb}
\end{eqnarray}
It can be checked (via L'Hospital's rule) that R\'{e}nyi's transfer $\alpha$-entropy reduces to Shannon TE in the $ \alpha \rightarrow 1$ limit, i.e. 
\begin{eqnarray}
    \lim_{\alpha\rightarrow 1}T^R_{\alpha, Y\rightarrow X}\ = \ T_{Y\rightarrow X}\, .
\end{eqnarray} 
From (\ref{RTE}) we see that $T^R_{\alpha,Y\rightarrow X}(k,l)$ may be intuitively interpreted as the degree of ignorance (or uncertainty) about  $X_{{t}_{n+1}}$ resolved by the past states $Y_n^{(l)}$ and $X_{n}^{(k)}$, over and above the degree of ignorance about $X_{{t}_{n+1}}$ already resolved by its own past state alone. Here the ignorance is quantified by R\'{e}nyi information measure (i.e. RE) of order $\alpha$. 

R\'{e}nyi TE can be also be negative  (unlike the Shannon TE). This means that uncertainty of the process $X_t$ becomes bigger knowing the past of $Y_t$, i.e. $H_{\alpha}(X_{{t}_{n+1}}|X_{n}^{(k)})\leq H_{\alpha}(X_{{t}_{n+1}}|X_{n}^{(k)},Y_n^{(l)})$. If $X_t$ and $Y_t$ are independent, then  $ T^R_{\alpha,Y\rightarrow X}= T^R_{\alpha,X\rightarrow Y} = 0$. However, in contrast to Shannon's case, the fact that $T^R_{\alpha,Y\rightarrow X}=0$ does necessarily imply the independence of the two underlying stochastic processes. Nonetheless, in Section~\ref{Sec.4.aa} we prove that in case of Gaussian (Wiener) processes $0$-valued RTE is a clear signature of independence. %General problem with RTE is that in its definition see (\ref{rte}) we explicitly don't have deviation from Markov property (\ref{markov_property}) as it is in Shannon case (\ref{8.cc}). Nevertheless, definition of Markov property (\ref{markov_property}) can be rewritten by means of \textit{filtrations} and applied to the concept of RTE. This issue will be discussed elsewhere.

%%%%%%%%%%%%%%%%%%%%%%%%%%%%%%%%%%%
\subsection{Escort distribution}
%%%%%%%%%%%%%%%%%%%%%%%%%%%%%%%%%%%%

Due to the non-linear way in which probability distributions enter in the definition of RE, cf. Eq.~(\ref{RE}), the RTE represents a useful measure of
transmitted information that quantifies dominant information flow 
between certain parts of underlying distributions. In fact, for $0<\alpha<1$ the corresponding information flow accentuates marginal events, while for $\alpha>1$ more probable (close-to-average) events are emphasized~\cite{JizbaRE}. In this respect one can zoom or amplify  different parts of probability density functions involved by merely choosing appropriate values of $\alpha$. This is particularly useful in studies of time sequences, where marginal events are of a crucial importance, for instance, in financial time series.

\begin{figure}
\begin{center}
\includegraphics[width=0.7\textwidth]{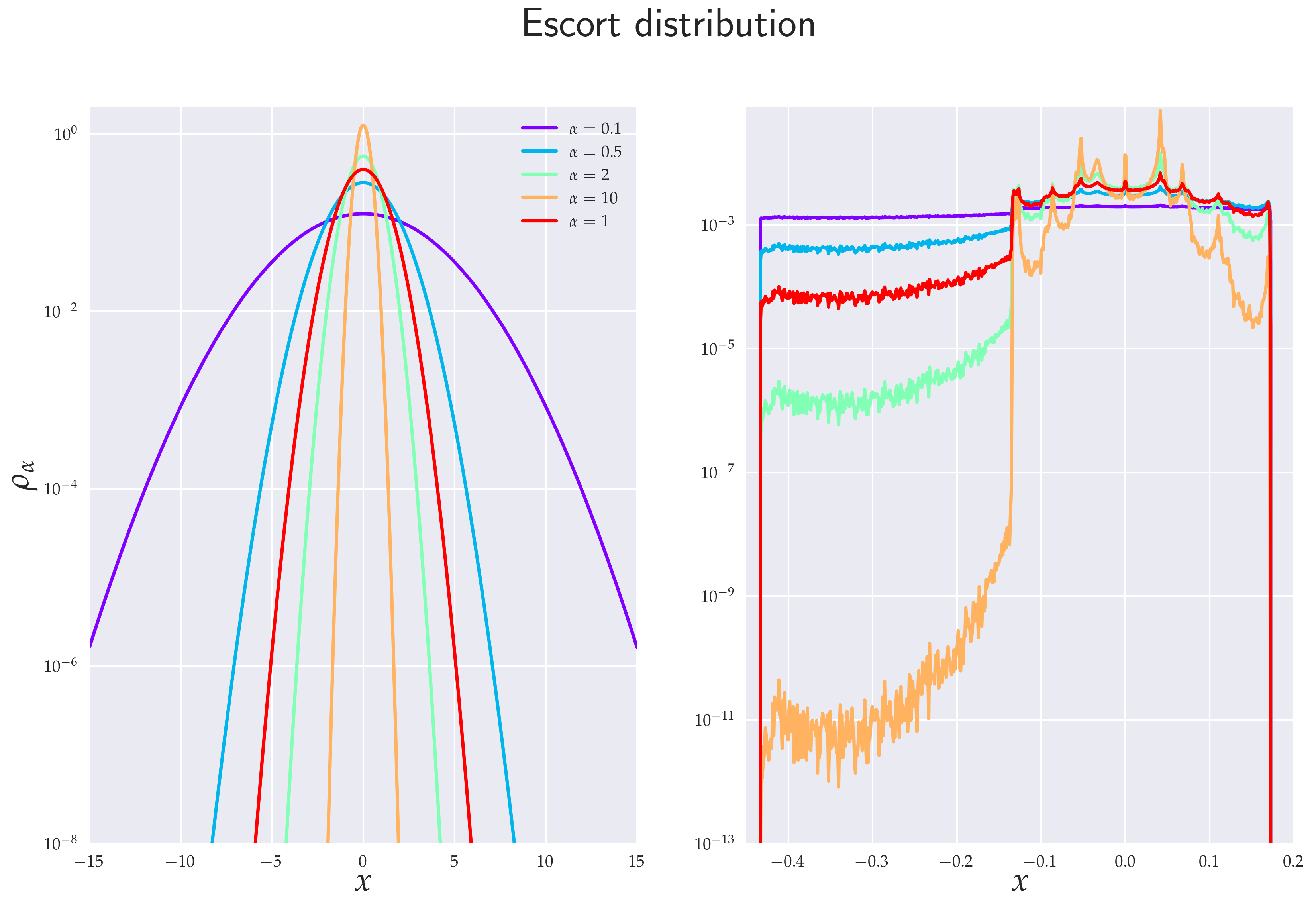}
\caption{Illustration of the concept of escort distribution $\rho_{\alpha}$ on histograms. The left figure depicts log-scaled normal distribution $\mathcal{N}(0,1)$, while on the right figure we show the log-scaled histogram for $x_1-$projection increments from R\"{o}ssler system (\ref{Rossler}). Both figures demonstrate that the escort distribution deforms the original distribution ($\alpha = 1$) so that for $0<\alpha<1$ less probable events are emphasized (the smaller $\alpha$ the greater emphasize) while high probable events are accordingly suppressed. For $\alpha>1$  the situation is reversed.}
\label{escort_distr}
\end{center}
\end{figure}

In order to better understand  the aforementioned  ``zooming'' property of RTE we rewrite (\ref{RTE}) in the form
\begin{equation}
    T^R_{\alpha,Y\rightarrow X}(k,l) \ = \ \frac{1}{1-\alpha}\log_2 \left( \frac{\sum \frac{p^{\alpha}(x_n^{(k)})}{\sum p^{\alpha}(x_n^{(k)})}p^{\alpha}(x_{n+1}|x_n^{(k)})}{\sum \frac{p^{\alpha}(x_n^{(k)},y_n^{(l)})}{\sum p^{\alpha}(x_n^{(k)},y_n^{(l)})}p^{\alpha}(x_{n+1}|x_n^{(k)},y_n^{(l)})}\right)\, .
    \label{rte}
\end{equation}
This particular representation shows how the underlying distribution changes (or deforms) with the change of parameter $\alpha$. Numerator and denominator inside the log-function contain the so-called \emph{escort} (or \emph{zooming}) \emph{distributions} $\rho_{\alpha}$ 
\begin{equation}
    \rho_{\alpha}\ \equiv \ \frac{p^{\alpha}(x)}{\sum p^{\alpha}(x)}\, ,
    \label{escortt}
\end{equation}
which emphasizes less probable events for $0<\alpha<1$ and more probable events when $\alpha>1$, see Fig.~\ref{escort_distr}.
Note also that 
$\rho_{\alpha}(x_n^{(k)},y_n^{(l)})$ is not the joint probability distribution of $X_n^{(k)}$ and $Y_n^{(l)}$ as it does not satisfy the Kolmogorov--de Finetti relation for conditional probabilities~\cite{Jizba:2017c}. 

%One can see, that numerator in (\ref{rte}) presents the average probability weighted with $\rho_{\alpha}(x_n^{(k)})$ and the denominator is average with respect to the $\rho_{\alpha}(x_n^{(k)},y_n^{(l)})$. 

In connection with (\ref{rte}) we may note that for $0<\alpha<1$ the multiplicative factor  is positive, and so the RTE is negative
%i. e. $I_{\alpha}(x_{n+1}:y_n^{(l)}|x_n^{(k)})<0$, 
if by learning $Y^{(l)}_n$ the rare events are (on average) more emphasized
than in the case when only 
$X^{(k)}_n$ alone is known.
Analogically, for $\alpha>1$ the RTE can be negative when, by learning $Y^{(l)}_n$, the more probable events are (on average) more accentuated
in comparison with the situation when $Y^{(l)}_n$ is not known.
It should be stressed that the 
analogous situation does not hold for Shannon's TE. This is because in the limit $\alpha \rightarrow 1$ we regain expression (\ref{8.cc}), which is nothing but relative entropy and as such it is always non-negative due to Gibbs inequality. At the same time, Shannon's TE is by its very definition also a mutual information. 
While RTE is also defined to be a mutual information, it is not relative entropy (in the RE case those two concepts do not coincide). It can be shown (basically via Jensen's inequality)~\cite{JA} that  the
relative entropy based on RE is also non-negative but this is not true for ensuing mutual information, which, however, serves as a conceptual basis for the definition of RTE.

\section{R\'{e}nyi transfer entropy and causality \label{Sec.4.aa}}
%%%%%%%%%%%%%%%%%%%%%%%%%%%%%%%%%%%%%%%%%%%%%%%

As already seen, R\'{e}nyi TE (analogously as Shannon TE) is a directional measure of information transfer.
Let us now  comment on the connection of the RTE with the causality concept.

%%%%%%%%%%%%%%%%%%%%%%%%%%%%%%%%%
\subsection{Granger causality --- Gaussian variables}
%%%%%%%%%%%%%%%%%%%%%%%%%%%%%%%%%%

The first general definition of causality, which could be quantified and measured computationally was given
by Wiener in 1956, namely ``{\em ... For two simultaneously measured signals, if we can predict the first signal better by using
the past information from the second one than by using the information without it, then we call the second signal causal
to the first one...}''~\cite{Wiener}.

The introduction of the concept of causality into the experimental practice, namely into analyses of data observed in consecutive time instants (i.e., time series) is, however, due to Nobel prize winner (economy, 2003) C.W.J.~Granger.
The so-called {\em Granger causality} is defined such that 
process $Y_{t}$ {\em Granger causes} another process $X_{t}$ if, in an appropriate
statistical sense, $Y_{t}$
assists in predicting the future of $X_t$
beyond the degree to which $X_t$ already predicts its own
future. 

%future values of $X_t$ can be better
%predicted using the past values of $Y_t$ and $X_t$ rather than only past values of $X_t$.

The standard test of Granger causality was developed by Granger himself~\cite{Granger} and it is based on a linear regression model, namely
\begin{eqnarray}
X_t \  = \ a_{0t} \ + \ \sum_{\ell=1}^k a_{1\ell} X_{t-\ell} \ + \ \sum_{\ell=1}^l a_{2\ell} Y_{t-\ell} \ + \ e_t\, ,
\label{Granger}
\end{eqnarray}
%
%
%\begin{equation}
 %   Y_t \ = \ a_0 +\sum_{i=1}^l a_i Y_{t-i} +\sum_{j=1}^k b_j X_{t-j}+ e_t\, ,
  %  \label{Granger}
%\end{equation}
%
where $a_{0\ell}, a_{1\ell}, a_{2\ell}$ are (constant) regression coefficients, $l$ and $k$ represent the maximum number of lagged observations included in the model (i.e., memory indices),
%a number of time lags,
$t$ is a discrete time with the time step $\tau$ ($\ell$ is also quantified in units of $\tau$)
%with possible discrete values $n+1,\ldots ,N$ 
and $e_t$ is uncorrelated random variable (residual) with zero mean and variance $\sigma^2$. The {\em null hypothesis} that $Y_t$ does not cause $X_t$ (in the sense of Granger) is accepted  if and only if  $a_{2\ell} =0$ for $\ell =1,\ldots, l$. In the latter case we will call the ensuing regression model as {\em reduced regression model}. 

It is not difficult to show that for Gaussian variables, the RTE and Granger causality are entirely equivalent. To see this we use the {\em standard measure} of Granger causality, which is defined as~\cite{Geweke:1982a}
\begin{eqnarray}
\mathcal{F}^{(k,l)}_{Y\rightarrow X} \ = \ \log_2 \frac{|\Sigma(e'_t)|}{|\Sigma(e_t)|}\, ,
\label{17.cvb}
\end{eqnarray}
where $\Sigma(\ldots)$ is the covariance matrix,  $|\ldots|$ denotes the matrix determinant and $e_t$, $e'_t$
are residuals in the full and reduced regression model, respectively. We have chosen the logarithm to the base $2$,  rather than  $e$ for a technical convenience. We now prove the following theorem:\\

\noindent {\bf{Theorem~3.1}} {\em If the joint process $X_t$, $Y_t$ is Gaussian then there is an exact equivalence between the Granger causality and RTE, namely
\begin{eqnarray}
\mathcal{F}^{(k,l)}_{Y\rightarrow X} \ = \ 2 T^R_{\alpha,Y\rightarrow X}(k,l)\, .
\label{18.aa}
\end{eqnarray}\\
}
This can be proved in the following way. We first define {\em partial covariance} as
\begin{eqnarray}
\Sigma(\bf{X}|\bf{Y}) \ = \ \Sigma(\bf{X}) \ - \ \Sigma(\bf{X},\bf{Y})\Sigma(\bf{Y})^{-1} \Sigma(\bf{X},\bf{Y})^{\top}\, ,
\end{eqnarray}
where ${\Sigma(\bf{X})}_{ij} = \mbox{cov}(X_i,X_j)$ and ${\Sigma(\bf{X},\bf{Y})}_{ij} =\mbox{cov}(X_i,Y_j)$
with ${\bf{X}}$ and ${\bf{Y}}$ being random vector (or multivariate) variables. Let ${\bf{X}}$ and ${\bf{Y}}$ be jointly distributed random vectors in the linear regression model  
\begin{eqnarray}
{\bf{X}} \ = \ {\mathbf{a}} \ + \ {\bf{Y}} \mathbb{A}  \ + \  {\bf{e}}\, .
\label{19.ccc}
\end{eqnarray}
Here ${\mathbf{a}}$ is a constant vector, $\mathbb{A}$ contains regression coefficients and  {\bf{e}} is a residual random vector with zero mean.

We now apply the least square method to the mean square error 
\begin{eqnarray}
{\mathcal{E}}^2 \ \equiv \  \sum_i \mathbb{E}(e_i^2)  \ = \ \sum_i \mathbb{E}\left[({\bf{X}} - {\bf{Y}} \mathbb{A} - {\mathbf{a}})_i^2 
 \right]\, ,\end{eqnarray}
Here $\mathbb{E}(\ldots)$ denotes average value.  The ensuing least square  equations
\begin{eqnarray}
\frac{\partial {\mathcal{E}}^2}{\partial \mathbb{A}_{ij}} \ = \ 0 \;\;\;\; \mbox{and} \;\;\;\; \frac{\partial {\mathcal{E}}^2}{\partial {a}_k} \ = \ 0\, ,
\end{eqnarray}
yield that
\begin{eqnarray}
&& a_l \ = \ { \mathbb{E}}(X_l) \ - \ \sum_k{\mathbb{E}}(Y_k)\mathbb{A}_{kl}\, ,  \label{22.cvb}\\[2mm]
&&\mathbb{A}_{li} \ = \ \sum_j[\Sigma({\bf{X}})]^{-1}_{lj} \ \! \Sigma({\bf{Y}},{\bf{X}})_{ji}\, .
\label{23.cvb}
\end{eqnarray}
From (\ref{19.ccc}) follows that 
\begin{eqnarray}
\mathbb{E}(X_i X_j) \ = \ \mathbb{E}\left[ ({\mathbf{a}} \ + \  {\bf{Y}} \mathbb{A}  \ + \  {\bf{e}})_i({\mathbf{a}} \ + \ {\bf{Y}} \mathbb{A}  \ + \  {\bf{e}})_j\right] \, ,
\end{eqnarray}
which after employing (\ref{22.cvb}) can be equivalently rewritten as
\begin{eqnarray}
{\mbox{cov}}(X_i,X_j) \ = \ \sum_{l,k}{\mbox{cov}}(Y_l,Y_k) \mathbb{A}_{li}\mathbb{A}_{kj}
 \ + \ \mbox{cov}(e_i,e_j)\, ,
\end{eqnarray}
or equivalently
\begin{eqnarray}
\Sigma({\bf{X}}) \ = \ \mathbb{A}^{\top} {\Sigma}({\bf{Y}}) \mathbb{A} \ + \ \Sigma({\bf{e}})\, .
\label{26.cc}
\end{eqnarray}
If we now insert (\ref{23.cvb}) to (\ref{26.cc}) we obtain 
\begin{eqnarray}
{\mbox{cov}}(e_i,e_j) \ = \ {\mbox{cov}}(X_i,X_j) \ - \ {\mbox{cov}}(X_i,Y_k) [{\mbox{cov}}(Y_k,Y_i)]^{-1} [{\mbox{cov}}(X_i,Y_j)]^{\top}\, ,
\end{eqnarray}
which might be equivalently written as
\begin{eqnarray}
\Sigma({\bf{e}}) \ = \  \Sigma(\bf{X}|\bf{Y}) \, .
\end{eqnarray}
If we now take ${\bf{X}} = (X_{t_{n+1}})$, ${\bf{a}} = (a_{0t_n})$,  ${\bf{Y}} = (X^{(k)}_n, Y^{(l)}_n)$, ${\mathbb{A}} = \mbox{diag}(a_{1n}^{(k)}, a_{2n}^{(l)})$  for full regression model and  ${\bf{Y}} = (X^{(k)}_n)$, ${\mathbb{A}} = \mbox{diag}(a_{1n}^{(k)})$ for reduced regression model,
we might write that
\begin{eqnarray}
\mathcal{F}^{(k,l)}_{Y\rightarrow X} \ = \ \log_2 \frac{|\Sigma(e'_t)|}{|\Sigma(e_t)|} \ = \ \log_2 \left(\frac{|\Sigma(X_{t_{n+1}}|X^{(k)}_n)|}{|\Sigma(X_{t_{n+1}}|X^{(k)}_n,Y^{(l)}_n)|} \right)\, .
\label{30.cv}
\end{eqnarray}
At this stage we can use the fact that RE of multivariate Gaussian variable ${\bf{X}}$ is~\cite{Jizba:2015} 
\begin{eqnarray}
H_{\alpha}({\bf{X}}) \ = \ \frac{1}{2} \log_2 |\Sigma({\bf{X}})| \ + \ \frac{D_{_{\bf{X}}}}{2} \log_2 \left(2\pi \alpha^{\alpha'/\alpha} \right)\, .
\label{31.bch}
\end{eqnarray}
Here $D_{_{\bf{X}}}$ is the dimension of ${\bf{X}}$ and $\alpha'$ is a H\"{o}lder dual variable to $\alpha$ (i.e., $1/\alpha + 1/\alpha' = 1$). In particular, for jointly multivariate Gausian variables ${\bf{X}}$ and ${\bf{Y}}$ we can use (\ref{12.gb}) to write
\begin{eqnarray}
H_{\alpha}({\bf{X}}|{\bf{Y}})  &=&  \left[\frac{1}{2}\log_2|\Sigma({\bf{X}}\oplus{\bf{Y}})| + \frac{D_{_{\bf{X}}} + D_{_{\bf{Y}}}}{2} \log_2 \left(2\pi \alpha^{\alpha'/\alpha} \right)\right]\nonumber\\[2mm] 
&-& \left[\frac{1}{2}\log_2|\Sigma({\bf{Y}})| + \frac{D_{_{\bf{Y}}}}{2} \log_2 \left(2\pi \alpha^{\alpha'/\alpha} \right)\right] 
\nonumber\\[2mm] 
&=& \frac{1}{2}\log_2|\Sigma({\bf{X}}|{\bf{Y}})| \ + \ 
\frac{D_{_{\bf{X}}}}{2} \log_2 \left(2\pi \alpha^{\alpha'/\alpha} \right)\, .
\end{eqnarray}
Here $\oplus$ denotes  direct sum. Employing finally the defining relation (\ref{RTE}), we obtain 
\begin{eqnarray}
T^R_{\alpha,Y\rightarrow X}(k,l) \ &=& \ H_{\alpha}(X_{t_{n+1}}|X_n^{(k)}) \ - \ H_{\alpha}(X_{t_{n+1}}|X_n^{(k)},Y_n^{(l)}) \nonumber \\[2mm]
&=& \frac{1}{2} \log_2 \left(\frac{|\Sigma(X_{t_{n+1}}|X^{(k)}_n)|}{|\Sigma(X_{t_{n+1}}|X^{(k)}_n,Y^{(l)}_n)|} \right)\, .
\label{33.vb}
\end{eqnarray}
This confirms the statement of Theorem~3.1. 
%
%The equation (\ref{Granger}) can be reformulated in terms of probability theory. Let us denote with $P(x_{t+1}|x_t^{(k)},y_t^{(l)})$ the cumulative distribution function of the process $X$ conditional on the joint $(k,l)$-history of both processes. Let $P(x_{t+1}|x_t^{(k)})$ denote the distribution function of  $X$ conditional on just its own $l$-history. Then the process $Y_t$ is said to Granger-cause process $X_t$ (with memory indices $k,l$) iff 
%
%\begin{equation}
%\label{GC}
    %P(x_{t+1}|x_t^{(k)},y_t^{(l)}) \ \neq \ P(x_{t+1}|x_t^{(k)})\, .
%\end{equation}
%
%This is basically a deviation from Markov order which lies in the definition of Shannon transfer entropy. Therefore, transfer entropy can be used as a statistic test for Granger causality. On the other hand, Granger causality can only detect linear dependencies between variables, thus transfer entropy holds as its generalization.
%
%**********
%
In addition, since the standard measure of Granger causality (\ref{17.cvb}) is typically defined only for univariate target and source variables $X_t$ and $Y_t$, we can omit $|\ldots|$ in (\ref{30.cv}) and (\ref{33.vb}).

Theorem~3.1 deserves two comments. First, the theorem is clearly true for any $\alpha$. In fact,  it is $\alpha$ independent, which means that for Gaussian processes we can employ any of RTE's to
test Granger causality. When TE is phrased in terms of Shannon entropy, it is typically easier to use various multivariate autoregressive model fitting techniques (e.g., Lewinson--Wiggins--Robinson algorithm or the least-squares linear regression approach~\cite{Seth:2010})
to derive 
%$\Sigma(e_t)$ and $\Sigma(e'_t)$, and thus also 
$\mathcal{F}^{(k,l)}_{Y\rightarrow X}$
more efficiently than by employing direct entropy/mutual information-based estimators. On the other hand, since the efficiency and robustness of RTE estimators crucially hinges on the parameter $\alpha$ employed~\cite{Jizba:2014c} (see also our discussion in Section~\ref{Sec.3.cc}), it might be in many cases easier to follow the information theoretic route to Granger causality
(provided the Gaussian framework is justified). One can even test the Gaussian assumption in actual time series by determining the RTE for various $\alpha$ parameters and check if the results are $\alpha$ independent.

Second, the exact equivalence between the Granger causality and RTE can be in the Gaussian case retraced to the fact that in Eq.~(\ref{31.bch}) the second additive term on the RHS is proportional to $D_{\bf{X}}$.
It is not difficult to see (by a direct inspection) that this proportionality 
will be preserved also in many other exponential distributions that satisfy Markov factorization property. In these cases the
equivalence  between the Granger causality and RTE statistics will also be preserved. However, for generic distributions the additive term in (\ref{31.bch}) will no longer be a linear function of $D_{\bf{X}}$ and hence it will not get cancelled. This, in turn, spoils the desired equivalence. 
In the following section we will discuss a possible generalization  of Theorem~3.1  in the context of heavy tailed distributions.

%In network theory one can confront \textit{Causal Markov condition} and so on..... More or less all of these concepts can be traced to the \textit{probabilistic causation} which states that causes raise the probability of their effects, i.e.
%
%\begin{equation}
    %Pr(\textrm{effect}|\textrm{cause}) > Pr(\textrm{effect}|\textrm{cause didn't appear})
%\end{equation}
%
%(for deterministic causation cause would raise %probability of effect to the value of 1). %Therefore, for causality description it is %convenient to use probabilistic approach, in %particular conditional probability. 

%We refer the reader to~\cite{Chicharro:2012a,Lizier:2010a,Ay:2008a} for further discussion of the complex relationship
%between concepts of information transfer and causality. 

%%%%%%%%%%%%%%%%%%%%%%%%%%%%%%%%%%%%%%%%%%%%%%%%%%%%%%%%%
\subsection{Granger causality --- heavy tailed variables}
%%%%%%%%%%%%%%%%%%%%%%%%%%%%%%%%%%%%%%%%%%%%%%%%%%%%%%%%

It is not difficult to find relations analogous to (\ref{33.vb}) also in more general setting. Here we will illustrate this point with heavy tailed (namely $\alpha$-Gaussian) random variables  where computations can be done analytically.

It is well known that if variance and mean are the only
statistical observables, then the conventional maximum entropy principle (MaxEnt) based on Shannon entropy yields
Gaussian distribution. Similarly, if the very same MaxEnt is applied to R\'{e}nyi entropy $H_{\alpha}$ one obtains the so-called $\alpha$-Gaussian distribution~\cite{JA}
\begin{eqnarray}
p_i \ = \ \frac{1}{Z_q} \left[1-\beta(\alpha-1)x_i^2\right]_+^{1/(\alpha-1)}\, ,
\label{34.vb}
\end{eqnarray}
that decays
asymptotically following power law. Here $[z]_+ = z$ if $z\ge 0$ and $0$ otherwise, $Z_q$ is  the normalization factor. It is more conventional to write (\ref{34.vb}) as
\begin{eqnarray}
p_i \ = \ Z_q^{-1}\ \! \exp_{\{2-\alpha\}}{(-\beta x^2_i)} \, ,
\end{eqnarray}
where 
\begin{eqnarray}
e^{x}_{\{\alpha\}} \ = \ \left[1 \ + \ (1-\alpha)x\right]^{1/(1-\alpha)}_+\, ,
\end{eqnarray}
is the Box--Cox $\alpha$-exponential~\cite{Tsallis:book}. $\alpha$-Gaussian distribution (\ref{34.vb}) has finite variance (and more generally covariance matrix) for $\frac{D}{2+D} < \alpha \le 1$.
%
%Here covariance matrix is finite only for $D < 2\alpha/(1-\alpha)$.
%
Let us now assume that Granger's linear  (full/reduced) regression model is described  by joint processes $X_t$ and $Y_t$ that are $\alpha$-Gaussian. We now prove the following theorem:\\

\noindent {\bf{Theorem~3.2}} {\em If the joint process $X_t$, $Y_t$ is $\alpha$-Gaussian with $\alpha \in \left(\frac{1 + k +l }{3+k+l},1\right]$ (i.e., finite covariance matrix region) then %there is an exact equivalence between 
$\mathcal{F}^{(k,l)}_{Y\rightarrow X} - 2T^R_{\alpha,Y\rightarrow X}(k,l)$ is a monotonically decreasing function of $\alpha$ (at fixed $k$ and $l$) with zero reached at a stationary point $\alpha =1$. The  leading-order correction to the Granger causality is ``$k$'' independent and has the form
\begin{eqnarray}
\mathcal{F}^{(k,l)}_{Y\rightarrow X}    \ = \   2T^R_{\alpha,Y\rightarrow X}(k,l)  \ + \ \frac{l(\alpha-1)^2}{4} \ + \ \mathcal{O}((\alpha-1)^3) \, .
\label{37.aa}
\end{eqnarray}
}

This result explicitly illustrates how certain ''soft''  heavy-tailed processes can be related to the concept of the Granger causality via universal type of corrections that are principally discernible in
data analysis.

Theorem~3.2 can be proved in a close analogy with our proof of Theorem~3.1. In fact,
all steps in the proof are identical up to Eq.~(\ref{30.cv}). For $D$-dimensional $\alpha$-Gaussian process the scaling property~(\ref{31.bch})  reads 
\begin{eqnarray}
H_{\alpha}({\bf{X}}) \ = \ \frac{1}{2} \log_2 |\Sigma({\bf{X}})| \ + \ H_{\alpha} ( {\bf{Z}}_{\alpha}^{{\mathbf{1}},D} )\, .
\label{38.bch}
\end{eqnarray}
Here ${\bf{Z}}_{\alpha}^{{\mathbf{1}},D}$ 
represents an $\alpha$-Gaussian random vector with zero mean and unit ($D\times D$) covariance  matrix.
Relation (\ref{38.bch}) results from the following chain of identities
\begin{eqnarray}
H_{\alpha}({\bf{X}}) \! \!&=&\!\!  H_{\alpha}( \sqrt{\Sigma({\bf{X}})} \ \! {\bf{Z}}_{\alpha}^{{\mathbf{1}},D}) \nonumber \\[2mm]
&=& \!\! \frac{1}{1-\alpha}\log_2 \int_{{\mathbb{R}}^D} d^D{\bf{y}}\left( \int_{{\mathbb{R}}^D} d^D{\bf{z}} \  \delta\left({\bf{y}} - \sqrt{\Sigma({\bf{X}})} \ \! {\bf{z}}\right) {\mathcal{F}}({\bf{z}}) \right)^{\!\alpha}\nonumber \\[2mm]
&=& \!\! \frac{1}{1-\alpha}\log_2\left[ |\Sigma({\bf{X}})|^{(1-\alpha)/2} \int_{{\mathbb{R}}^D} d^D{\bf{y}} \ \! {\mathcal{F}}^{\alpha}({\bf{y}})\right] \nonumber \\[2mm]
&=&\! \!\frac{1}{2} \log_2 |\Sigma({\bf{X}})| \ + \ H_{\alpha} ( {\bf{Z}}_{\alpha}^{{\mathbf{1}},D} )\, ,
\label{39.jk}
\end{eqnarray}
which is clearly valid for any non-singular covariance matrix. 
In the derivation  ${\mathcal{F}}(\ldots)$ denoted the $\alpha$-Gaussian probability density function with unit covariance matrix and zero mean. 
We can now use the simple fact that
\begin{eqnarray}
 &&\mbox{\hspace{-15mm}}H_{\alpha} ( {\bf{Z}}_{\alpha}^{{\mathbf{1}},D} )  =  \log_2 \left[ \left(\frac{\pi}{\mathfrak{b}(1-\alpha)}  \right)^{D/2} \ \! \frac{\Gamma\left(\frac{1}{1-\alpha}-\frac{D}{2} \right)}{\Gamma\left(\frac{1}{1-\alpha} \right)} \ \!  \left(1- \frac{D}{2\alpha}(1-\alpha)\right)^{1/(\alpha-1)}\right]\nonumber \\[2mm]
 &&\mbox{\hspace{-15mm}}= \frac{D}{2} \log_2\left[ {2\pi\alpha}\right] \ + \ \log_2 \left[ \frac{\Gamma\left(\frac{1}{1-\alpha}-\frac{D}{2} \right)}{{(1-\alpha)^{D/2}}\Gamma\left(\frac{1}{1-\alpha} \right)}\right]  \ + \ 
 \log_2\left[\left(1 - \frac{D}{2\alpha} (1-\alpha)  \right)^{\frac{D}{2} - \frac{1}{1-\alpha}}\right] ,
 \label{41.cg}
\end{eqnarray}
(where $\mathfrak{b} = [2\alpha -  D(1-\alpha)]^{-1}$) to write
\begin{eqnarray}
H_{\alpha}({\bf{X}}|{\bf{Y}})  &=& \frac{1}{2}\log_2|\Sigma({\bf{X}}|{\bf{Y}})| \ + \ H_{\alpha}( {\bf{Z}}_{\alpha}^{{\mathbf{1}},D_{_{\bf{X}}}+ D_{_{\bf{Y}}}} ) 
\ - \ H_{\alpha} ( {\bf{Z}}_{\alpha}^{{\mathbf{1}},D_{_{\bf{Y}}}} ) 
\, .
\end{eqnarray}
At this stage we note that
\begin{eqnarray}
H_{\alpha} ( {\bf{Z}}_{\alpha}^{{\mathbf{1}},D_{_{\bf{X}}}+ D_{_{\bf{Y}}}} )
\!  &-& \! H_{\alpha} ( {\bf{Z}}_{\alpha}^{{\mathbf{1}},D_{_{\bf{Y}}}} )  \ - \ H_{\alpha} ( {\bf{Z}}_{\alpha}^{{\mathbf{1}},D_{_{\bf{X}}}} ) \nonumber \\[2mm] &=& \! H_{\alpha} ( {\bf{Z}}_{\alpha}^{{\mathbf{1}},D_{_{\bf{X}}}}| {\bf{Z}}_{\alpha}^{{\mathbf{1}},D_{_{\bf{Y}}}}) \ - \  H_{\alpha} ( {\bf{Z}}_{\alpha}^{{\mathbf{1}},D_{_{\bf{X}}}} )\, , 
\label{43aa}
\end{eqnarray}
which is not zero as it was in the case of Gaussian distribution. In fact, from the foregoing discussion it is clear that for the $\alpha$-Gaussian random variables we can write the RTE in the form
\begin{eqnarray}
&&\mbox{\hspace{-13mm}}T^R_{\alpha,Y\rightarrow X}(k,l) \ = \ H_{\alpha}(X_{t_{n+1}}|X_n^{(k)}) \ - \ H_{\alpha}(X_{t_{n+1}}|X_n^{(k)},Y_n^{(l)}) \nonumber \\[2mm]
&&= \ \frac{1}{2} \log_2 \left(\frac{\Sigma(X_{t_{n+1}}|X^{(k)}_n)}{\Sigma(X_{t_{n+1}}|X^{(k)}_n,Y^{(l)}_n)} \right) %\nonumber \\[2mm]  &&
\ + \ H_{\alpha} ( {\bf{Z}}_{\alpha}^{{{1}},1}| {\bf{Z}}_{\alpha}^{{\mathbf{1}},k}) \ - \ H_{\alpha} ( {\bf{Z}}_{\alpha}^{{{1}},1}| {\bf{Z}}_{\alpha}^{{\mathbf{1}},k+l}) \nonumber \\[2mm]
&&= \ \frac{1}{2}\mathcal{F}^{(k,l)}_{Y\rightarrow X} \ + \ I_{\alpha}({\bf{Z}}_{\alpha}^{{{1}},1}:{\bf{Z}}_{\alpha}^{{\mathbf{1}},l}|{\bf{Z}}_{\alpha}^{{\mathbf{1}},k})\, .
\label{44.vbc}
\end{eqnarray}
Here we have set ${\bf{Z}}_{\alpha}^{{{1}},1}$ to correspond to the random variable $X_{t_{n+1}}$ with unit variance. Similarly, ${\bf{Z}}_{\alpha}^{{\mathbf{1}},k}$ and ${\bf{Z}}_{\alpha}^{{\mathbf{1}},l}$ correspond to unit covariance random variables $X^{(k)}_n$ and $Y^{(l)}_n$, respectively.  

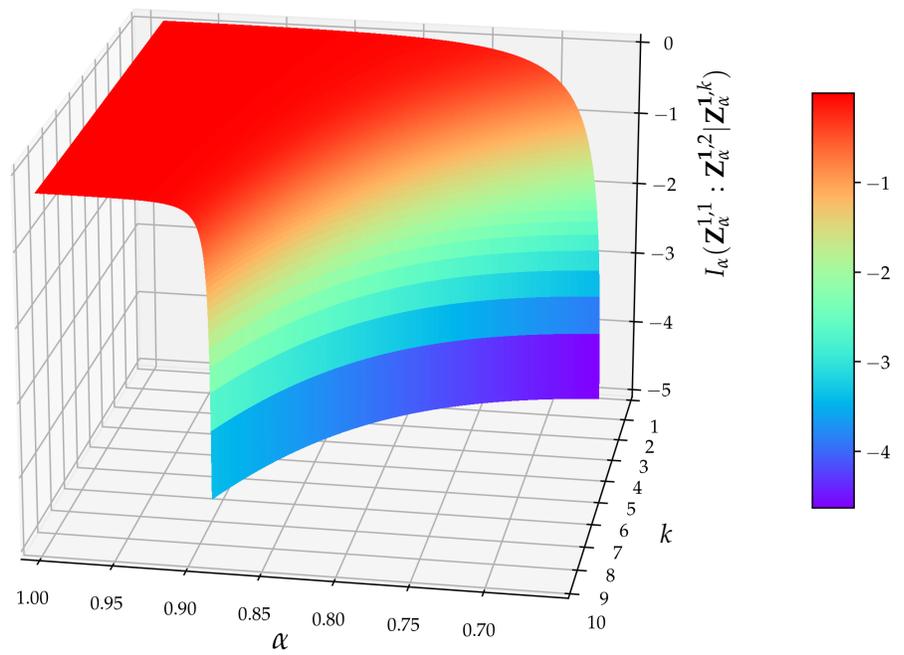
\begin{figure}
\begin{center}
\includegraphics[width=0.7\textwidth]{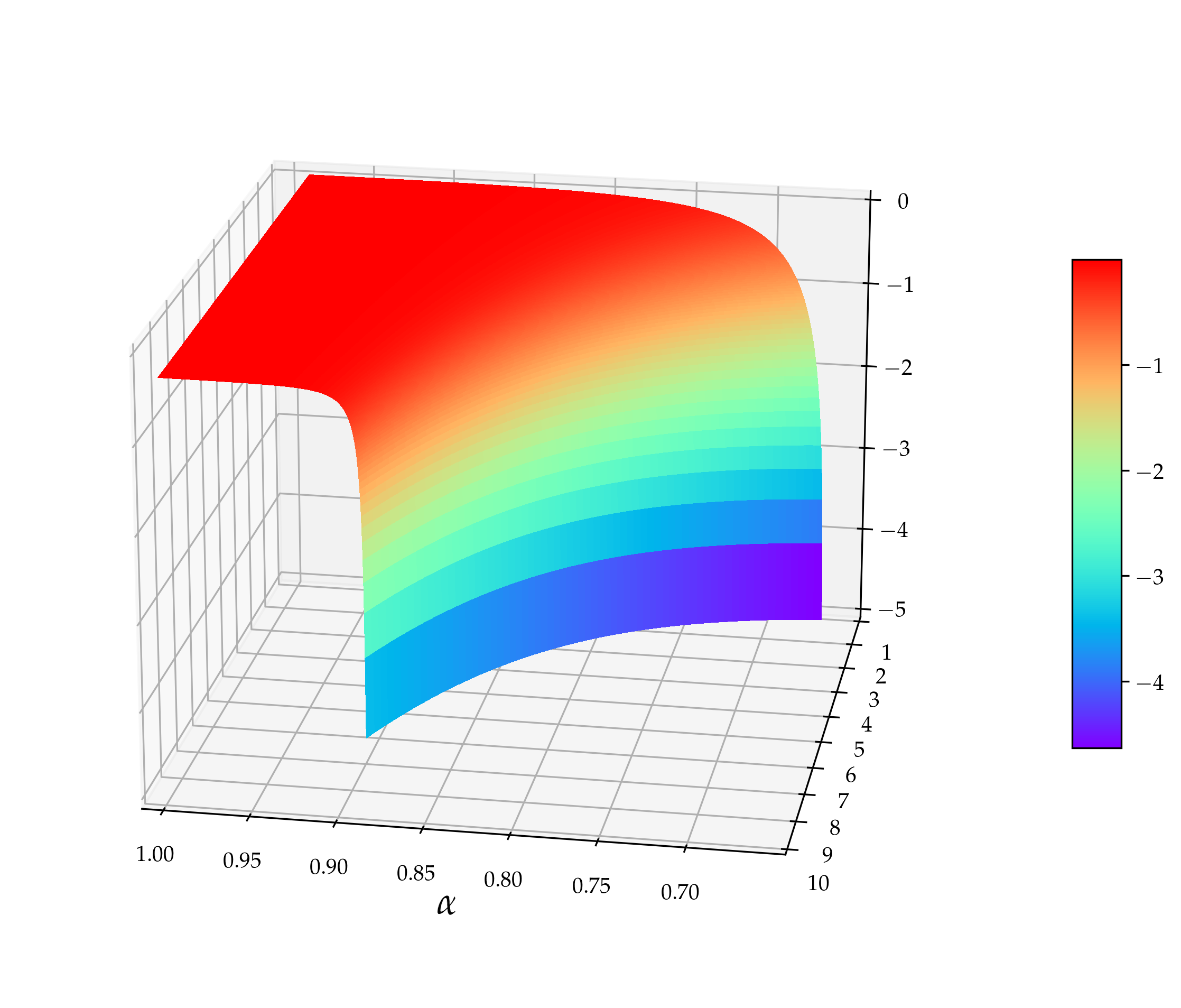}
\begin{picture}(10,4)
\put(-95,70){{$k$} } \put(-81,170){{\rotatebox{90}{$I_{\alpha}({\bf{Z}}_{\alpha}^{{{1}},1}:{\bf{Z}}_{\alpha}^{{\mathbf{1}},2}|{\bf{Z}}_{\alpha}^{{\mathbf{1}},k})$}} } 
\end{picture}
\vspace{-5mm}
\caption{Example of $I_{\alpha}({\bf{Z}}_{\alpha}^{{{1}},1}:{\bf{Z}}_{\alpha}^{{\mathbf{1}},l}|{\bf{Z}}_{\alpha}^{{\mathbf{1}},k})$ for $l=2$ and $k = 1,2, \ldots, 10$. Range validity of $\alpha$ is thus between $\frac{3 + k}{5 +k}$ and $1$. }
\label{fig.2.cc}
\end{center}
\end{figure}

Clearly, when $Y_t$ and $X_t$ processes are independent (and hence {\em not causal} in Granger sense), their joint distribution factorizes and thus  $H_{\alpha} ( {\bf{Z}}_{\alpha}^{{\mathbf{1}},D_{_{\bf{X}}}+ D_{_{\bf{Y}}}} ) \mapsto H_{\alpha} ( {\bf{Z}}_{\alpha}^{{\mathbf{1}},D_{_{\bf{X}}}}\times  {\bf{Z}}_{\alpha}^{{\mathbf{1}},D_{_{\bf{Y}}}})$. Additivity of the RE then ensures that $H_{\alpha} ( {\bf{Z}}_{\alpha}^{{{1}},1}| {\bf{Z}}_{\alpha}^{{\mathbf{1}},k}) \! =\! H_{\alpha} ( {\bf{Z}}_{\alpha}^{{{1}},1}| {\bf{Z}}_{\alpha}^{{\mathbf{1}},k+l})$  and hence $I_{\alpha}({\bf{Z}}_{\alpha}^{{{1}},1}:{\bf{Z}}_{\alpha}^{{\mathbf{1}},l}|{\bf{Z}}_{\alpha}^{{\mathbf{1}},k})$ is zero. In other words, when two processes are not Granger causal their RTE is zero. Actually, it is not difficult to see that this is true irrespective of a specific form of the distribution involved. Opposite is, however, not true since $I_{\alpha}({\bf{Z}}_{\alpha}^{{{1}},1}:{\bf{Z}}_{\alpha}^{{\mathbf{1}},l}|{\bf{Z}}_{\alpha}^{{\mathbf{1}},k})$ might be (unlike in  Shannon's case) negative, and consequently $T^R_{\alpha,Y\rightarrow X}(k,l) $ can be zero even if $\mathcal{F}^{(k,l)}_{Y\rightarrow X}$ is not. To understand this point better we explicitly evaluate  $I_{\alpha}({\bf{Z}}_{\alpha}^{{{1}},1}:{\bf{Z}}_{\alpha}^{{\mathbf{1}},l}|{\bf{Z}}_{\alpha}^{{\mathbf{1}},k})$ for our $\alpha$-Gaussian random variables. Using (\ref{41.cg}) we can write
\begin{eqnarray}
\mbox{\hspace{-10mm}}I_{\alpha}({\bf{Z}}_{\alpha}^{{{1}},1}:{\bf{Z}}_{\alpha}^{{\mathbf{1}},l}|{\bf{Z}}_{\alpha}^{{\mathbf{1}},k}) &=&  \log_2\left[\frac{\Gamma\left(\frac{1}{1-\alpha} - \frac{1+k}{2}  \right)}{\Gamma\left( \frac{1}{1-\alpha} - \frac{k}{2}  \right)}
\frac{\Gamma\left( \frac{1}{1-\alpha} - \frac{k+l}{2}  \right))}{\Gamma\left(\frac{1}{1-\alpha} - \frac{1+k+l}{2}   \right)}
\right]\nonumber \\[2mm]
&+& \log_{2} \left[\frac{\left(\frac{\alpha}{1-\alpha} - \frac{1+k}{2}\right)^{\frac{1+k}{2} - \frac{1}{1-\alpha}}}{\left(\frac{\alpha}{1-\alpha} - \frac{k}{2}\right)^{\frac{k}{2} - \frac{1}{1-\alpha}}} 
\frac{\left(\frac{\alpha}{1-\alpha} - \frac{k+l}{2}\right)^{\frac{k+l}{2} - \frac{1}{1-\alpha}}}{\left(\frac{\alpha}{1-\alpha} - \frac{1+k+l}{2}\right)^{\frac{1+k +l}{2} - \frac{1}{1-\alpha}}} 
\right].
\label{45.hj}
\end{eqnarray}
By setting $\zeta = \frac{1}{1-\alpha} -\frac{k}{2}$ and $\xi = \frac{1}{1-\alpha} - \frac{k+l}{2}$, we can rewrite (\ref{45.hj}) as
\begin{eqnarray}
I_{\alpha}({\bf{Z}}_{\alpha}^{{{1}},1}:{\bf{Z}}_{\alpha}^{{\mathbf{1}},l}|{\bf{Z}}_{\alpha}^{{\mathbf{1}},k}) &=& \log_2\left[\frac{\Gamma\left(\zeta-\frac{1}{2}\right)}{\Gamma\left(\zeta \right)} \frac{(\zeta -1)^{\zeta} }{\left(\zeta - \frac{3}{2}\right)^{\zeta - \frac{1}{2}}}  
\frac{\Gamma\left( \xi\right)}{\Gamma\left( \xi - \frac{1}{2}\right)} \frac{\left(\xi -\frac{3}{2}\right)^{\xi -\frac{1}{2}}}{(\xi -1)^{\xi}}
\right]\nonumber \\[2mm]
&=& \log_2\left[\frac{\Gamma\left(\zeta-\frac{3}{2}\right)}{\Gamma\left(\zeta -1\right)} \frac{(\zeta -1)^{\zeta-1} }{\left(\zeta - \frac{3}{2}\right)^{\zeta - \frac{3}{2}}}  
\frac{\Gamma\left( \xi-1\right)}{\Gamma\left( \xi - \frac{3}{2}\right)} \frac{\left(\xi -\frac{3}{2}\right)^{\xi -\frac{3}{2}}}{(\xi -1)^{\xi-1}}
\right]\nonumber \\[2mm]
&\leq& -\frac{1}{2} \log_2\left[ \frac{(\xi -1)}{\left(\xi -\frac{3}{2}\right)}  \right] \ \leq \ 0\, ,
\end{eqnarray}
where on the last line we have used Ke\v{c}ki\'{c}--Vasi\'{c} inequality~\cite{Keckic:1971c}
\begin{eqnarray}
\frac{(x+1)^{x+1}}{(x+s)^{x+s}}\ \!e^{s-1} \ \leq \ \frac{\Gamma(x+1)}{\Gamma(x+s)} \ \leq \  \frac{(x+1)^{x+\frac{1}{2}}}{(x+s)^{x+s-\frac{1}{2}}} \ \! e^{s-1}\, ,
\end{eqnarray}
valid for $s \in (0,1)$. In addition, it can be numerically checked that $\frac{d I_{\alpha}({\bf{Z}}_{\alpha}^{{{1}},1}:{\bf{Z}}_{\alpha}^{{\mathbf{1}},l}|{\bf{Z}}_{\alpha}^{{\mathbf{1}},k})}{d\alpha} >0$, for all $l,k$ from the definition, so the maximum of $I_{\alpha}({\bf{Z}}_{\alpha}^{{{1}},1}:{\bf{Z}}_{\alpha}^{{\mathbf{1}},l}|{\bf{Z}}_{\alpha}^{{\mathbf{1}},k})$ is attained at $\alpha = 1$, see Fig.~\ref{fig.2.cc}. 
When $\alpha$ is close to $1$ then one can employ the asymptotic relation $\Gamma[x + \gamma] \sim \Gamma[x]x^{\gamma}$ valid for $x \gg 1$, $\gamma \in \mathbb{C}$, and rewrite (\ref{41.cg}) in the form $(D/2)\log_{2}[2\pi\alpha e^{\alpha}]$. In this case (\ref{45.hj}) tends to zero and we obtain equivalence between TE and Granger causality.  This result should not be so surprising because in the limit $\alpha \rightarrow 1$ RE tends to Shannon's entropy and $\alpha$-Gaussian distribution tends to Gaussian distribution.
The leading order behavior near $\alpha = 1$ can be obtained directly from (\ref{45.hj}). Ensuing Taylor expansion gives
\begin{eqnarray}
I_{\alpha}({\bf{Z}}_{\alpha}^{{{1}},1}:{\bf{Z}}_{\alpha}^{{\mathbf{1}},l}|{\bf{Z}}_{\alpha}^{{\mathbf{1}},k}) \ = \  -\frac{l(\alpha-1)^2}{8} \ + \ \mathcal{O}((\alpha-1)^3)  \, ,
\end{eqnarray}
so, the point $\alpha =1$ is a {\em stationary point} of $I_{\alpha}({\bf{Z}}_{\alpha}^{{{1}},1}:{\bf{Z}}_{\alpha}^{{\mathbf{1}},l}|{\bf{Z}}_{\alpha}^{{\mathbf{1}},k})$.
This closes the proof.

\section{Estimation of R\'{e}nyi entropy \label{Sec.3.cc}}
%%%%%%%%%%%%%%%%%%%%%%%%%%%%%%%%%%%%%%
%%%%%%%%%%%%%%%%%%%%%%%%%%%%%%%%%%%%%%%%%%%%%%%%%%%%%%
\subsection{RTE and derived concepts}
%%%%%%%%%%%%%%%%%%%%%%%%%%%%%%%%%%%%%%%%%%%%%%%%%%%%%%%%%

From the data analysis point of view it is not very practical to use the full joint processes
$X_n^{(k)}$ and $Y_n^{(l)}$ (cf. the defining relation~(\ref{RTE})) because (possibly) high values of $k$ and $l$ negatively influence accuracy of estimation of RTE. In the following sections we will thus switch to more expedient definition of RTE given by
\begin{eqnarray}
    T^R_{\alpha,Y\rightarrow X}(\{k\},\{m\},\{l\}) \ &=& \ H_{\alpha}(X_{t_{_{n+m}}}|X_n^{\{k\}}) \ - \ H_{\alpha}(X_{t_{_{n+m}}}|X_n^{\{k\}},Y_n^{\{l\}})\nonumber \\[2mm]
    &=& \ I_{\alpha}(X_{t_{_{n+m}}}:Y_n^{\{l\}}|X_n^{\{k\}})\, ,
    \label{RTE_subset}
\end{eqnarray}
where $X_n^{\{k\}}$ is a subset of past values of $X_t$ up to the time $n$ with number of elements equal to $k$, such that $\{k\}=\{\kappa_1,...,\kappa_k\}$ is a set of indices and $X_n^{\{k\}} \equiv X_{t_{{n-\kappa_1}}},X_{t_{{n-\kappa_2}}},\ldots,X_{t_{{n-\kappa_k}}}$ is a selected subsequence of $X_n^{(k)}$. The same notational convention applies to $Y_n^{\{l\}}$. In the definition (\ref{RTE_subset}) we have also added a third parameter $m$ --- the so-called {\em future step}. Though such a parametrization is often used in the literature on Shannon's TE, cf. e.g., Ref.~\cite{Vejmelka}, we will in the following employ only $m=1$ so as to conform with the definition~(\ref{RTE}). In such a case we will often omit the middle index in $T^R_{\alpha,Y\rightarrow X}(\{k\},\{1\},\{l\})$. 
%The notation allows to exclude certain indices so dimension of dataset to calculate R\'{e}nyi entropy and RTE decreases.

%%%%%%%%%%%%%%%%%%%%%%%%%%%%%%%%%%%%%%%%%%%%%%
\subsubsection{Balance of transfer entropy} 
%%%%%%%%%%%%%%%%%%%%%%%%%%%%%%%%%%%%%%%%%%%%%%

In order to compare RTE that flows in the direction from $Y \rightarrow X$ with the RTE that flows in opposite direction $X \rightarrow Y$ we define {\em balance of transfer entropy} 
\begin{equation} 
T^{R,\rm{\footnotesize{~balance}}}_{\alpha,Y\rightarrow X}(\{k\},\{l\}) \ = \ T^R_{\alpha,Y\rightarrow X}(\{k\},\{l\}) \ - \ T^R_{\alpha,X\rightarrow Y}(\{k\},\{l\})\, .
\label{balance transfer entropy}
\end{equation}

%%%%%%%%%%%%%%%%%%%%%%%%%%%%%%%%%%%%%%%%%%%%%
\subsubsection{Effective transfer entropy} 
%%%%%%%%%%%%%%%%%%%%%%%%%%%%%%%%%%%%%%%%%%%%%

To mitigate finite size effects we employ the idea of surrogate time series. To this end we define {\em effective transfer entropy}
\begin{equation} 
T^{R,\rm{\footnotesize{~effective}}}_{\alpha,Y\rightarrow X}(\{k\},\{l\}) \ = \  T^R_{\alpha,Y\rightarrow X}(\{k\},\{l\}) \ - \  T^R_{\alpha,Y^{(\rm{\footnotesize{sur}})}\rightarrow X}(\{k\},\{l\})\, ,
\label{effective transfer entropy}
\end{equation}
where $Y^{(\rm{\footnotesize{sur}})}$ stands for randomized (reordered) time series --- the surrogate data sequence.
Such a series has the same
mean, the same variance, the same autocorrelation function, and therefore the same power spectrum as the original
sequence, but (nonlinear) phase relations are destroyed. In effect, all the potential correlations between $X_n^{\{k\}}$ and $Y_n^{\{l\}}$ are removed, which means that $T^R_{\alpha,Y^{(\rm{\footnotesize{sur}})}\rightarrow X}(\{k\},\{l\})$ should be zero. In practice, this is not  the case, despite
the fact that there is no obvious structure in the data. 
The non-zero value of $T^R_{\alpha,Y^{(\rm{\footnotesize{sur}})}\rightarrow X}(\{k\},\{l\})$  must then be a byproduct of
the finite data set. Definition (\ref{effective transfer entropy}) then ensures that spurious effects caused by finite $k$ and $l$ are removed. For more technical exposition see, e.g. Ref.~\cite{Theiler:1992,Schreiber:1996,schreiber:2000}.

%%%%%%%%%%%%%%%%%%%%%%%%%%%%%%%%%%%%%%%%%%%%%%%%%%%%%%%%
\subsubsection{Balance of effective transfer entropy}
%%%%%%%%%%%%%%%%%%%%%%%%%%%%%%%%%%%%%%%%%%%%%%%%%%%%%%%%

Finally, we combine both previous definitions to form {\em balance effective transfer entropy} 
\begin{eqnarray}
T^{R,\rm{\footnotesize{~balance,~ effective}}}_{\alpha,Y\rightarrow X}(\{k\},\{l\}) \ &=& \ T^{R,\rm{\footnotesize{~effective}}}_{\alpha,Y\rightarrow X}(\{k\},\{l\}) \ - \ T^{R,\rm{\footnotesize{~effective}}}_{\alpha,X\rightarrow Y}(\{k\},\{l\}) \nonumber  \\[2mm] 
&=& \ T^R_{\alpha,Y\rightarrow X}(\{k\},\{l\}) \ - \ T^R_{\alpha,Y^{(\rm{\footnotesize{sur}})}\rightarrow X}(\{k\},\{l\})\nonumber \\[2mm] 
&-& \ T^R_{\alpha,X\rightarrow Y}(\{k\},\{l\}) \ + \ T^R_{\alpha,X^{(\rm{\footnotesize{sur}})}\rightarrow Y}(\{k\},\{l\})\, ,
\label{balance effective transfer entropy}
\end{eqnarray}
to quantify direction of flow of transfer entropy without finite size effects.

%%%%%%%%%%%%%%%%%%%%%%%%%%%%%%%%%%%%%%%%%
\subsection{Estimators employed}
%%%%%%%%%%%%%%%%%%%%%%%%%%%%%%%%%%%%%%%%%

%Typical straightforward solution to calculate RTE would use formula~(\ref{rte}) or its more pragmatic version (\ref{RTE_subset}). If we would follow the path and we would use high $k$ and $l$ memories of timeseries $X_t$ and $Y_t$ then using histogram as an estimator of probability distributions. General pitfall of estimators based on histograms is that they are only effective for low dimensional datasets. In high dimensional cases three kinds of problems can emerge: firstly, allocated memory for the histogram exceeds available memory of a computer; secondly, histogram boxes contain too many boxes with zeros; thirdly, boundaries of boxes can influence estimation. These drawbacks can be sidestepped by estimators based on distance of datapoints. 

Finding good estimators for RE's is still an open research area. First estimators for Shannon entropy based on $\ell$-nearest-neighbor in one-dimensional spaces were studied in statistics already almost 60 years ago by Dobrushin~\cite{Dobrushin} and  Va\v{s}\'{i}\v{c}ek~\cite{Vasicek}. Disadvantage of these estimators is that they can not be easily generalized to higher dimensional spaces and so they are inapplicable to the TE calculations. Nowadays, there are many usable frameworks --- most of them, of course, in Shannonian setting (see, e.g. Ref.~\cite{Kanz-book}, for recent review).  In is, however, important to stress 
that the naive estimation of TE by partitioning of the state space is problematic~\cite{Schreiber} and that such estimators frequently fail to converge to the correct result~\cite{Kaiser:02}. In
practice, more sophisticated techniques such as kernel~\cite{Silverman:86}
or $\ell$–nearest neighbor estimators~\cite{Kraskov:04,Frenzel:07} need to be
utilized. The latter techniques may, however, bring about their own
assumptions about the empirical distribution of the data
(see~\cite{Kaiser:02} for a good discussion of the issues involved).

In our work we use the $\ell$-nearest-neighbor entropy estimator for higher-dimensional spaces introduced by Leonenko {\em et al.}~\cite{Leonenko_Prozanto_Savani,Leonenko_Prozanto}. This estimator  
is suitable for RE and it can be effectively adapted and implemented by using formulas from the above mentioned papers. In particular, the approach is based on an estimator of the RE from a finite sequence of $N$ points that is defined as
\begin{eqnarray} 
 \widehat{H}_{N,\ell,\alpha} \ = \  \left\{ 
 \begin{array}{ll} 
 \alpha\neq 1 &~~~~ 
 \frac{1}{1-\alpha} \left[\log_{B} \left( \left( N - 1\right) \cdot \frac{\Gamma \left(\ell \right)}{\Gamma \left(\ell + 1 - \alpha \right)} \cdot V_{m} \right) \right.\\[3mm] 
 &~~~~ \left. + \ \log_{B} \left(
 \frac{1}{N} \sum_{i=1}^{N} \left( \rho_{\ell}^{(i)} \right)^{m \left(1-\alpha \right)} \right) \right]\, \\[7mm]
 \alpha = 1 &~~~~
  \log_{B} \left( \left( N-1 \right) \cdot \exp \left( - \psi(k)\right) \cdot V_m \right) \\[3mm]
  &~~~~ + \ \log_{B} \left( \frac{m}{N}\sum_{i=1}^{N} \log_{B} \left( \rho_{\ell}^{(i)} \right) \right) 
 \end{array}
 \right.
 .
 \label{eq.11.a}
\end{eqnarray}
Here $\Gamma \left(x \right)$ is Euler's gamma function, $\psi(x) = - {\Gamma '\left(x \right)}/{\Gamma \left(x \right)}$ is (negative) digamma function, $m=\dim{X_t}$ is the dimension of the dataset space $X_t$ and $\rho_{\ell} ^{(i)}$ is distance from data $i$ to $\ell$-th nearest data counterpart using a metric in the space $X_t$. $V_m$ is size of ball in space $X_t$ defined via the same metric. Finally, $\log_{B}$ is logarithm with base $B$ (we typically use $B=e$). In our computations we employ Euclidean metric which has $V_m = {\pi^{\frac{m}{2}}}/{\Gamma \left( \frac{m}{2} + 1 \right)}$.  Note in particular, that the estimator thus basically depends on $N$, i.e., the number of data in a dataset and on $\ell$, i.e., the rank of the nearest neighbor used. 

Advantage of the estimator (\ref{eq.11.a}) in contrast to the standard histogram method is:
\begin{itemize}
    \item relative accuracy for small dataset;
    \item applicability for high dimensional data;
    \item combination of the set estimators provides statistics for estimation.
\end{itemize}
We can also stress that in contrast to other RE estimators, such as {\em fixed-ball} estimator~\cite{Kanz-book}, the estimator (\ref{eq.11.a}) is not confined to any specific range of $\alpha$ values, though the efficiency of the estimator is, of course, $\alpha$ dependent. We will comment more on this point in Section~\ref{Sec.6.cc}. On the other hand, disadvantage of the method is computational complexity of the algorithm and complicated data container.
%However, we believe that advantages beats disadvantages.

To calculate RTE  and the  related quantities (\ref{balance transfer entropy}), (\ref{effective transfer entropy}) and (\ref{balance effective transfer entropy}) we apply the estimator formula (\ref{eq.11.a}).  Ensuing estimators to (\ref{RTE_subset}),  (\ref{balance transfer entropy}), (\ref{effective transfer entropy}) and (\ref{balance effective transfer entropy}) --- let call them generically $\cal{X}$, become dependent on $\ell$ (i.e., the nearest neighbor rank). We exploit this feature and define the mean value $\overline{\cal{X}}$ and standard deviation $\sigma_{_{\cal{X}}}$ with Bessel correction, respectively as
\begin{eqnarray}
    &&\overline{\cal{X}} \ = \ \frac{\sum_{\ell={n_{min}}}^{n_{max}} {\cal{X}}_{\ell}} {n_{max} - n_{min} + 1} \, ,\\[2mm]
    &&\sigma_{_{\cal{X}}} \ = \ \sqrt{ \frac{  \sum_{\ell=1}^{n} \left( {\cal{X}}_{\ell} \ - \ \overline{\cal{X}} \right)^2 }{n_{max} - n_{min}} }\, .
\end{eqnarray}
Here $n_{max}$ and $n_{min}$ is the highest and the lowest order of the nearest data counterpart respectively. Theoretically, we should use $n_{max} = M$, where $M$ stands for number of samples but such a setup would require enormous amount of computer memory to hold the distances. 

In our calculations we used $n_{max}=50$ which turned out to be a good compromise between accuracy and computer time.  On the other hand, as for $n_{min}$ we are little bit restricted by the fact that $n_{min}$ influences the interval of convergence of the estimator for various $\alpha$ (cf. discussion and proof in~\cite{Leonenko_Prozanto}). For instance, for $\ell=1$ the estimator converges in the interval $\alpha \in [0, 1+ \frac{1}{2\dim{(X_t)}}]$, while for $\ell>1$ one has $\alpha \in [0, \frac{\ell+1}{2}]$. For our particular purpose it will suffice to set $n_{min} = 5$, so that the interval of convergence will be $\alpha \in [0, 3]$. This will fully suit our needs.

%%%%%%%%%%%%%%%%%%%%%%%%%%%%%%%%%%%%%%%%%%%%%%%%%
\section{R\"{o}ssler system \label{Sec.5.cc}}
%%%%%%%%%%%%%%%%%%%%%%%%%%%%%%%%%%%%%%%%%%%%%%%%%

%%%%%%%%%%%%%%%%%%%%%%%%%%%%%%%%%%%%%%%%%%
\subsection{Equations for master system}
%%%%%%%%%%%%%%%%%%%%%%%%%%%%%%%%%%%%%%%%%%

In order to illustrate the use of RTE, we consider here two unidirectionally coupled 
R\"{o}ssler systems (oscillators). 
These often serve as a testbed for various  measures of synchronization including Shannon's TE~\cite{Palus2018,Rosenblum:1996,Cheng:2017}. R\"{o}ssler's system is described by three non-linearly coupled partial differential equations
\begin{eqnarray}
\label{Main_system}
    &&\Dot{x}_1 \ = \ -\omega_1\ \! x_2 \ - \ x_3\, ,\nonumber \\[2mm]
    &&\Dot{x}_2 \ = \ \omega_1\ \! x_1 \ + \ a x_2\, ,\nonumber \\[2mm]
    &&\Dot{x}_3 \ = \ b \ + \ x_3 (x_1\ - \ c)\, ,
    \label{Rossler}
\end{eqnarray}
with 4 coefficients $\omega_1$, $a$, $b$ and $c$. 
Strictly speaking,  only 3 coefficients are independent as $\omega_1$ can be set to one by appropriately re-scaling $x_2$.
RS was invented in 1976 by O.E.~R\"{o}ssler~\cite{Rossler:1976} and it represents probably the most elementary geometric construction of chaos in continuous systems.
In fact, since the Poincar\'{e}--Bendixson theorem precludes the existence of other than steady, periodic, or quasiperiodic
attractors in autonomous systems defined in one- or two-dimensional manifolds, the minimal dimension for chaos is three~\cite{Rossler:1976b}. Simplicity of the RS is bolstered by the fact that it has only one non-linear (quadratic) coupling.

RS classifies as  \textit{continuous (deterministic) chaotic system},
and more specifically as {\em chaotic attractor}. The word  ``attractor'' refers to the fact that whatever is the initial condition for the solution of the differential equations (\ref{Coupled_system}), the trajectory ${\mathbf{x}}(t)$ ends up (after a short transient period) at the same geometrical structure (see Fig.~\ref{Roessler_system}), which is  neither  a  fixed  point  nor  a  limit  cycle.  This attractive geometrical structure is known as R\"{o}ssler attractor.

For a future convenience we will call the RS (\ref{Main_system}) as {\em driving} or {\em master system} and denote it as $\{X\}$. 

%%%%%%%%%%%%%%%%%%%%%%%%%%%%%%%%%%%%%%%%%%%%%%%
\subsection{Equations for slave system}
%%%%%%%%%%%%%%%%%%%%%%%%%%%%%%%%%%%%%%%%%%%%%%%

In the following we investigate RTE between two R\"{o}ssler systems 
that are unidirectionally coupled in the variable $x_1$ via small adjustable parameter $\varepsilon$. The corresponding second RS --- {\em driven} or \textit{slave system} is defined as 
\begin{eqnarray}
\label{Coupled_system}
    &&\Dot{y}_1\ = \ -\omega_2\ \! y_2 \ - \ y_3 \ + \ \varepsilon(x_1 \ - \ y_2)\, , \nonumber \\[2mm]
    &&\Dot{y}_2 \ = \ \omega_2 \ \! y_1 \ + \ a y_2\, ,  \nonumber\\[2mm]
    &&\Dot{y}_3 \ = \ b \ + \  y_3 (y_1 \ - \ c)\, .
    \label{Rossler2}
\end{eqnarray}
Here we fix the coefficients so that $a=0.15$, $b=0.2$, $c=10.0$ and frequencies $\omega_1=1.015$ and $\omega_2=0.985$ and initial condition $(x_1(0), x_2(0), x_3(0))=(0, 0, 0)$ and $(y_1(0), y_2(0), y_3(0)) = (0, 0, 1)$. 
This parametrization is adopted from Ref.~\cite{palusRossler} where Shannon's TE between systems (\ref{Rossler}) and ({\ref{Rossler2}}) was studied. In the following we will denote the slave system also as $\{Y\}$. 
 
%%%%%%%%%%%%%%%%%%%%%%%%%%%%%%%%%%%%%%%%%%%%%%%%%%%%
\subsection{Numerical experiments with coupled RSs}
%%%%%%%%%%%%%%%%%%%%%%%%%%%%%%%%%%%%%%%%%%%%%%%%%%%%

Before we embark on the RTE analysis let us first take a look at the phenomenology of the coupled RSs (\ref{Main_system})-(\ref{Coupled_system})  by means of simple numerical experiments. In our numerical treatment
we simulate coupled RSs by using integration method, which is implemented in package SciPy named \texttt{solve\_ivp} with option LSODA that exploits Addams/BDF method, see, e.g. Ref.~\cite{2020SciPy-NMeth}. Projections of the $\varepsilon$-dependent RSs dynamics to various planes are presented in Fig.~\ref{Roessler_system}. For visualization purposes we use toolkit Matplotlib \cite{Matplotlib} that exploits toolkit NumPy \cite{2020NumPy-Array}. 
The resulting data-set analysed  consisted of 100000 data points.
To gain a better insight into the transient region  we performed a higher frequency sampling in the region $0.1 \leq \varepsilon \leq 0.15$, namely $0.001$, in contrast to standard $0.01$. In parallel we have displayed in  Fig.~\ref{Lyapunov} behavior of corresponding Lyapunov exponents, which we have adapted from \cite{Palus:2007a} and which helped to elucidate our discussion. 

%%%%%%%%%%%%%%%%%%%%%%%%%%%%%%%%%%%%%%%%%%%
\subsubsection{Projections}
%%%%%%%%%%%%%%%%%%%%%%%%%%%%%%%%%%%%%%%%%%

Instead of a conventional stereoscopic plotting we find for our purposes more convenient (and also illuminating) to focus on various plane projections of the coupled RSs. 
First of all, we notice on Fig.~\ref{Roessler_system} that the projections of RSs on the  $x_2$-$x_1$, $x_3$-$x_2$ and $x_1$-$x_3$ planes do not depend on the coupling between systems (i.e., they are $\varepsilon$ independent) as it should be expected because the slave system (\ref{Coupled_system}) does not influence dynamics of the master system (\ref{Main_system}), which is autonomous (irrespective of $\varepsilon$). It is, however, clear that signatures of the interaction between non-symmetrically coupled RSs (\ref{Main_system})-(\ref{Coupled_system})
will show up in projections on the $x_i$-$y_j$ and $y_i$-$y_j$ planes.

Secondly, when the RSs are not coupled (i.e., when $\varepsilon=0$) we have two autonomous RSs --- in fact, two strange attractors that differ only by values of their frequency coefficients and initial values.  
The autonomy of respective RSs is  clearly seen in projections on $x_i$-$x_j$ and $y_i$-$y_j$ planes (cf. Fig.~\ref{Roessler_system}). A different density of trajectories (in a given time window $t = 100000$) can be ascribed to the frequency mismatch.
Projections on the $x_1$-$y_1$ and $x_2$-$y_2$ planes show 
how the ensuing chaotic and (component-wise) uncorrelated  trajectories
fill their support regions. In particular, we can observe that on the background of densely packed chaotic trajectories appear clear vertical stripes of dominantly visited regions in the slave system.
Vertical stripes are clearly visible because limit cycles in the autonomous slave system are far more
localized than in the master system. 
%We also see that the more inner the orbits in the slave RS are the more often they are  visited. 
Projection on the $x_3$-$y_3$ plane indicates that most of the time the master system orbits venture to $x_3$ direction the slave system orbits are in the vicinity of the $y_1$-$y_2$ plane and vice versa.

By continuously increasing
the coupling strength  $\varepsilon$ from zero value
we can observe that already a small interaction significantly changes evolution of the slave system. For instance, in Fig.~\ref{Roessler_system} we see that when $\varepsilon=0.01$  then the diffusive term $\varepsilon(x_1 -  y_2)$ significantly disperses limit cycles in the
slave system. This is  reflected 
not only in all projections on the  $y_i$-$ y_j$ planes but also in projections on the $x_1$-$y_1$ and $x_2$-$y_2$ planes. In the latter two cases the diffusion causes that horizontal stripes completely disappear. 
Finally, projection on the $x_3$-$y_3$ plane does not change significantly from the $\varepsilon=0$ case. 

When we further increase $\varepsilon$, we see that 
the behavior of the slave system starts to qualitatively depart 
from that of the master system.  For $\varepsilon$ around $0.1$, the slave system orbit diffuses to the region around origin that is basically not visited (apart from an initial transient orbit) by the master system orbit (cf. projections on the $y_i$-$y_j$ planes). In addition, projections on the $x_1$-$y_1$ and $x_2$-$y_2$ planes disclose that ensuing support areas are not anymore filled. 
In fact, we can see a development of a slant stripe structure. On the other hand, projection on the $y_3$-$x_3$ plane reveals that the slave system orbits stop to visit regions farther from $y_3=0$. 
For yet higher $\varepsilon$ (around $0.14$) orbit of the system $\{Y\}$ first converge to a single limit cycle  before it makes again a transition into a chaotic regime. Finally, we can observe that at $\varepsilon \sim 0.14$ the slave system rarely deviates far from $y_3=0$ and spends most of its time in the close vicinity of the $y_1$-$y_2$ plane --- its evolution is ``flattened''. 

\begin{figure}
\begin{center}
\includegraphics[width=0.6\textwidth]{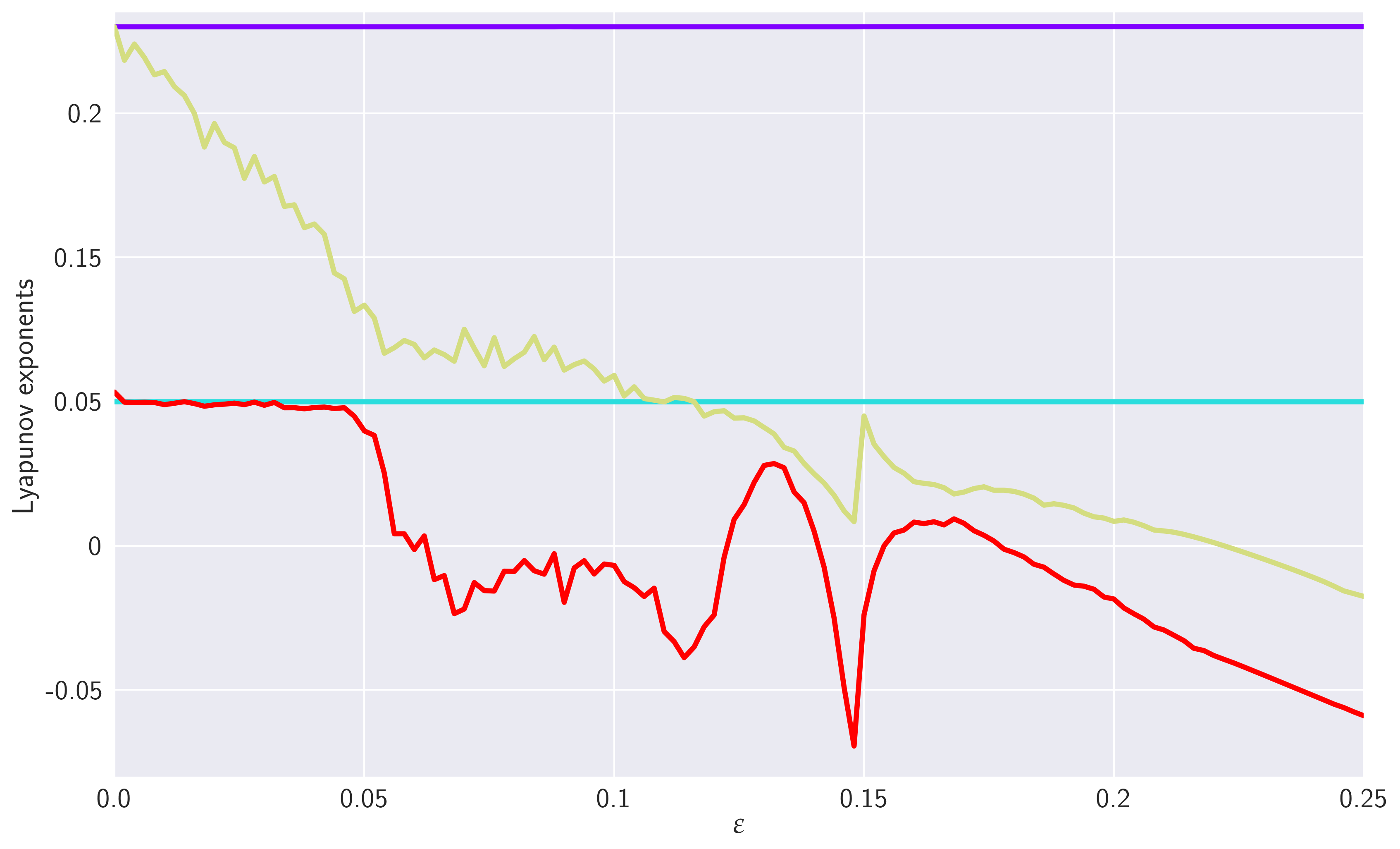}
\end{center}
%\vspace{-45mm}
\caption{Two largest Lyapunov exponents of the masters system (constant --- violet and green) and the slave system (decreasing --- read and yellow). So, for small $\varepsilon$ the  signature of LEs is $+\!+00-\!-$ while after synchronization we end up with the signature $+0-\!-\!-\!-$. After
synchronization there is a ``collaps'' of the dimension, in the sense that the slave system is completely dependent on the master system so, that there is only one dimension (direction) in which there is an expansion. Accordingly, there is only one LE with positive sign. The LEs are measured in nats per a time unit.}
\label{Lyapunov}
\end{figure}

\begin{figure}
\begin{center}
    
\includegraphics[width=\figuresizeRoessler]{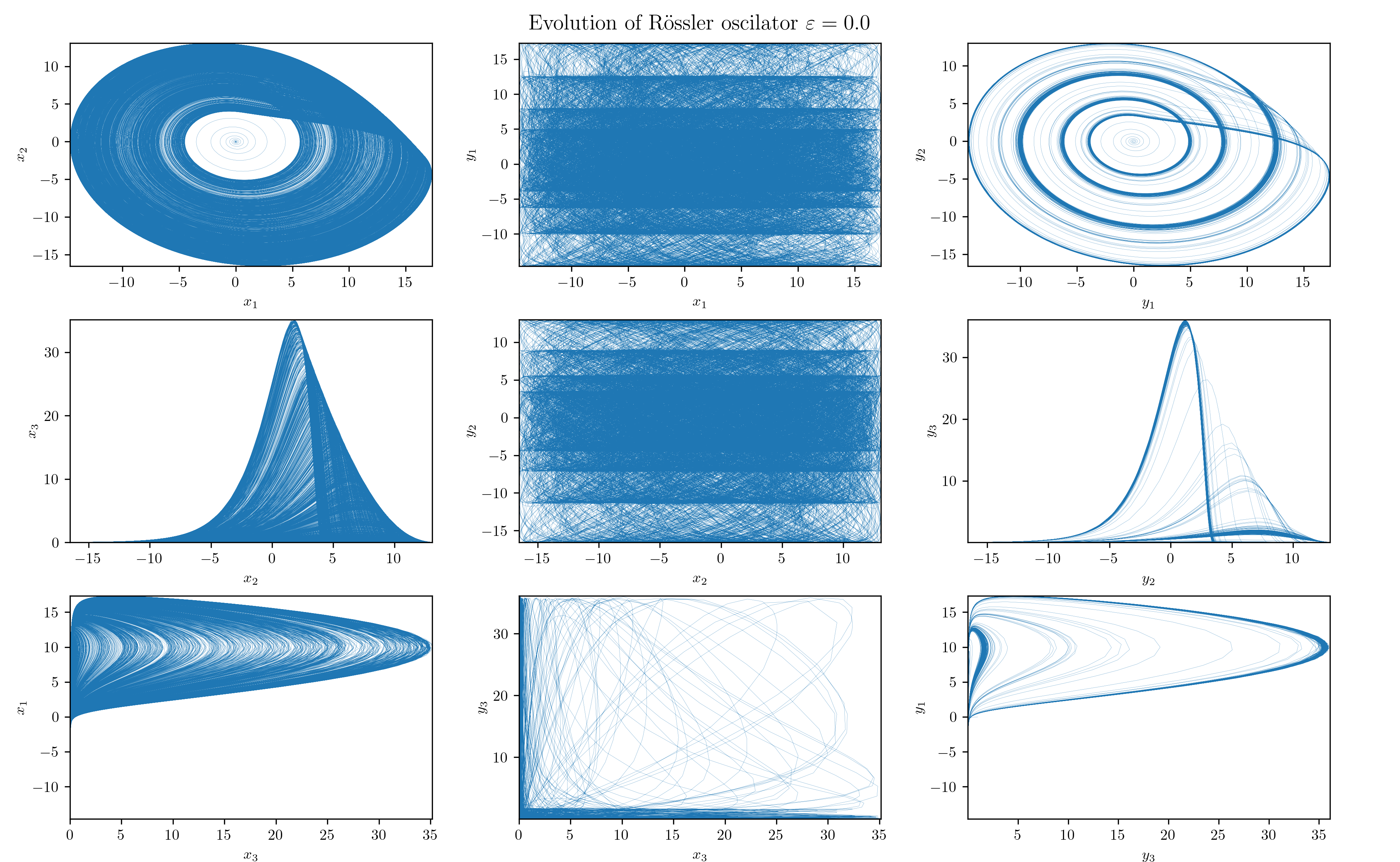}
\includegraphics[width=\figuresizeRoessler]{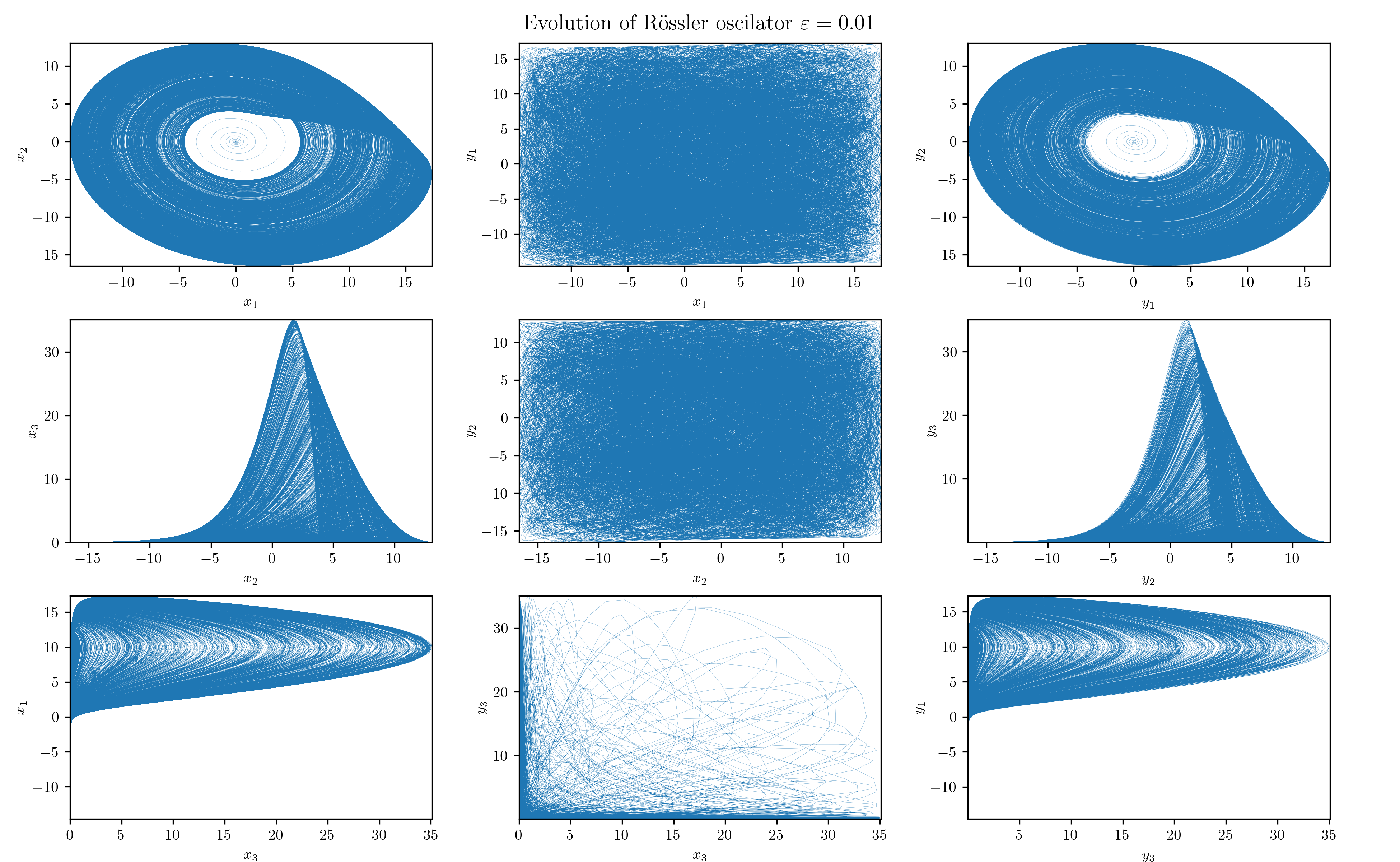}

\includegraphics[width=\figuresizeRoessler]{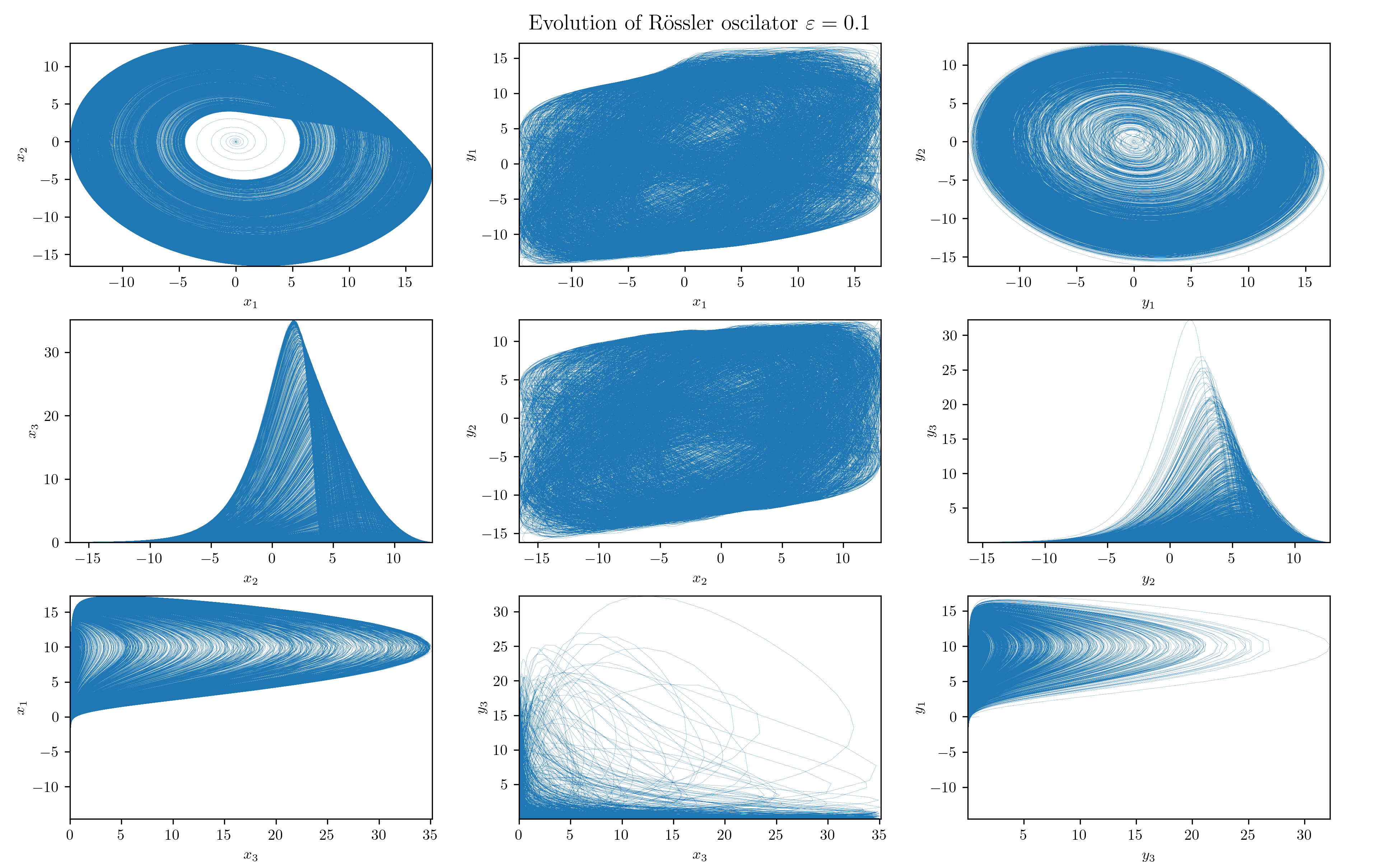}
\includegraphics[width=\figuresizeRoessler]{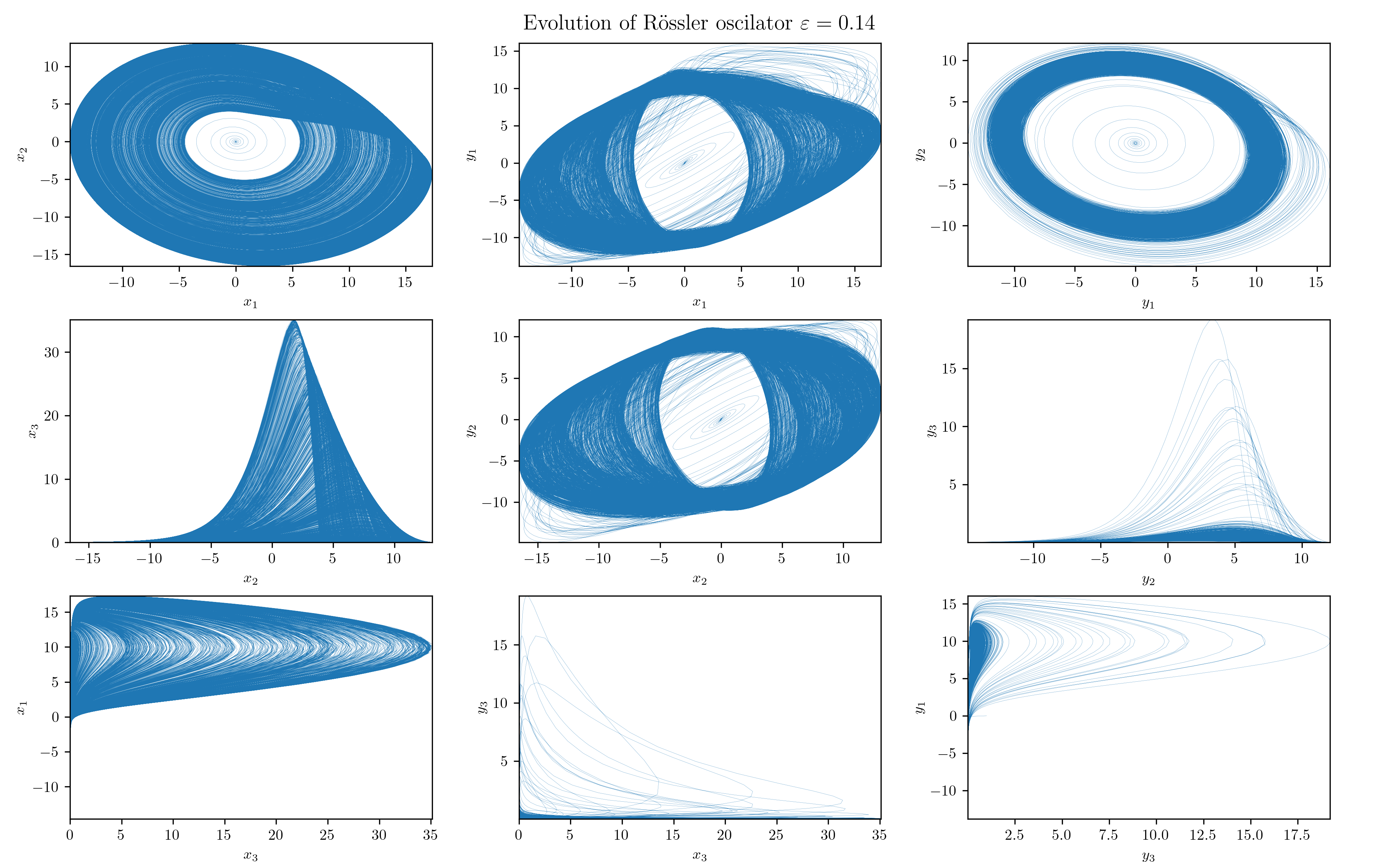}

\includegraphics[width=\figuresizeRoessler]{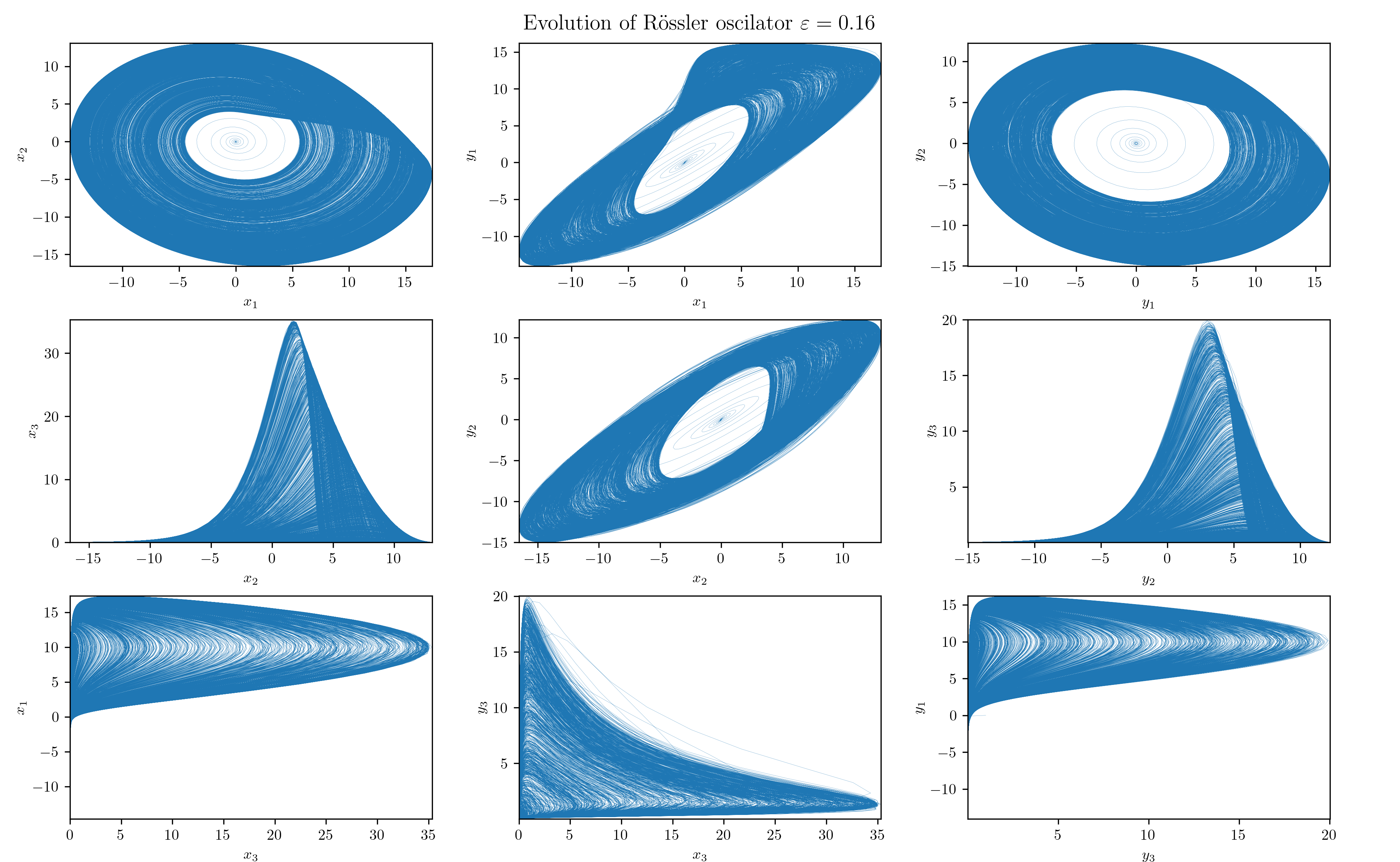}
\includegraphics[width=\figuresizeRoessler]{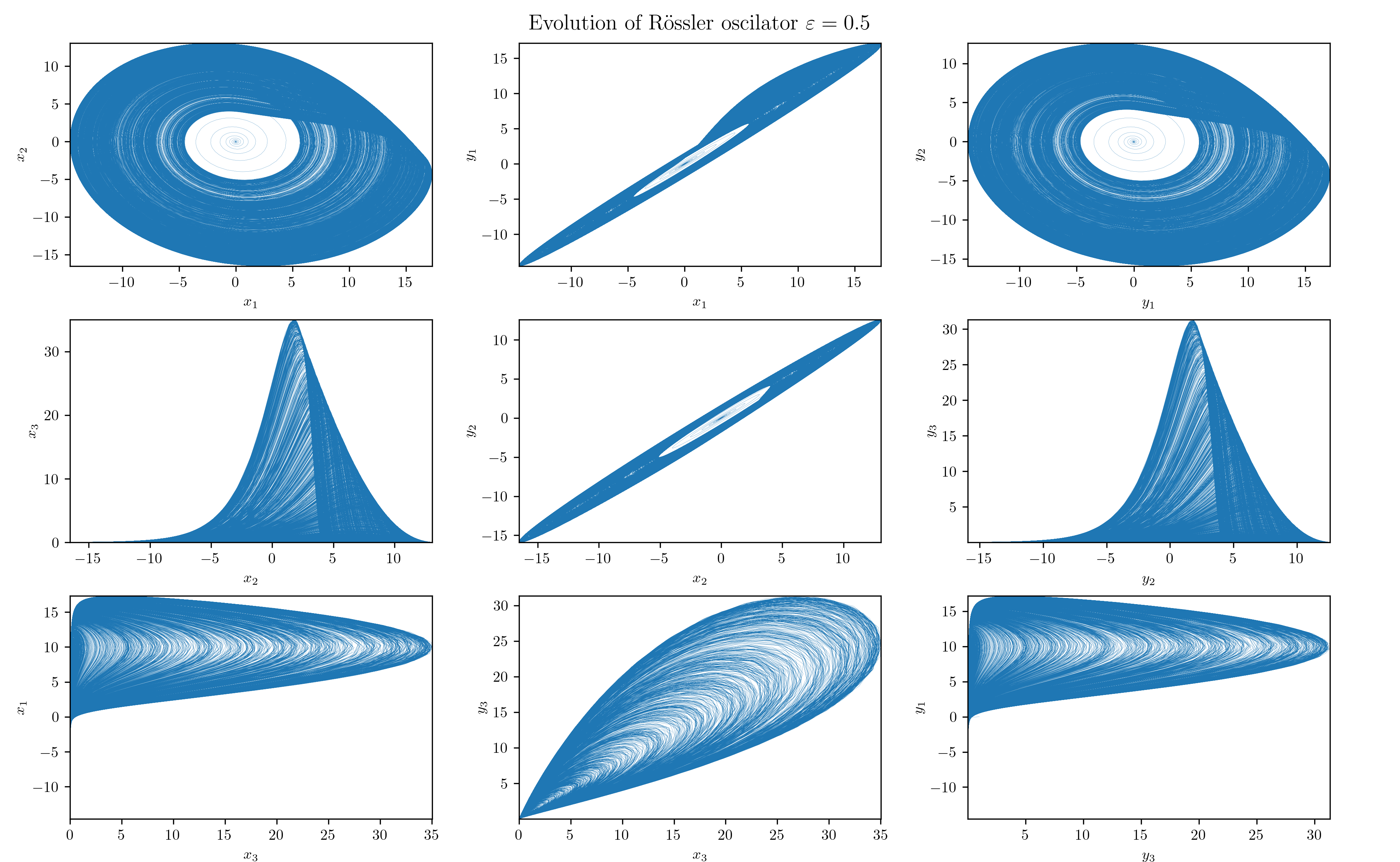}

\caption{Projections of the RSs (\ref{Rossler}) and (\ref{Rossler2}) on various planes. For each fixed $\varepsilon$ we depict 9 figures that correspond (from top to bottom and left to right) to projections on the $x_2$-$x_1$, $x_3$-$x_2$, $x_1$-$x_3$, $x_1$-$y_1$, $x_2$-$y_2$, $x_3$-$y_3$, $y_2$-$y_1$, $y_3$-$y_2$ and $y_1$-$y_3$ planes. In figure we display altogether 9 values of $\varepsilon$ corresponding (from left to right from top to bottom) to $\varepsilon = 0, 0.01, 0.1, 0.14, 0.16$ and  $0.5$.  Initial values are chosen as $x_1(0), x_2(0), x_3(0)= 0$, $y_1(0),y_2(0)=0$ and $y_3(0)=1$. Further projections for the transient region $0.12 \lesssim \varepsilon \lesssim 0.15$ are shown in Fig.~\ref{Roessler_system_II}. All RSs are depicted in the time window $t= 10000$.}
\label{Roessler_system}

\end{center}
\end{figure}

Moreover, at $\varepsilon \sim 0.14$  we can also notice that projections on the $y_1$-$x_1$ and $y_2$-$x_2$ planes underwent a change in topology (in fact, this happens already at around $\varepsilon \sim 0.12$). Onset of this ``topological phase transition'' is closely correlated with the behavior of the largest Lyapunov exponent (LE) of the slave system. In fact, coupled RSs have altogether six Lyapunov exponents.  When $\varepsilon = 0$ one has two autonomous RSs each with three LEs --- one positive, one zero and one negative (signature $+0-$ is a typical hallmark of a strange ettractor in 3 dimensions). While at $\varepsilon = 0$, signature of LEs is $+\!+00-\!-$, with increasing $\varepsilon$ all three LEs associated with $\{Y\}$ decrease initially monotonically, cf. Fig.~\ref{Lyapunov}. After a
transient negativity and a return to zero (red curve in Fig.~\ref{Lyapunov}), the originally positive LE of the slave system becomes monotonically decreasing
and negative for $\varepsilon \gtrsim 0.15$.  In particular we see that the critical value $\varepsilon \sim 0.12$ at which the ``topological phase transition'' happens coincides with value at which the largest LE of the system $\{Y\}$ crosses zero. 
%This is also the onset value at which the two LSs enter a synchronized regime --- synchronization threshold.
%~\cite{Palus:2007a}.

%can be affiliated with the onset value for the {\em amplitude synchronization}.
 
Particularly noteworthy is an abrupt (non-analytic)  change in the behavior of LEs at the value $\varepsilon \sim 0.145$.
At  this value the LE changes direction and starts to increase with increasing $\varepsilon$. The increase stops at $\varepsilon \sim 0.15$ when the yellow-colored LE in Fig.~\ref{Lyapunov} reaches (approximate) value zero after which it monotonically decreases. Such a decrease starts also for the second red-colored LE but at slightly different value of $\varepsilon$. 

%In the next section we will see that the critical value $\varepsilon \sim 0.12$ at which the topological transition happens and both LEs have a simultaneous non-analyticity can be affiliated with the onset value for the {\em amplitude synchronization}.
%In passing, we can mention that  it is difficult  from the behavior of LEs alone to deduce the onset of phase synchronization. 

For stronger interactions with  $0.15 \lesssim \varepsilon\lesssim 0.2$ we see (cf. Fig.~\ref{Roessler_system} with $\varepsilon=0.16$) that the slave system starts to approach the structure of the master-system strange attractor (cf. $x_i$-$x_j$ and $y_i$-$y_j$ projections). From the tilt and thinning of projections on the $x_1$-$y_1$ and $x_2$-$y_2$ planes one may deduce that the amplitude synchronization in $x_1$ and $y_1$  (as well as $x_2$ and $y_2$) directions increases. Projection on the $x_3$-$y_3$ plane shows that amplitudes  in $x_3$ and $y_3$ directions are also synchronized (being roughly a half-cycle behind each other).   

Finally, for very strong interactions, e.g. for $\varepsilon \sim 0.5$ the synchronization is almost complete: the system $\{Y\}$ basically fully emulates master-system's behavior with both systems being now structurally identical (cf.  $x_i$-$x_j$ and $y_i$-$y_j$ projections). Full synchronization is nicely seen in projections on the $x_1$-$y_1$ and $x_2$-$y_2$ planes. 
Note, that also amplitudes in the $x_3$ and $y_3$ directions start to synchronize.
%This is also confirmed by the study of LEs for the driven system. ****

%%%%%%%%%%%%%%%%%%%%%%%%%%%%%%%%%%%%%%%%%%%%%%%%%%%%%%%%%%%%%%%%%%%%
\section{Numerical analysis of RTE for coupled RSs\label{Sec.6.cc}}
%%%%%%%%%%%%%%%%%%%%%%%%%%%%%%%%%%%%%%%%%%%%%%%%%%%%%%%%%%%%%%%%%%%%

In the previous section we learned some essentials about the coupled RS (\ref{Rossler})-(\ref{Rossler2}). 
%
%will provide information about how the underlying dynamics changes (e.g. when phase or amplitude synchronization occurs) when $\varepsilon$ changes. 
%
In order to demonstrate
the inner workings of the RTE and to gain a further insight into the way how the two RSs approach synchronization, we compute here the RTE for various salient situations, such as the RTE between $x_1$- and $y_1$-component, between $x_1$- and $y_3$-component or RTE between full master and slave system.
\begin{figure}
    
\includegraphics[width=0.37\textwidth]{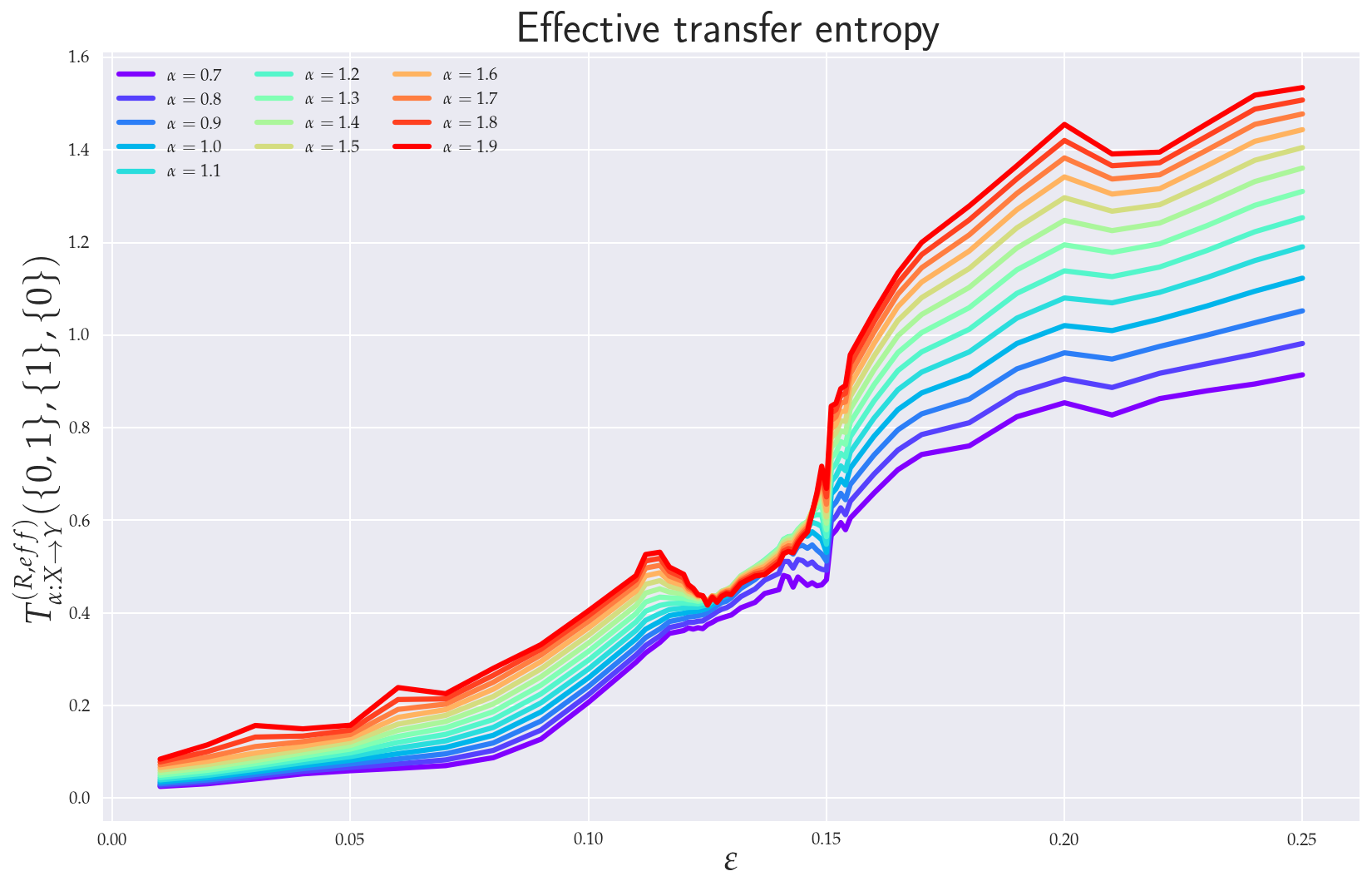}
\includegraphics[width=0.37\textwidth]{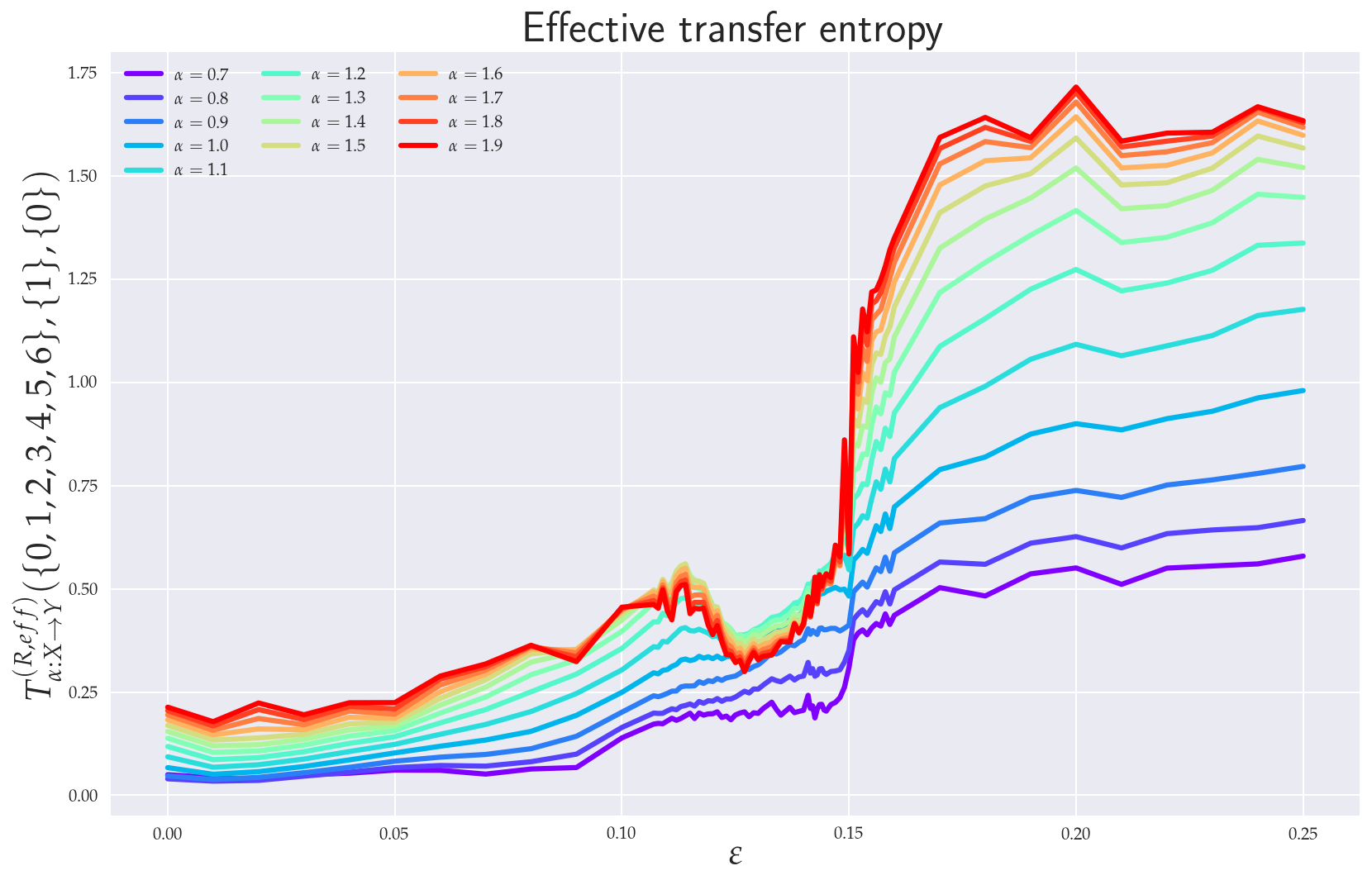}

\includegraphics[width=0.37\textwidth]{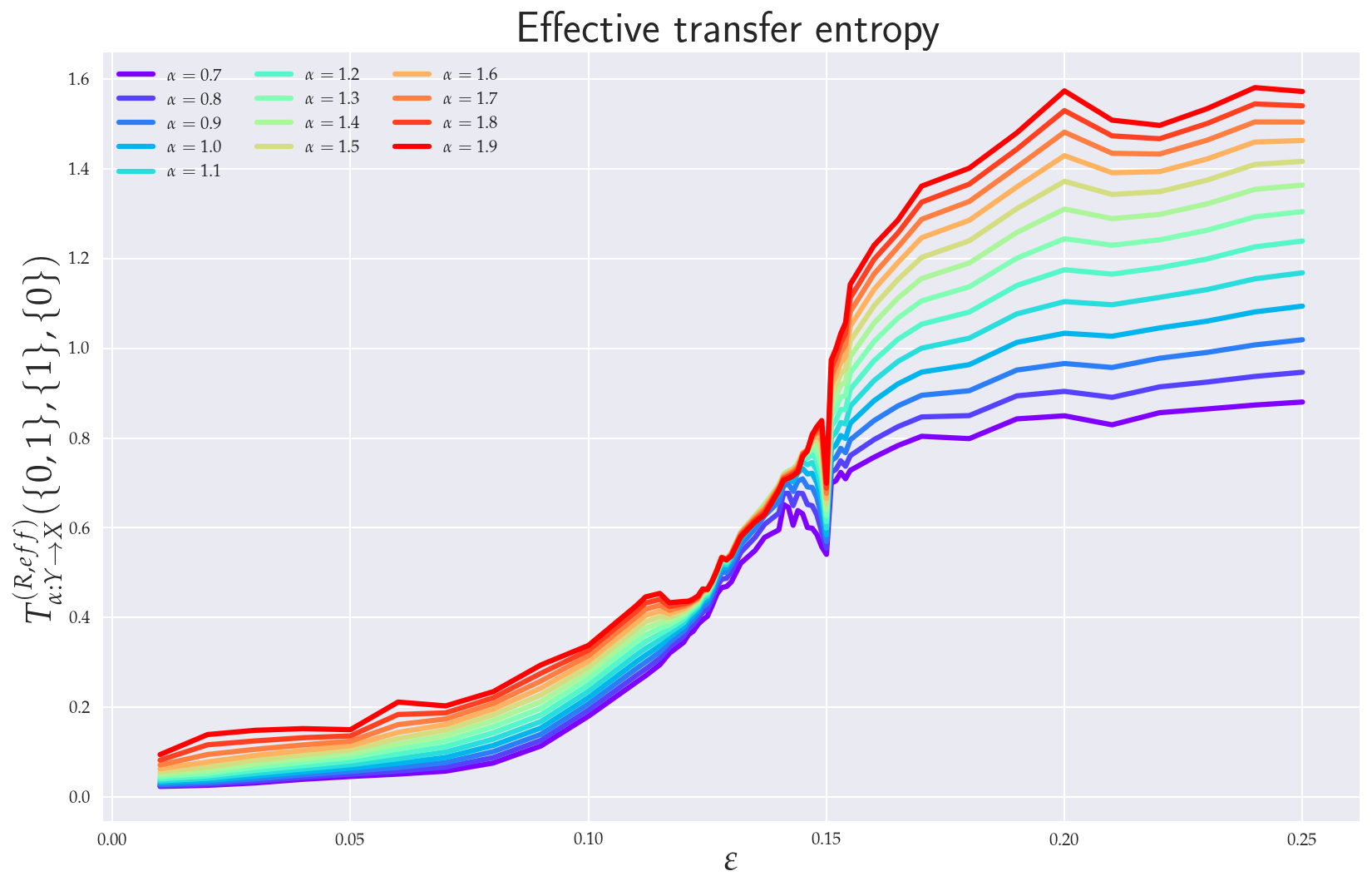}
\includegraphics[width=0.37\textwidth]{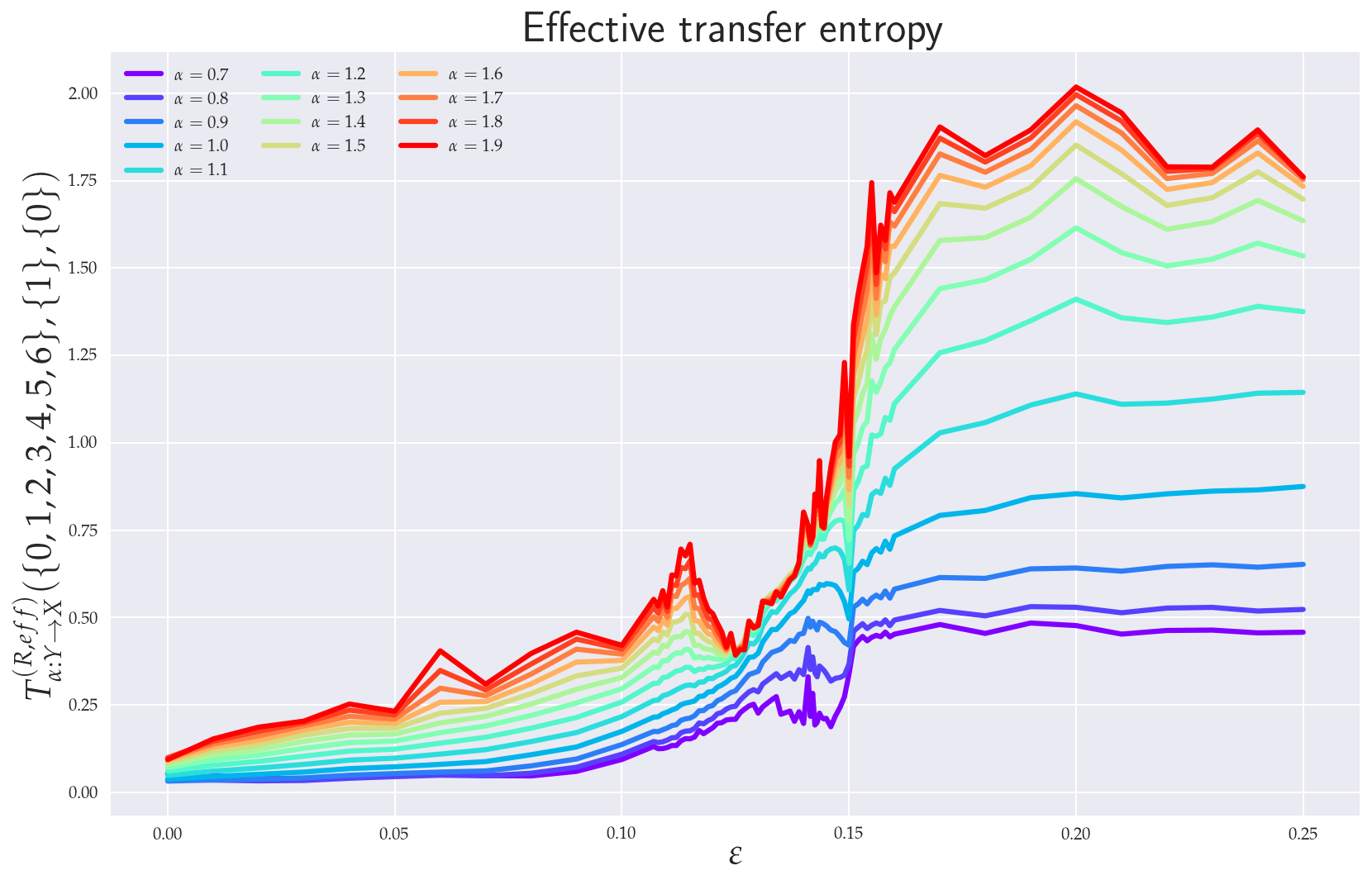}

\caption{Effective RTE between $x_1$ and $y_1$ for two different histories of $x_1$, i.e. $T^{R,\rm{\footnotesize{~effective}}}_{\alpha,x_1\rightarrow y_1}(\{0, 1\}, \{1 \},\{0\})$, $T^{R,\rm{\footnotesize{~effective}}}_{\alpha,x_1\rightarrow y_1}(\{0, 1, 2 ,3 , 4, 5, 6\}, \{1 \}, \{0 \})$, $T^{R,\rm{\footnotesize{~effective}}}_{\alpha,y_1\rightarrow x_1}(\{0, 1\}, \{1 \},\{0\})$, $T^{R,\rm{\footnotesize{~effective}}}_{\alpha,y_1\rightarrow x_1}(\{0, 1, 2 ,3 , 4, 5, 6\}, \{1 \}, \{0 \})$, respectively, from left to right and top to bottom. RTE is measured in nats.}
\label{Transfer_entropy_1d}

\end{figure}

%%%%%%%%%%%%%%%%%%%%%%%%%%%%%%%%%%
%\subsection{Effective RTE between $x_1$ and $y_1$ directions}
%%%%%%%%%%%%%%%%%%%%%%%%%%%%%%%%%%%%%%
\subsection{Effective RTE between \texorpdfstring{x\textsubscript{1}}{\texttwoinferior} and \texorpdfstring{y\textsubscript{1}}{\texttwoinferior} directions}
\begin{figure}[t]
\begin{center}
    
\includegraphics[width=\figuresizeRoessler]{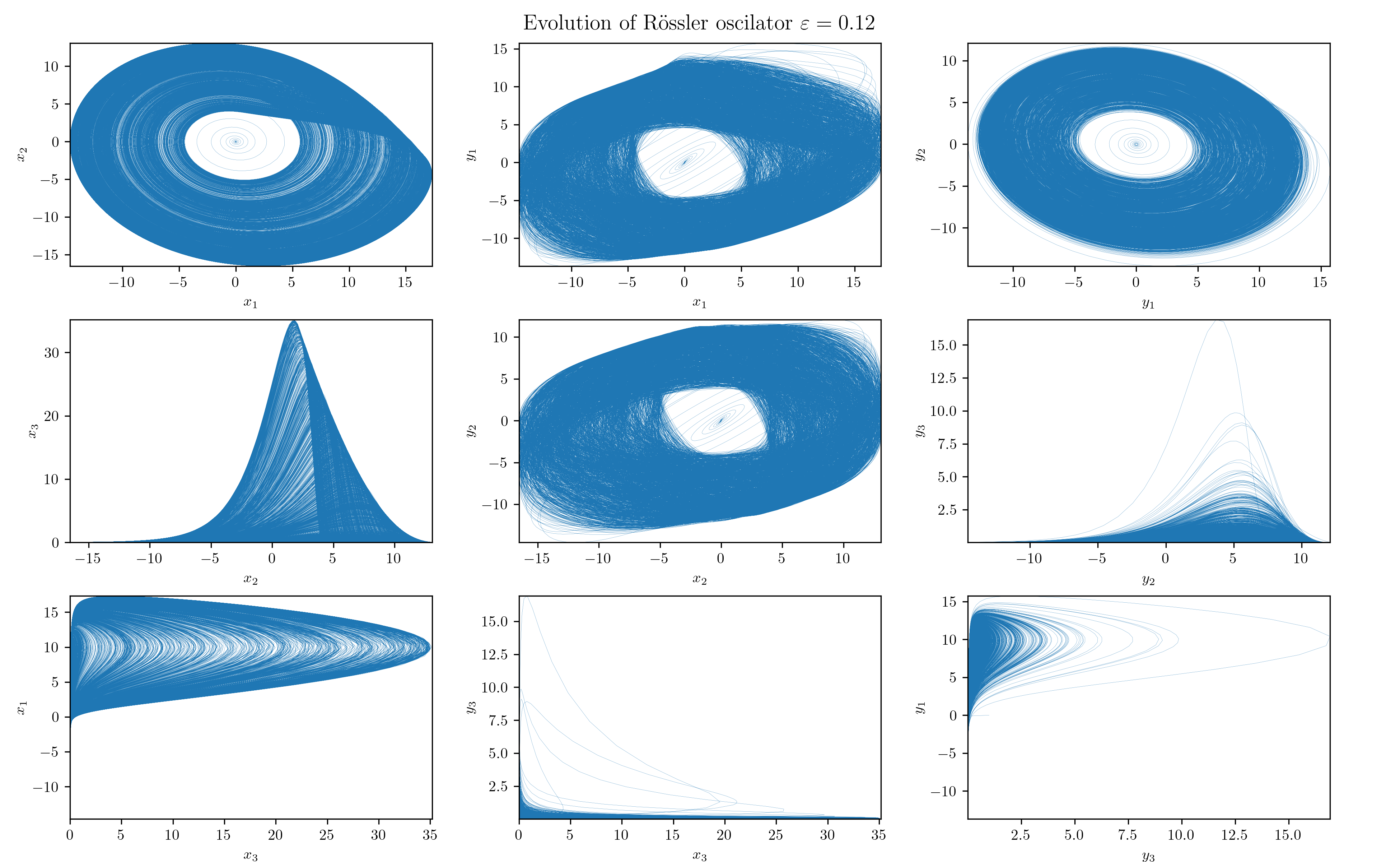}
\includegraphics[width=\figuresizeRoessler]{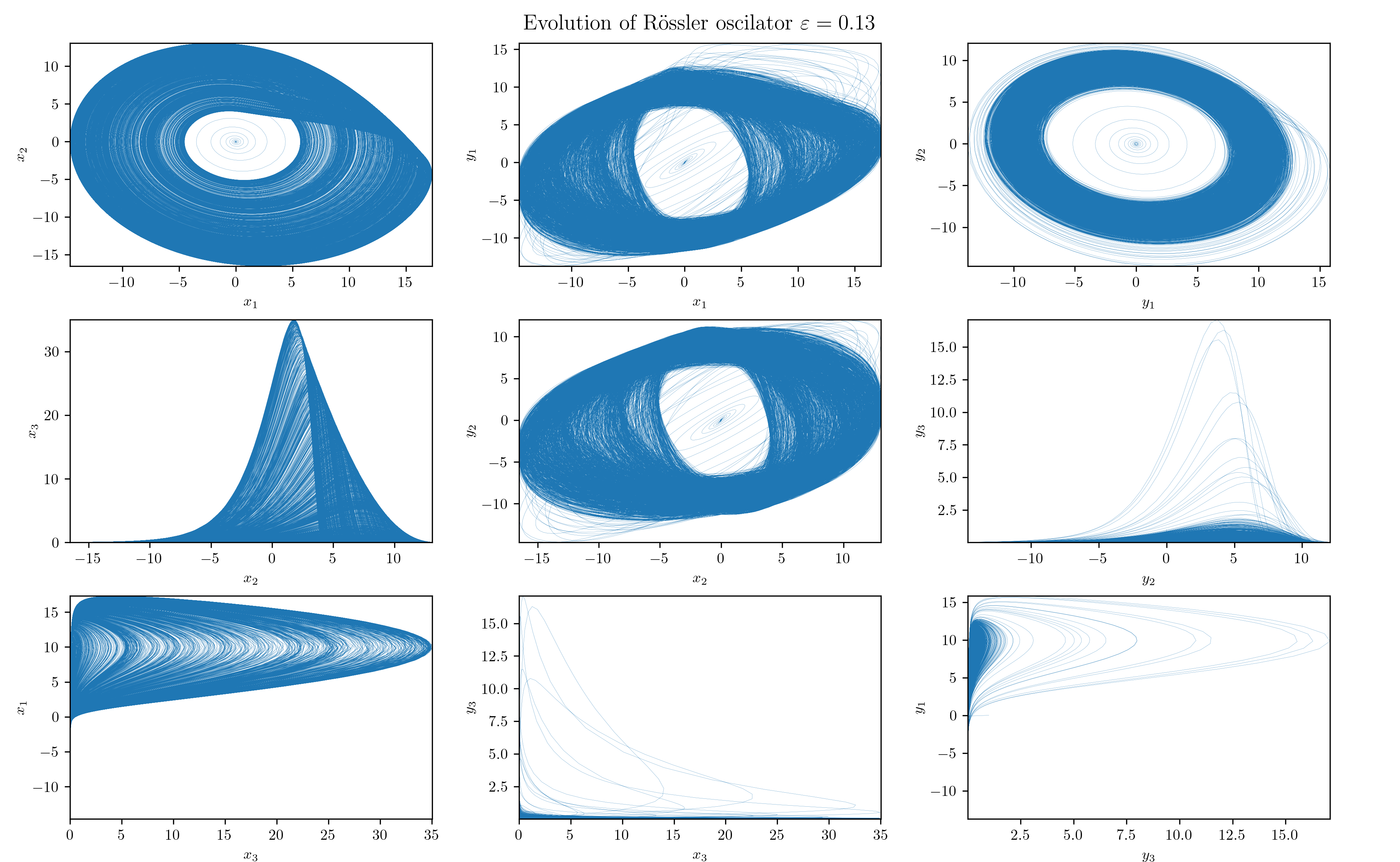}

\includegraphics[width=\figuresizeRoessler]{plots/roessler-0.14.png}
\includegraphics[width=\figuresizeRoessler]{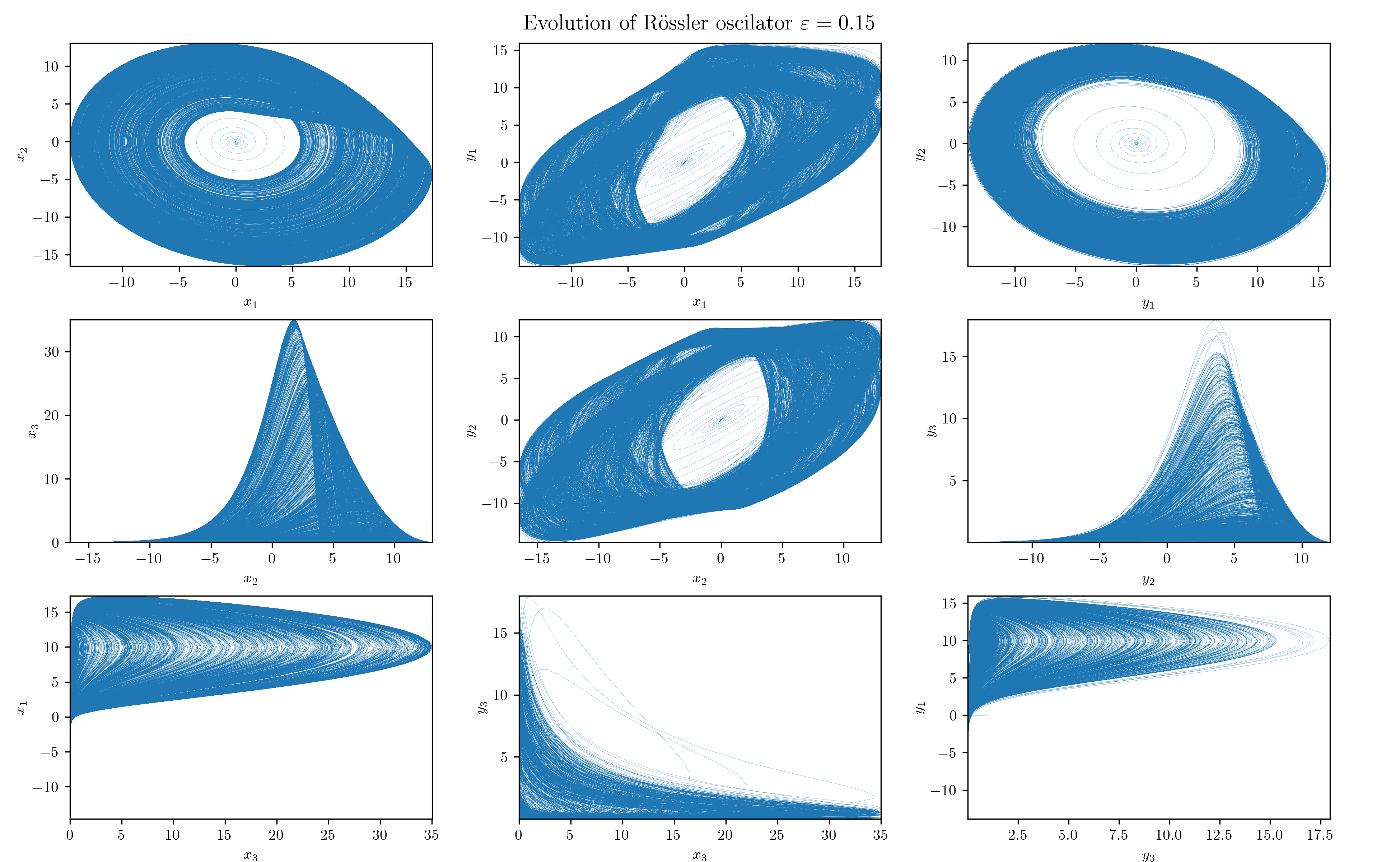}
\caption{Four projections of the RSs (\ref{Rossler}) and (\ref{Rossler2}) in the transient region $0.12 \lesssim \varepsilon \lesssim 0.15$. Depicted are projections (from left to right, from top to bottom) with $\varepsilon = 0.12$, $0.13$, $0.14$ and $0.15$.  With increasing $\varepsilon$ one can observe contraction (measure concentration) of the regions with denser orbit population in the slave system. At the critical point $\varepsilon = 0.15$ the contracted orbit regions abruptly expand
and the slave system starts its way toward full synchronization with the master system (cf. also Fig.~\ref{Roessler_system}).  All RSs are depicted in the time window $t= 10000$.}
\label{Roessler_system_II}
\end{center}
\end{figure}

%%%%%%%%%%%%%%%%%%%%%%%%%%%%%%%%%%%%
In order to understand the dynamics of the two coupled nonlinear dynamical systems  (\ref{Rossler}) and (\ref{Rossler2}) on their route to synchronization we first analyze the effective RTE between $x_1$ and $y_1$ components. Corresponding plots for different coupling strength $\varepsilon$ and different order $\alpha$ are depicted in Fig.~\ref{Transfer_entropy_1d}. We can observe first that the effective RTE from $x_1$ to $y_1$ gradually increases with the increasing coupling strength till $\varepsilon \sim 0.12$.  The regime between $\varepsilon \sim 0.12$ and $\varepsilon \sim 0.15$ as seen from Fig.~\ref{Roessler_system} corresponds to a transient synchronization behavior which stabilizes only after $\varepsilon \sim 0.15$. This can also be seen from the behavior of the LEs at Fig.~\ref{Lyapunov}.
It should also be noted that the behavior of effective RTEs in the transient regime is apparently almost identical for all $\alpha$ in both  $T^{R,\rm{\footnotesize{~effective}}}_{\alpha,x_1\rightarrow y_1}(\{0, 1\}, \{1 \},\{0\})$ and $T^{R,\rm{\footnotesize{~effective}}}_{\alpha,y_1\rightarrow x_1}(\{0, 1\}, \{1 \},\{0\})$. This would, in turn, indicate that the information transfer is the same across all sectors of the underlying probability distributions. Upon closer inspection though, such a  highly correlated behavior will disappear when more historic data  on $\{X\}$ and $\{Y\}$ are included (cf. $T^{R,\rm{\footnotesize{~effective}}}_{\alpha,x_1\rightarrow y_1}(\{0, 1, 2 ,3 , 4, 5, 6\}, \{1 \}, \{0 \})$ and $T^{R,\rm{\footnotesize{~effective}}}_{\alpha,y_1\rightarrow x_1}(\{0, 1, 2 ,3 , 4, 5, 6\}, \{1 \}, \{0 \})$ on Fig.~\ref{Transfer_entropy_1d}). The same conclusion can be reached when the effective RTE for the full 6-dimensional systems is considered, cf. Fig~\ref{Transfer_entropy_6d}. Nevertheless, from Fig.~\ref{Transfer_entropy_1d} it can  clearly be inferred that in the transient region strong correlations do exist albeit not for all $\alpha$s. In particular, one starts with correlated flow for  $\alpha \gtrsim 1.2$  that gets
stronger as $\varepsilon$ increases. On the other hand, as $\varepsilon$ approaches $0.15$ the information flow decreases for $\alpha \lesssim 1$. This can be seen clearly on both Fig.~\ref{Transfer_entropy_1d} and Fig.~\ref{Transfer_entropy_6d}. At $\varepsilon = 0.15$ the information flow abruptly increases
for all $\alpha$. This is similar to a first order phase transition  in statistical physics. In this respect our ``topological phase transition'' would be more like a second order phase transition
due to a smooth change in the entropic flow across the critical point $\varepsilon = 0.12$. 
This scenario is also supported by Fig.~\ref{Roessler_system_II} where
the actual behavior of the RS  between 
the two critical points for 4 selected values of $\varepsilon$'s is  
depicted. Note in particular, 
how the increase in the RTE  for $\alpha \gtrsim 1.2$ (as well as the decrease of RTE for  $\alpha \lesssim  1$) is reflected in the contraction (measure concentration) of the regions with denser orbit population in the slave system. This, in turn, reinforces the picture that RTEs with higher $\alpha$s describe the transfer of information between more central parts of underlying distributions, which in this case relate to a higher occupation density of the $\{Y\}$ system orbit.  From Fig.~\ref{Roessler_system_II} we can also note that 
at the critical point $\varepsilon = 0.15$ the contracted orbit regions abruptly expand
and the slave system starts its way toward a full synchronization with the master system. This is again compatible with the fact that the RTE abruptly increases for all $\alpha$ at this point --- i.e., all parts of underlying distributions participate in this transition and consequently the occupation density of the $\{Y\}$ system orbit spreads. In this respect the point $\varepsilon = 0.15$ represents {\em threshold to full synchronization} while the point $\varepsilon = 0.12$ denotes {\em threshold to transient behavior prior full synchronization}. The latter can be identified with a phase synchronization threshold, which should be at (or very close to) this point~\cite{Palus:2007a}.

After the critical point $\varepsilon \sim 0.15$, both RS enter full synchronization. In fact, the full synchronization starts when the information flow from all sectors of underlying distributions (i.e., for all $\alpha$s) starts to be (almost) $\varepsilon$ independent and when  $T^{R,\rm{\footnotesize{~balance,~ effective}}}_{\alpha,X\rightarrow Y}$ approach zero --- 
so there is a one-to-one relation between the states of the systems and time series of the $\{X\}$ system can be predicted from time series $\{Y\}$ system, and vice versa. Indeed, from Fig.~\ref{Transfer_entropy_1d} (cf. also Fig.~\ref{Transfer_entropy_6d} and Fig.~\ref{Balance_standard_effective_transfer}) we see that all 
$T^{R,\rm{\footnotesize{~effective}}}_{\alpha,Y\rightarrow X}$ proceed in a slow increase toward their asymptotic values in the fully synchronized state. 

%%%%%%%%%%%%%%%%%%%%%%%%%%%%%%%%%%%%%%%%
%\subsection{Effective RTE between $x_3$ and $y_3$ directions}
%%%%%%%%%%%%%%%%%%%%%%%%%%%%%%%%%%%%%%%%

%%%%%%%%%%%%%%%%%%%%%%%%%%%%%%%%%%%%%%%%%%
\subsection{Effective RTE between \texorpdfstring{x\textsubscript{3}}{\texttwoinferior} and \texorpdfstring{y\textsubscript{3}}{\texttwoinferior} directions}
%%%%%%%%%%%%%%%%%%%%%%%%%%%%%%%%%%%%%%%%%%

\begin{figure}[h]
\begin{center}
    
\includegraphics[width=0.36\textwidth]{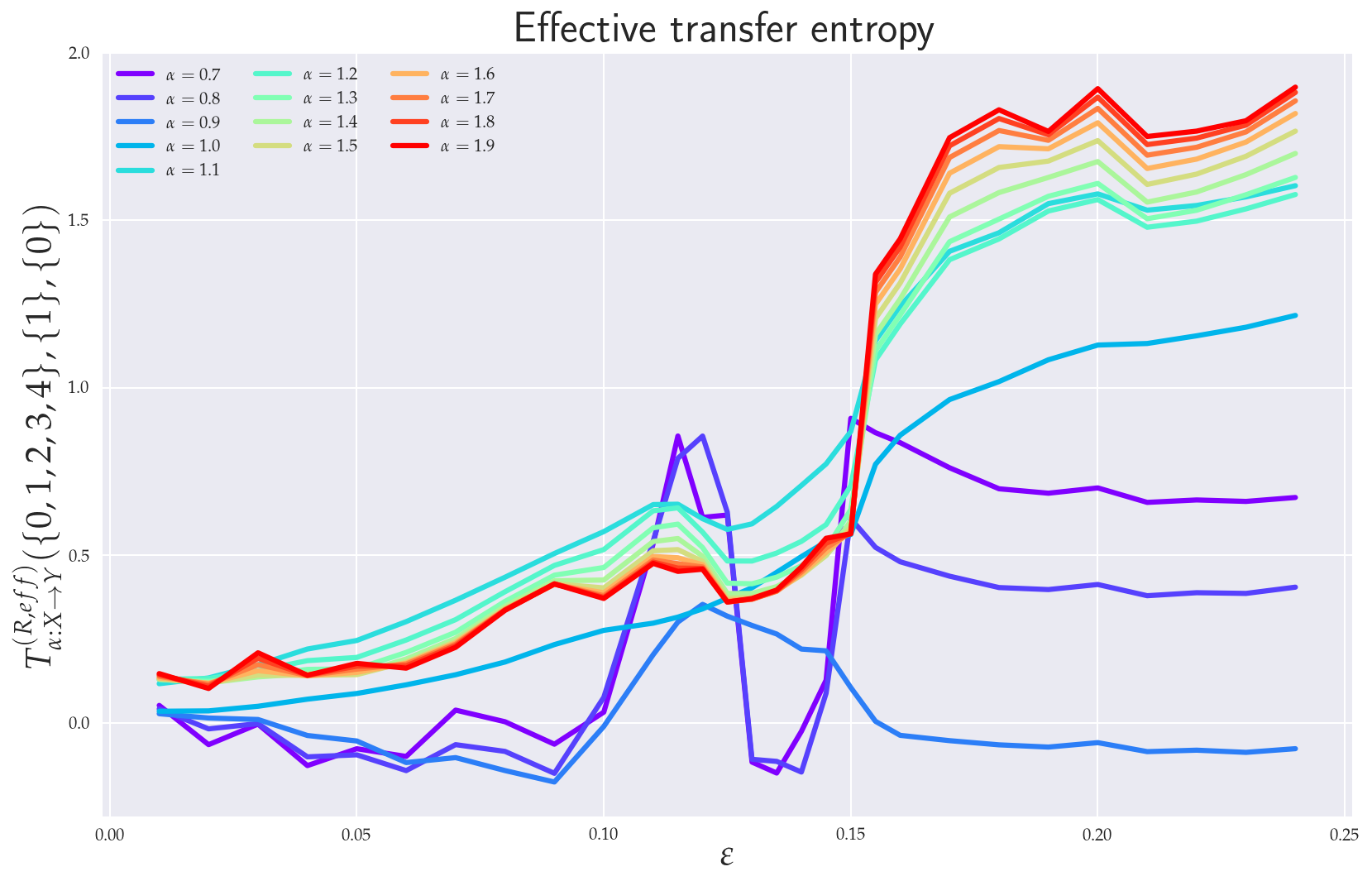}
\includegraphics[width=0.36\textwidth]{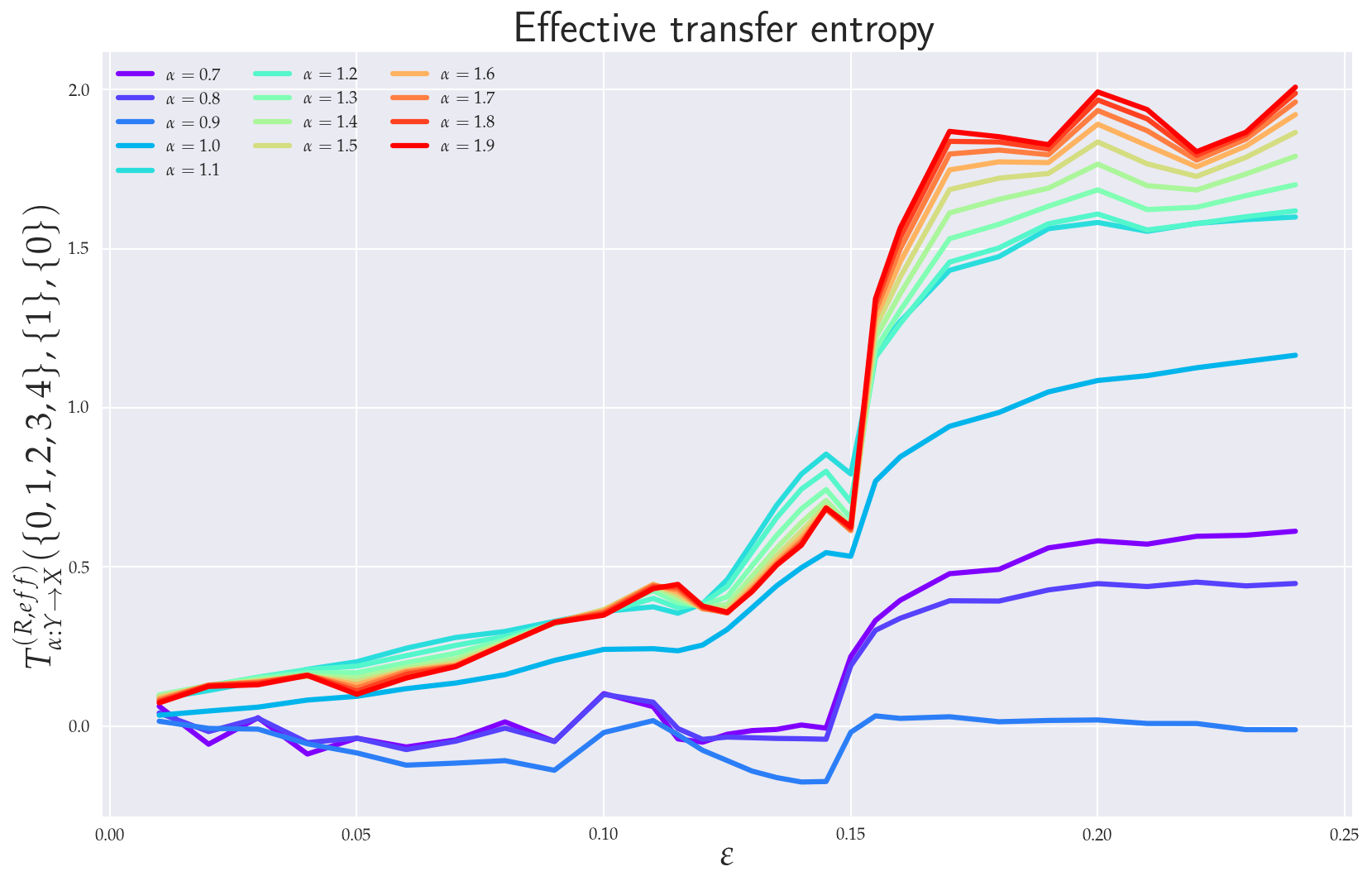}
\caption{Effective RTE between $x_3$ and $y_3$ directions. From left to right: $T^{R,\rm{\footnotesize{ effective}}}_{\alpha,x_3\rightarrow y_3}(\{0, 1, 2, 3, 4\}, \{1\}, \{0\})$  and 
$T^{R,\rm{\footnotesize{ effective}}}_{\alpha,y_3\rightarrow x_3}(\{0, 1, 2, 3, 4\}, \{1\}, \{0\})$. Note a sudden increase in entropy transfer from master to slave system at $\varepsilon = 0.12$ (i.e. threshold to transient behavior)  for $\alpha <1$. RTE
is measured in nats.
%Balance of effective RTE from $x_1$ to $y_3$  $T^{R,\rm{\footnotesize{~balance,~ effective}}}_{\alpha,x_1\rightarrow y_3}(\{0, 1, 2, 3, 4\}, \{1\}, \{0, 1, 2, 3, 4\})$ (left) and $T^{R,\rm{\footnotesize{~balance,~ effective}}}_{\alpha,x_3\rightarrow y_3}(\{0, 1, 2, 3, 4\}, \{1\}, \{0, 1, 2, 3, 4\})$ (right).
} 
\label{Balance_standard_effective_transfer_projections}

\end{center}
\end{figure}

As already seen from Figs.~\ref{Roessler_system} and \ref{Roessler_system_II},  particularly distinct are  projections on the $x_3$-$y_3$ plane. In Fig.~\ref{Balance_standard_effective_transfer_projections}
we the ensuing  effective RTE between $x_3$ and $y_3$ directions.  

Particularly noticeable is a sudden increase in entropy transfer from master to slave system at $\varepsilon = 0.12$ (i.e. at the threshold to transient behavior)  for $\alpha <1$. No comparable increase is observed from slave to master. This, might be explained
as an influx of information needed to organize the chaotically correlated regime that exists prior the (correlated) transient regime (cf. $x_i$-$y_i$ projections in Figs.~\ref{Roessler_system} and \ref{Roessler_system_II}). It should also be noticed that ordinary Shanonnian TE ($\alpha =1$) is completely blind to such an information transfer.   

As for the the transient region we can observe that the effective RTE has qualitatively very similar behavior as the effective RTE between $x_1$ and $y_1$, namely a distinct decrease  in information transfer for $\alpha <1$ and  increase for $\alpha > 1$. This again reveals a measure concentration. In this case the orbit occupation density concentrates around the $y_1$-$y_2$ plane of the slave systems, cf. projections depicted in Fig.~\ref{Roessler_system_II}. 
Situation abruptly changes at the synchronization threshold $\varepsilon = 0.15$ after which the effective RTE approaches for each $\alpha$ a fixed asymptotic value that turns out to be the same both for  $T^{R,\rm{\footnotesize{ effective}}}_{\alpha,x_3\rightarrow y_3}$ and $T^{R,\rm{\footnotesize{ effective}}}_{\alpha,y_3\rightarrow x_3}$.

%with a negligible cross-correlations between $x_3$ and $y_3$ directions.

%%%%%%%%%%%%%%%%%%%%%%%%%%%%%%%%
\subsection{Effective RTE  for the full system}
%%%%%%%%%%%%%%%%%%%%%%%%%%%%%%%%

\begin{figure}
    
\includegraphics[width=0.37\textwidth]{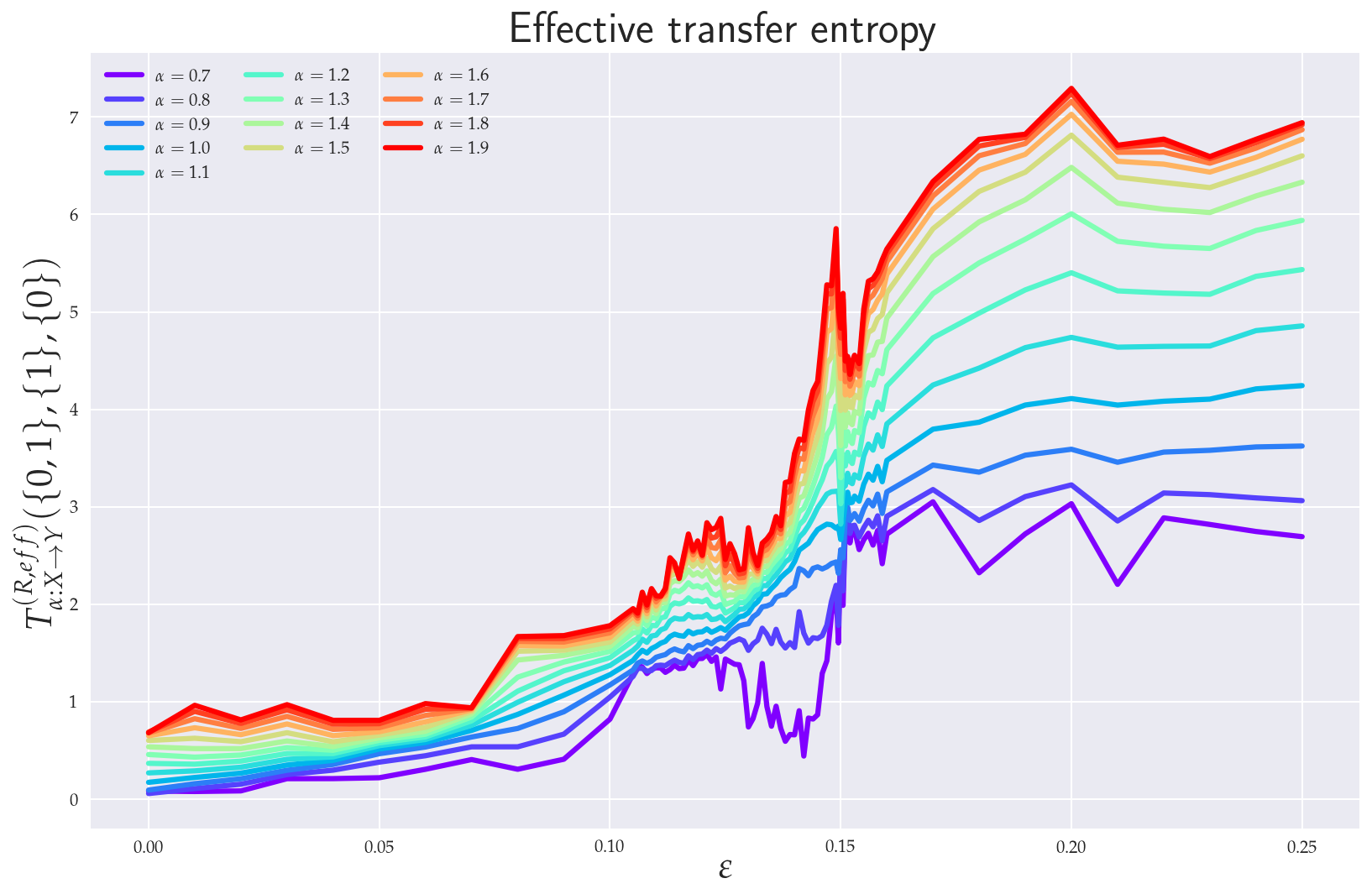}
\includegraphics[width=0.37\textwidth]{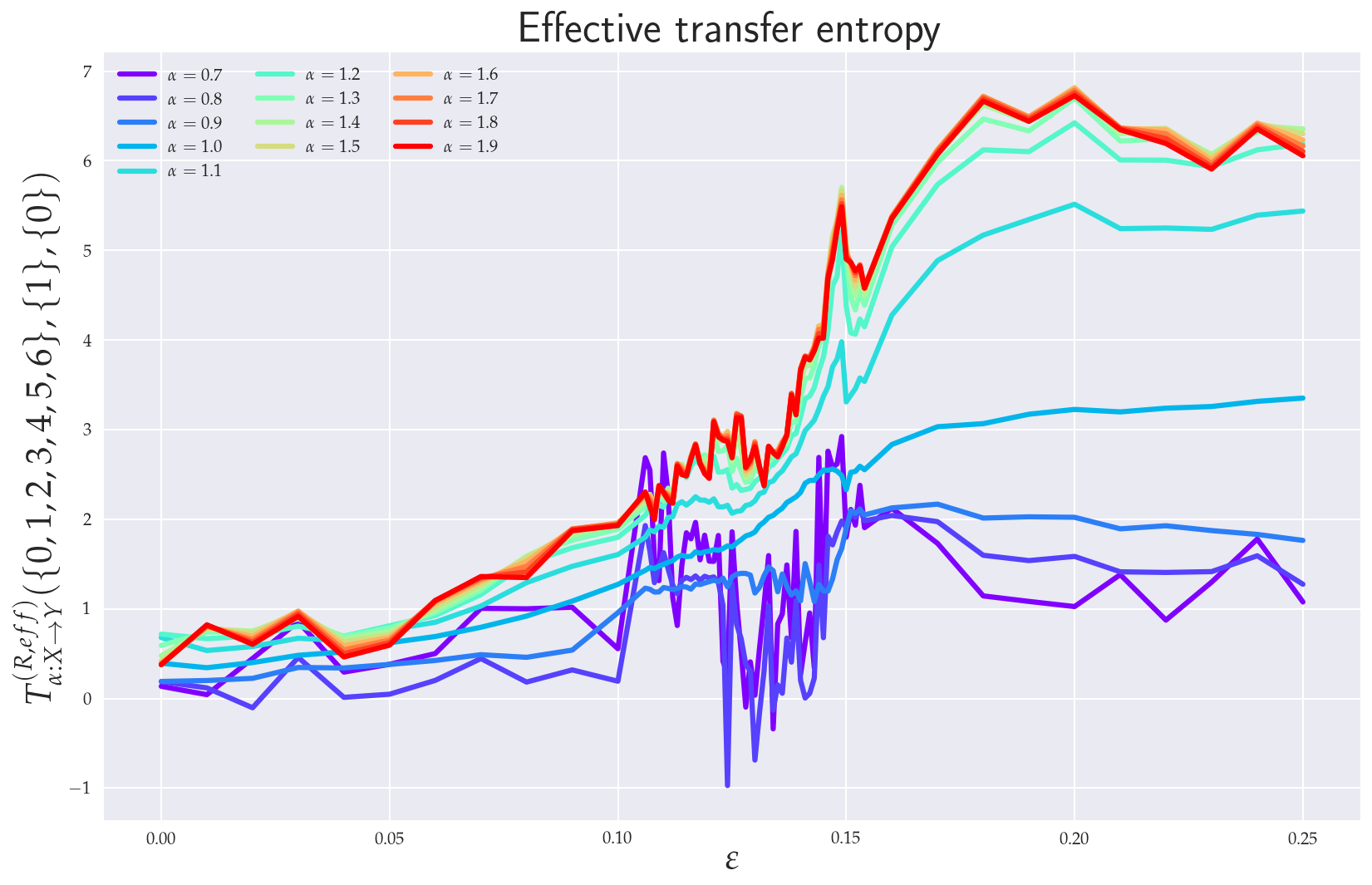}

\includegraphics[width=0.37\textwidth]{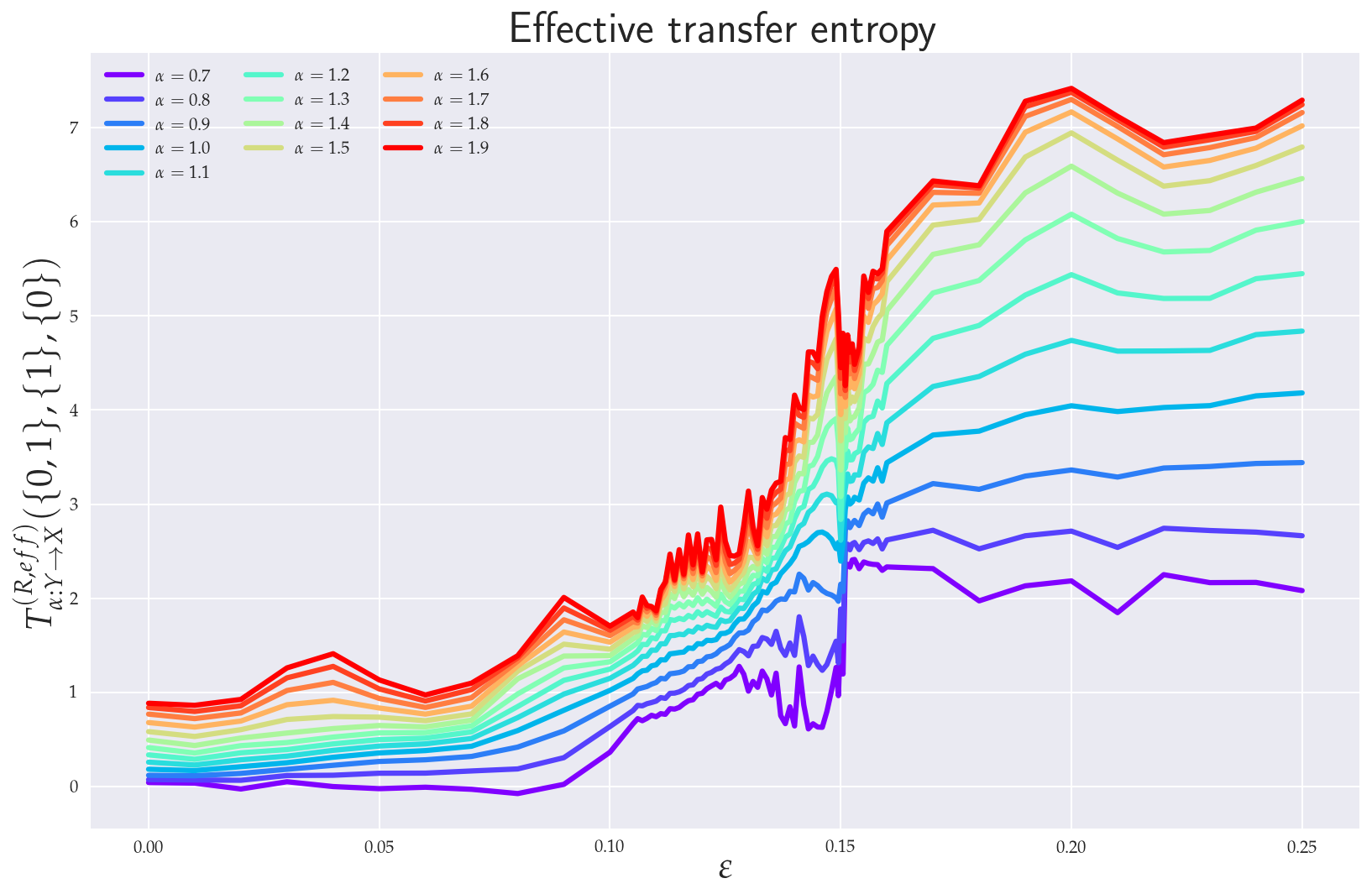}
\includegraphics[width=0.37\textwidth]{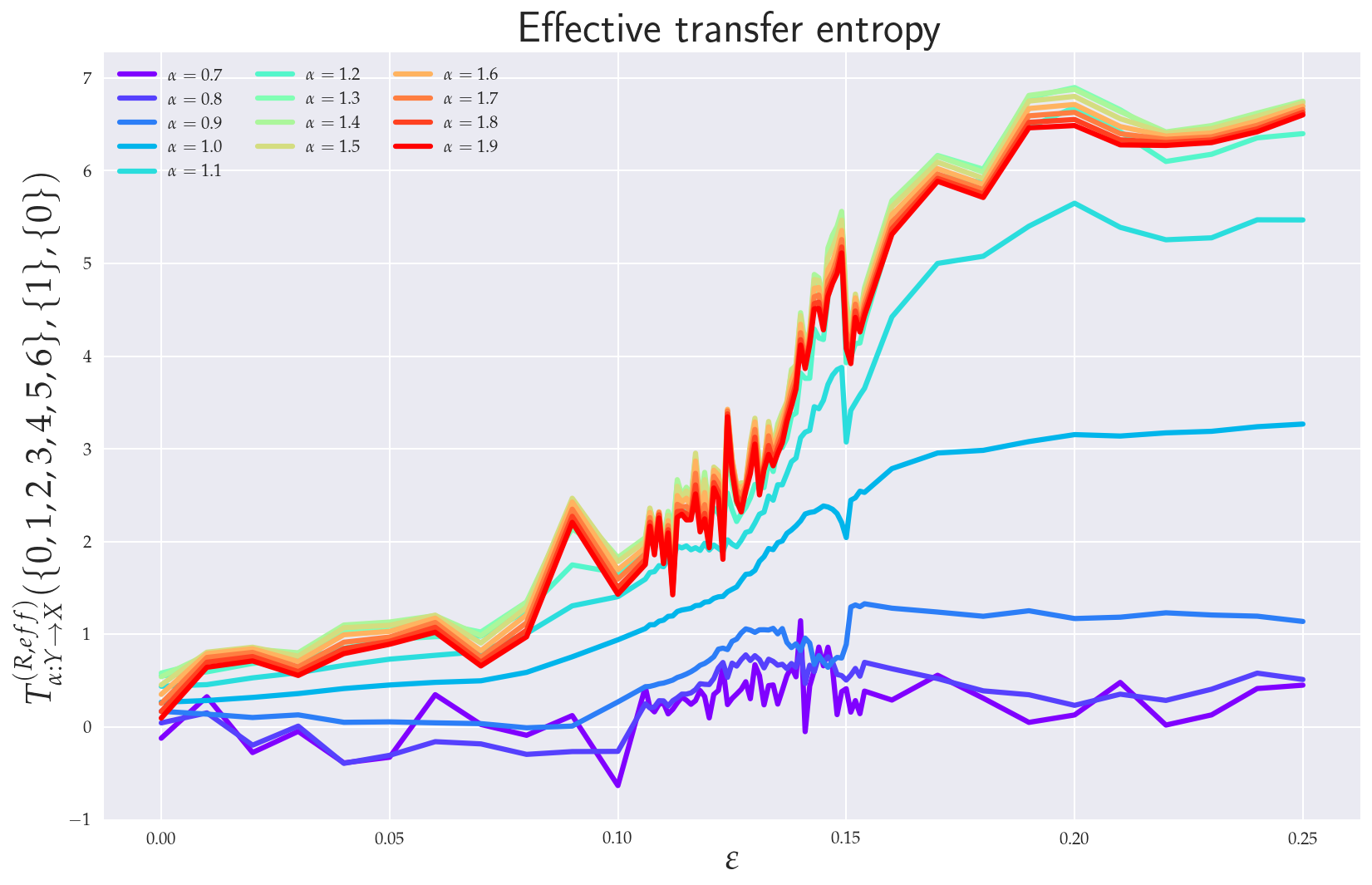}
\caption{Effective transfer entropy for the full system (6 dimensions)  and for different values of $\alpha$ as a function of the coupling $\varepsilon$. On  the {\em first line} we depict $T^{R,\rm{\footnotesize{~effective}}}_{\alpha,X\rightarrow Y}(\{0, 1\}, \{1 \},\{0\})$ and $T^{R,\rm{\footnotesize{~effective}}}_{\alpha,X\rightarrow Y}(\{0, 1, 2 ,3 ,4, 5, 6\}, \{1 \}, \{0 \})$, while on the {\em second line} we depict   $T^{R,\rm{\footnotesize{~effective}}}_{\alpha,Y\rightarrow X}(\{0, 1\}, \{1 \},\{0\})$ and $T^{R,\rm{\footnotesize{~effective}}}_{\alpha,Y\rightarrow X}(\{0, 1, 2 ,3 ,4, 5, 6\}, \{1 \}, \{0 \})$. RTE is measured in nats.}
\label{Transfer_entropy_6d}

\end{figure}

In general, for a reliable inference it is desirable that the conditioning variable in the definition or RTE (\ref{RTE})  contains all relevant information about future values of the system or processes generating this variable in the uncoupled case. So it should be a full 3-dimensional vector $X$ or $Y$ in the case of RS. To this end we display in Fig.~\ref{Transfer_entropy_6d} the effective RTE for the full 6-dimension RS with information transfers in both $X \rightarrow Y$ and $Y \rightarrow X$ directions. Corresponding plots 
are depicted for different coupling strength $\varepsilon$, different order $\alpha$ and different memories.

In particular, we can see that the information flow in the transient region starts, after a brief decrease at around $\varepsilon \sim 0.12$,  sharply increase (in both directions) for $\alpha \gtrsim 1.2$. This, in turn implies that there is an increase in correlating activity 
in between regions with a higher occupation density in both REs. 
Behavior depicted in Fig.~\ref{Roessler_system_II} can help us to better understand this situation. In particular, we see that in the transient region the  $\{Y\}$ system  reshapes its  
orbit occupation density so that the ensuing measure concentrates more around its peak  while its tail parts are thinner.  
In fact, Fig.~\ref{Roessler_system_II} also shows that this measure concentration increases till almost $\varepsilon \sim 0.15$. The measure concentration behavior is reflected by the decrease of the RTE for $\alpha \lesssim 1$, i.e., decreasing information transfers between tail parts. This situation is even more pronounced when more memory is included in the effective RTEs, cf. both right pictures in Fig.~\ref{Transfer_entropy_6d}.

At the synchronization threshold $\varepsilon = 0.15$ the information flow abruptly changes for all $\alpha$'s with a particularly strong increase for $\alpha \lesssim 1$. This indicates that  the orbit occupation density
of the $\{Y\}$ system abruptly reshapes by lowering measure concentrated around its peak and broadening it in tails so that also tail parts may enter the full synchronization regime.

%%%%%%%%%%%%%%%%%%%%%%%%%%%%%%%%%%%%%%%
\subsection{Balance of effective RTE}
%%%%%%%%%%%%%%%%%%%%%%%%%%%%%%%%%%%%%%%

%\subsubsection{Projection to $x_1$ and $y_1$}

\begin{figure}[t]

\includegraphics[width=0.37\textwidth]{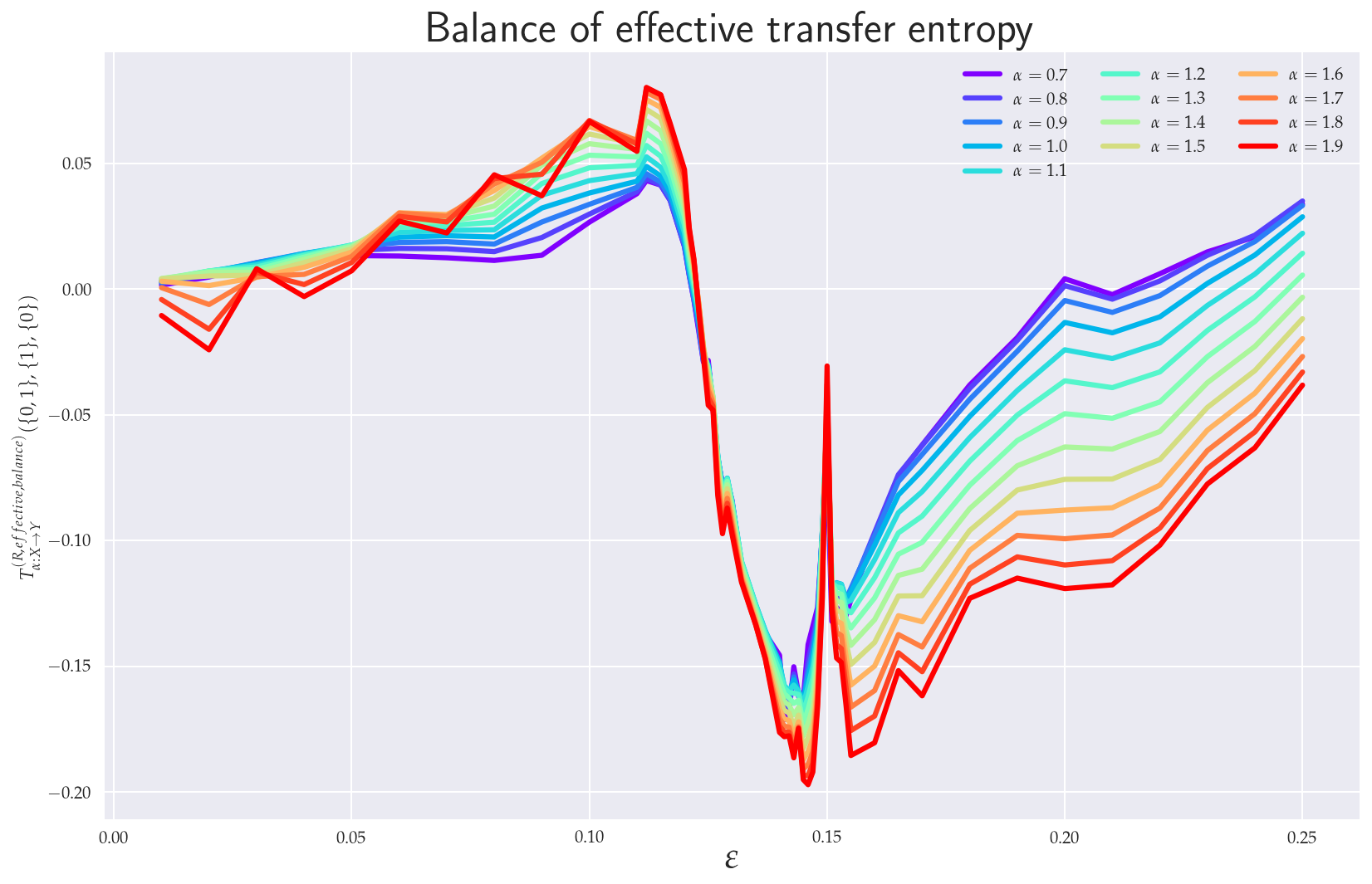}
\includegraphics[width=0.37\textwidth]{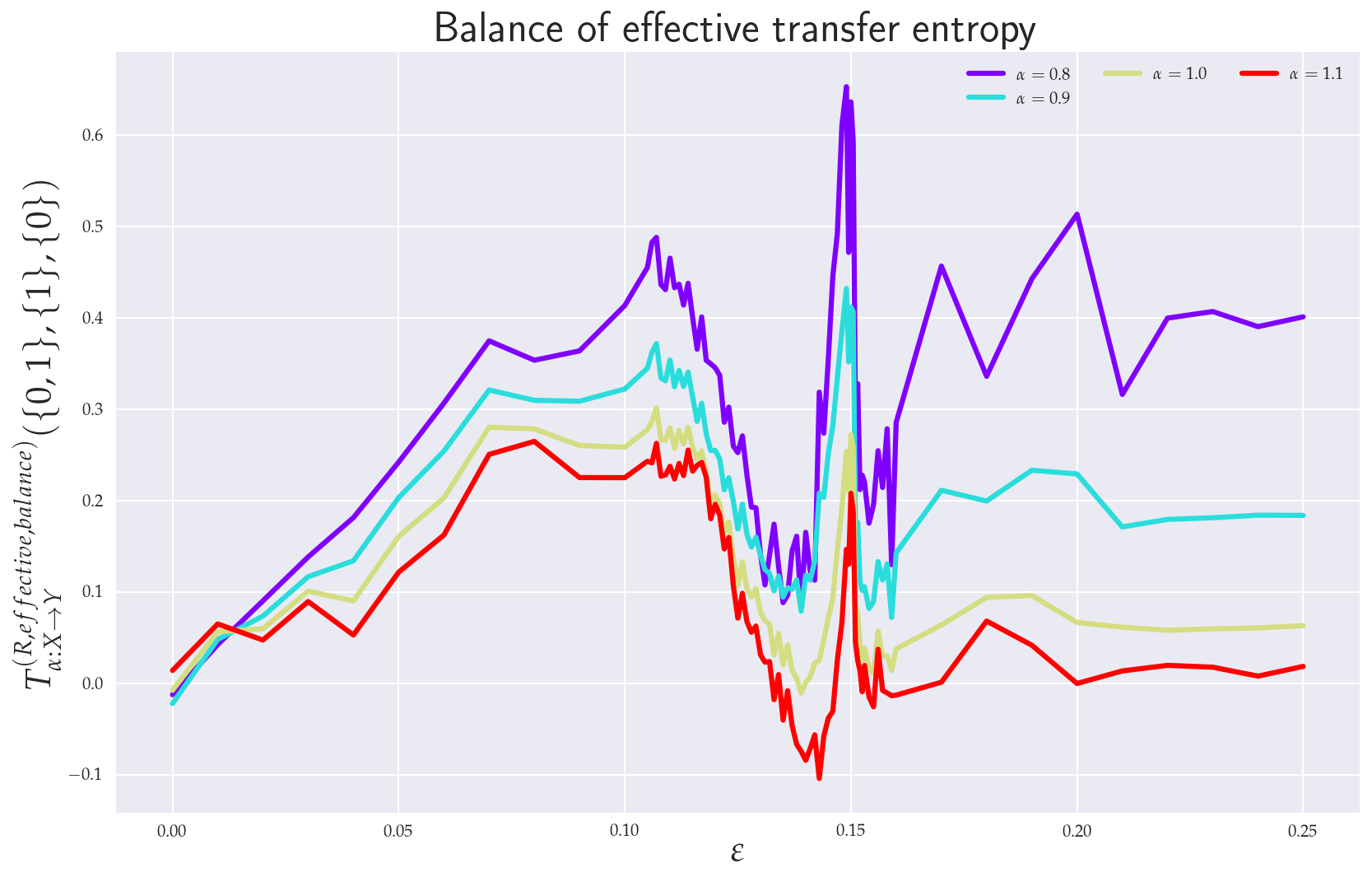}
\caption{Balance of effective RTEs from $x_1$ to $y_1$  $T^{R,\rm{\footnotesize{~balance,~effective}}}_{\alpha,x_1\rightarrow y_1}(\{0, 1\}, \{1\}, \{0\})$ (left) and balance of effective RTE  for the full system  $T^{R,\rm{\footnotesize{~balance,~ effective}}}_{\alpha,X\rightarrow Y}(\{0, 1\}, \{1\}, \{0\})$ (right).}%Balance of transfer entropy (left) with history of initial system $\{1, 2\}$ and balance of effective transfer entropy (right) with history of initial system $\{0, 1\}$ for projection to $x_1$ and $y_1$ (top) and the full system (bottom).}
\label{Balance_standard_effective_transfer}
    
\end{figure}

\begin{figure}
\begin{center}
    
\includegraphics[width=0.36\textwidth]{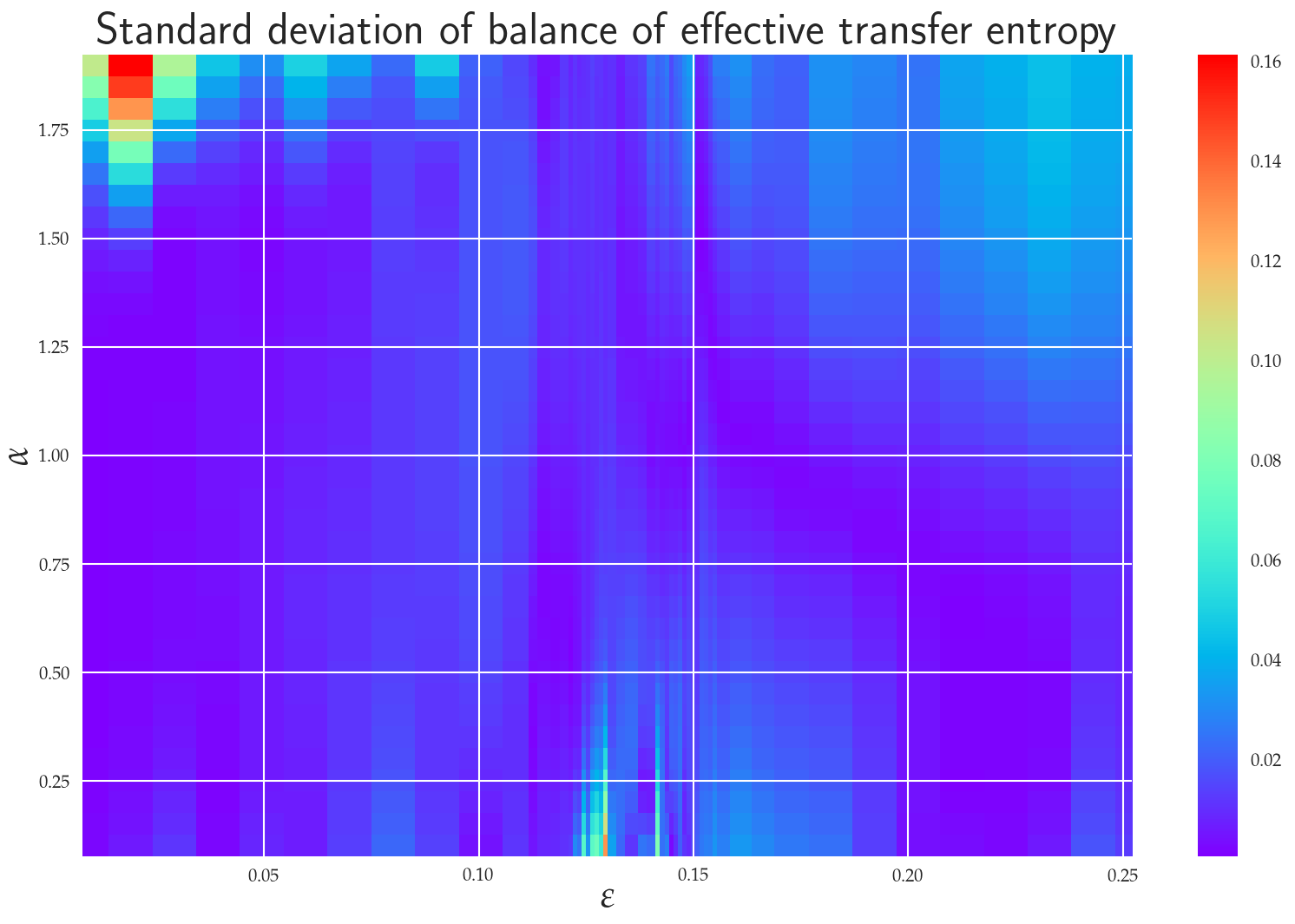}
%{figures/1D/balance_effective_transfer_entropy_3,4_1_0_X->Y_implot_std.png}
\includegraphics[width=0.36\textwidth]{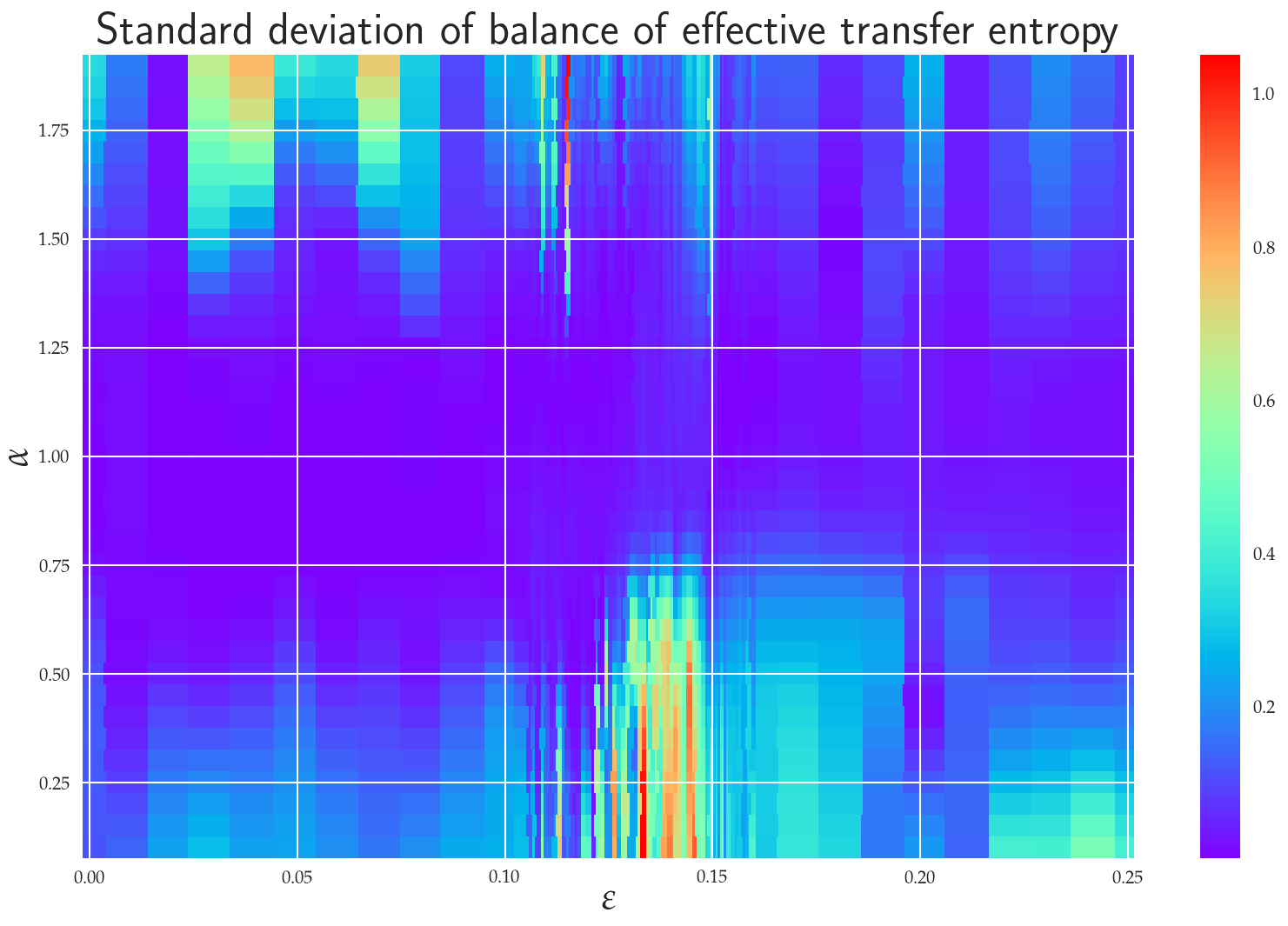}

%{figures/full_system/balance_effective_transfer_entropy_3,4_1_0_X->Y_implot_std}

\caption{Dependence of standard deviation of the balance of effective RTEs $T^{R,\rm{\footnotesize{~balance,~effective}}}_{\alpha,x_1\rightarrow y_1}(\{0, 1\}, \{1\}, \{0\})$ (left) and $T^{R,\rm{\footnotesize{~balance,~ effective}}}_{\alpha,X\rightarrow Y}(\{0, 1\}, \{1\}, \{0\})$ (right).}
\label{Balance_standard_effective_transfer_std}
\end{center}
\end{figure}

In order to quantify  the difference between coupled ($X\rightarrow Y$) and uncoupled direction ($Y\rightarrow X$) information flow directions we depict in Fig.~\ref{Balance_standard_effective_transfer}
balance of effective RTEs between  $T^{R,\rm{\footnotesize{~effective}}}_{\alpha,X\rightarrow Y}$ and $T^{R,\rm{\footnotesize{~effective}}}_{\alpha,Y\rightarrow X}$  for two different situations. Let us first concentrate on the balance of effective RTE $T^{R,\rm{\footnotesize{~balance,~effective}}}_{\alpha,x_1\rightarrow y_1}(\{0, 1\}, \{1\}, \{0\})$. There we can clearly see that before synchronization threshold (``topological phase transition'') i.e. for $\varepsilon \lesssim 0.12$ we have 
$T^{R,\rm{\footnotesize{~effective}}}_{\alpha,x_1\rightarrow y_1} > T^{R,\rm{\footnotesize{~effective}}}_{\alpha,y_1\rightarrow x_1}$, which indicates the correct direction of coupling.  The fact that for  $\alpha>1.6$ and $\varepsilon \lesssim 0.04$ one has $T^{R,\rm{\footnotesize{~balance,~effective}}}_{\alpha,x_1\rightarrow y_1}(\{0, 1\}, \{1\}, \{0\}) < 0$ can be attributed to smaller reliability of the estimator in this region, cf. Fig.~\ref{Balance_standard_effective_transfer_std}  for estimation of ensuing standard deviations. We can also observe that at the  synchronization threshold $T^{R,\rm{\footnotesize{~balance,~effective}}}_{\alpha,x_1\rightarrow y_1}(\{0, 1\}, \{1\}, \{0\})$ changes sign and slowly returns back to positive values in the fully synchronized regime. A similar behavior was reported in~\cite{Palus:2007a} for Shannon's TE. Moreover, in this transient region the effective RTEs have the same values irrespective of $\alpha$ or, in other words, information transfer is the same across all sectors of the underlying probability distributions.  This is akin to the behavior, which in statistical physics is typically associated with phase transitions --- except for the fact that now we have a critical line rather than a critical point. However, as we have already mentioned in the previous two paragraphs, this degeneracy is only spurious and will be removed by considering either the effective RTE for the full (6 dimensional) RS or longer memory. 

After  $\varepsilon \sim 0.15$ the approach to full synchronization proceeds at slightly different rates for different $\alpha$s. 
This can equivalently be restated as saying that different parts of the underlying distributions enter synchronization differently. 
Dependence of the balance of effective RTE for the full  (6 dimensional) system is shown on the right figure in Fig.~\ref{Balance_standard_effective_transfer}. Here the behavior is less reliable for larger values of $\alpha$ ($\alpha \gtrsim1.2$) and for smaller $\alpha$s ($\alpha \lesssim 0.8$), cf. Fig.~\ref{Balance_standard_effective_transfer_std}. In the region of reliable $\alpha$s the behavior is qualitatively similar to that of  $T^{R,\rm{\footnotesize{~balance,~effective}}}_{\alpha,x_1\rightarrow y_1}(\{0, 1\}, \{1\}, \{0\})$. On the other hand, apart from the region of a transient synchronization we clearly have  $T^{R,\rm{\footnotesize{~effective}}}_{\alpha,X\rightarrow Y} > T^{R,\rm{\footnotesize{~effective}}}_{\alpha,Y\rightarrow X}$, which implies the correct direction of coupling. Approach to full synchronization is also easily recognized --- the RTEs saturate to constant values (i.e.,  information transfer is $\varepsilon$ independent) and both $T^{R,\rm{\footnotesize{~effective}}}_{\alpha,X\rightarrow Y}$ and  $T^{R,\rm{\footnotesize{~effective}}}_{\alpha,Y\rightarrow X}$ start to approach each other. In this respect RTEs with lower $\alpha$s enter the synchronization regime slower than RTEs with larger $\alpha$s. In other words, events described by the tail parts of the distributions $p(x_{n+1}|x_n^{(k)})$ and $p(x_{n+1}|x_n^{(k)},y_n^{(l)})$ (corresponding to $\alpha < 1$) will fully synchronize at higher values of $\varepsilon$ than corresponding events described by central parts ($\alpha >1$).

In passing we might notice that since  
both $T^{R,\rm{\footnotesize{~effective}}}_{\alpha,X\rightarrow Y}$ and  $T^{R,\rm{\footnotesize{~effective}}}_{\alpha,Y\rightarrow X}$ approach each other in the fully synchronized state,  both $\{X\}$ and $\{Y\}$ system have to have the same underlying distributions (due to reconstruction theorem for REs~\cite{JA,JizbaRE})  and hence they are indistinguishable, as one would expect.

%

%%%%%%%%%%%%%%%%%%%%%%%%%%%%%%%%%%%%%%%%%%%%%%%%%%%%
%\subsection{Standard deviation of transfer entropy}
%%%%%%%%%%%%%%%%%%%%%%%%%%%%%%%%%%%%%%%%%%%%%%%%%%%%

%Standard deviation of balance of effective transfer entropy for projected system to $x_1$ and $y_1$ and full system are in Fig. \ref{Balance_standard_effective_transfer_std}. Projected system has local maximum in region $\alpha\leq0.7$ and $0.12 \leq \varepsilon \leq 0.16$ with interruption at $\varepsilon = 0.15$ and $\varepsilon \geq 0.22$ and $\alpha \geq 1.25$. For the full system, there is region of local maximum at $\alpha\leq0.7$ and $0.12 \leq \varepsilon \leq 0.16$ with interruption at $\varepsilon = 0.15$. Other regions for either $\alpha \leq 0.4$ and $\alpha \geq 1.6$ are caused by instability of the method. Regarding difference of standard deviation we observe that highest maximum are $10x$ of lowest minimum but they are marginal in comparison to the mean and they are higher of the full system.

%Finally, the results of the simulations are stable in wide region around $\alpha = 1$. However, standard deviations increase as an numeric effects for low and high simulated $\alpha$s. Beside that in region $0.12 \leq \varepsilon \leq 0.16$ or $\varepsilon \geq 0.22$ show systematic increase of standard deviation.

%%%%%%%%%%%%%%%%%%%%%%%%%%%%%%%%%%%%%%%%%%%%%%%%%%%%%%
\section{Discussion and Conclusions \label{Sec.8.cc}}
%%%%%%%%%%%%%%%%%%%%%%%%%%%%%%%%%%%%%%%%%%%%%%%%%%%%%%
\subsection{Theoretical results}
%%%%%%%%%%%%%%%%%%%%%%%%%%%%%%%%%%%%%%%%%%%%

In this paper we have analyzed the concept of RTE for two unidirectionally coupled R\"{o}ssler systems.  The idea was to 
illustrate how the RTE can deal with such issues as synchronization and, more generally,  causality. Despite the earlier applications of the RTE in bivariate (mostly financial) time series, many questions still has remained about how to properly qualify and quantify results thus obtained. Here we went some way towards this goal. 

First, we have shown that the concept of Granger causality is exactly equivalent to the RTE for Gaussian processes, which may in turn be used as a test of Gaussianity. This is because  RTEs are in the Gaussian framework all the same, and hence the results should be $\alpha$ independent. On the other hand, since the efficiency and robustness of RTE estimators crucially hinges on the parameter $\alpha$ employed, it might be in many cases easier to follow the information theoretic route to Granger causality
(provided the Gaussian framework is justified).

Second, we have demonstrated that 
the equivalence between the Granger causality and RTE can also be established for certain 
heavy-tailed processes --- for instance, for soft $\alpha$-Gaussian processes. Particularly, in this latter case one could clearly see the connection between Granger causality, R\'{e}nyi's parameter $\alpha$ and the heavy-tail power.

%%%%%%%%%%%%%%%%%%%%%%%%%%%%%%%%%%%%%%%%
\subsection{Numerical analysis of RTE for  R\"{o}ssler systems}
%%%%%%%%%%%%%%%%%%%%%%%%%%%%%%%%%%%%%%%%

In order to estimate the RTE, we have employed the $\ell$-nearest-neighbor entropy estimator  of Leonenko {\em et al.}~~\cite{Leonenko_Prozanto_Savani,Leonenko_Prozanto}. The latter is not only suitable for RE evaluation but it can also be easily numerically implemented to RTEs so that these can be computed almost in real time, which is relevant, e.g., in finance in various risk-aversion decisions. 
Spurious effects caused by finite size of dataset were taken into account by working with effective RTEs.

In order to gain further insight into a practical applicability and efficiency of the RTE
we have tested it on two unidirectionally coupled R\"{o}ssler systems --- master and slave system.   To have a clear idea what to expect we have first  looked at the phenomenology of the coupled RSs by means of simple numerical simulations (presented in Fig.~\ref{Roessler_system}). This was also accompanied by comparison with Lyapunov exponents computed in Refs.~\cite{Palus:2007a,palusRossler} and reproduced in Fig.~\ref{Lyapunov}. In particular, we could clearly observe how the RSs synchronize with the increasing value of coupling strength. In this connection, we have also identified  critical values of coupling strengths
at which both {\em threshold to transient behavior} (or ``topological phase transition'') and {\em threshold to full synchronization}  occurred.

More specifically, we were particularly interested in the transient region between
chaotic correlations regime and full synchronization, which had not as yet been discussed in the literature. To gain a better understanding about this region we have employed in the range $\varepsilon \in [0.1, 0.15]$ a higher frequency sampling, namely $0.001$, in contrast to standard $0.01$ one used for other $\varepsilon$'s. Threshold to transient behavior was identified at the scale $\varepsilon = 0.12$ where the positive LE crossed to negative values and where the projection on the $x_1$-$y_1$ and $x_2$-$y_2$ planes underwent a change of topology (cf.  Fig.~\ref{Roessler_system}). From the point of view of RTEs this threshold behavior was reflected in peaking the information flow between various directions. Particularly  pronounced was the increase in the effective RTE between $x_1$ and $y_1$ (in both directions) for $\alpha >1$, which reflected the increase in orbit occupation density around the peak in the $y_1$-$y_2$ plane in the slave system. Even more marked was the high peak in information flow from $x_3$ to $y_3$ for $\alpha < 1$ (see Fig.~\ref{Balance_standard_effective_transfer_projections}), which described an influx of information needed to ``organize'' chaotic  correlations that exist between $x_3$ and $y_3$ directions  prior $\varepsilon \lesssim 0.12$.
Furthermore, the RTE was especially instrumental in understanding the measure concentration phenomenon in the transient regime. 
%This crossover was already observed in \cite{Palus:2007a,Palus2018,palusRossler}. 
Finally, after a sharp ``firsts-order-type'' transition at the threshold of synchronization the effective RTEs
approached slowly their asymptotic values (distinct for each $\alpha$) in the synchronized state. In addition, in the synchronized state
both $T^{R,\rm{\footnotesize{~effective}}}_{\alpha,X\rightarrow Y}$ and  $T^{R,\rm{\footnotesize{~effective}}}_{\alpha,Y\rightarrow X}$ approach each other, which reveals that both $\{X\}$ and $\{Y\}$ system have the same underlying distributions and hence they are indistinguishable.

As for causality issue, we could reliably infer the coupling direction from the RTE only till $\varepsilon \lesssim 0.12$, i.e., till the threshold to transient behavior.
After this value the RSs started to synchronize, first partially (in the transient regime) and then fully after $\varepsilon = 0.15$. In fact, the full synchronization started when the information flow from all sectors of underlying distributions (i.e., for all $\alpha$s) began to be (almost) $\varepsilon$ independent and when  $T^{R,\rm{\footnotesize{~balance,~ effective}}}_{\alpha,X\rightarrow Y}$ approach zero --- 
so there was a one-to-one relation between the states of the systems and time series of the $\{X\}$ system can be predicted from time series $\{Y\}$ system, and vice versa, hence one could not make any statement about coupling direction.

We should also finally re-emphasize  that the standard deviation of the RTEs importantly depends on $\alpha$, cf. Eq.~(\ref{Balance_standard_effective_transfer_std}). 
%For instance, the balance effevtive RTE between $x_1$ and $y_1$ directions is quite noisy around $\varepsilon \sim 0.15$ (i.e., onset of synchronization) and the minimal noise value is not attained at $\alpha = 1$ (Shannon transfer entropy). In fact, higher values of $\alpha$ feature a lower level of fluctuations and hence  
%they are more reliable for the description around the critical coupling. 
%
For instance, the balance effective RTE for full system is around the transient region quite reliably described by $0.8 \lesssim \alpha \lesssim 1.25$, though
the minimal noise value is not attained at $\alpha = 1$ (Shannon transfer entropy) but at $\alpha = 1.16$. Clearly, the $\alpha$-dependence of fluctuations is generally dynamics dependent and in many interesting  real-world processes  it is simply more reliable to utilize non-Shannonian TEs.

%%%%%%%%%%%%%%%%%%%%%%%%%%%%%%%%%%%%%%%%%%
\subsection{Conclusions \label{Sec.8}}
%%%%%%%%%%%%%%%%%%%%%%%%%%%%%%%%%%%%%%%%%%

In this paper we have discussed the issue of  R\'{e}nyi transfer entropy in the context of causality with a particular emphasize on the issue of synchronization. Notably, we proved that the Granger causality is entirely equivalent to the RTE for Gaussian processes and showed how the Granger causality and the RTE are related in the case of heavy-tailed (namely $\alpha$-Gaussian) processes. These results allow to bridge the gap between autoregressive and  R\'{e}nyi entropy based information-theoretic approaches.

To put some flesh on the bare bones, we have illustrated some inner workings of the RTE  by analyzing RTE between bivariate time series generated from two unidirectionally coupled R\"{o}ssler systems that undergo synchronization.
The route to synchronization was scrutinized by considering the effective RTE (and other derived concepts)  between various master-slave components as well as between the full master and slave systems. We observed that the effective RTE could clearly identify  a transient synchronization region (in the coupling strength), i.e., regime between chaotic (master-slave) correlations  and the synchronization threshold. In the transient region the effective RTE allowed to infer measure concentration for orbit occupation density that cannot be deduced from Shannon's TE alone.

We have also seen that the direction of coupling and hence causality could be reliably inferred only for coupling strengths $\varepsilon < 0.12$ (onset of transient regime), i.e. when two RSs were coupled, but not yet fully.
This is in agreement with an earlier observation in Ref.~\cite{Palus:2007a}. As soon as the RSs
are synchronized, they produce identical time series,
and there is no way to infer the correct causality relation solely 
from the measured data.

We can conclude with a general observation that
a clear conceptual advantage of information theoretic measures in general and RTE in particular, as compared to standard Granger causality, is that they are sensitive to nonlinear signal properties as they do not rely on linear regression models. On the other hand, a clear limitation of the RTE, in comparison to Granger causality, is that they are by their very formulation restricted to bivariate situations (though multivariate generalization is possible, it substantially increases dimensionality in the estimation problem, which might be hard to
solve with a limited amount of available data). In addition, the RTEs often require substantially more data than regression methods.

%\subsection{Outlooks}

%%%%%%%%%%%%%%%%%%%%%%%%%%%%%%%%%%%%%%%%%%
\vspace{6pt} 

%%%%%%%%%%%%%%%%%%%%%%%%%%%%%%%%%%%%%%%%%%
%% optional
%\supplementary{The following are available online at \linksupplementary{s1}, Figure S1: title, Table S1: title, Video S1: title.}


%%%%%%%%%%%%%%%%%%%%%%%%%%%%%%%%%%%%%%%%%%
%%%%%%%%%%%%%%%%%%%%%%%%%%%%%%%%%%%%%%%%%%
\authorcontributions{Conceptualization, P.J.; Formal analysis, H.L. and Z.T.; Methodology, P.J., H.L. and Z.T.; Validation, H.L. and Z.T.; Software design, data structures, computer calculation and visualization, H.L.; Writing—original draft, P.J.; Writing—review \& editing, P.J., H.L. and Z.T. All authors have read and agreed to the published version of the manuscript.}

\funding{P.J., H.L. and Z.T. were supported by the Czech Science Foundation Grant No. 19-16066S.}
\institutionalreview{Not applicable.}
\informedconsent{Not applicable.}
\dataavailability{Not applicable.} 
\acknowledgments{We thank 
 Milan Palu\v{s} for helpful comments and discussions and for providing us source code for Fig.~\ref{Lyapunov}.}
\conflictsofinterest{The authors declare no conflict of interest.} 
%% Optional
%\sampleavailability{Samples of the compounds ... are available from the authors.}
%%%%%%%%%%%%%%%%%%%%%%%%%%%%%%%%%%%%%%%%%%
%% Only for journal Encyclopedia
%\entrylink{The Link to this entry published on the encyclopedia platform.}
%%%%%%%%%%%%%%%%%%%%%%%%%%%%%%%%%%%%%%%%%%
%%%%%%%%%%%%%%%%%%%%%%%%%%%%%%%%%%%%%%%%%%
%% Optional
\abbreviations{The following abbreviations are used in this manuscript:\\

\noindent 
\begin{tabular}{@{}ll}
RE & R\'{e}nyi entropy \\
TE & Transfer entropy \\
RTE & R\'{e}nyi transfer entropy \\
PDF & Probability density function\\
ITE & Information-theoretic entropy \\
RS & R\"{o}ssler system \\
KSE & Kolmogorov--Sinai entropy rate\\
LE & Lyapunov exponent
\end{tabular}}

%%%%%%%%%%%%%%%%%%%%%%%%%%%%%%%%%%%%%%%%%%
%% Optional
%\appendixtitles{no} % Leave argument "no" if all appendix headings stay EMPTY (then no dot is printed after "Appendix A"). If the appendix sections contain a heading then change the argument to "yes".
%\appendixstart
%\appendix
%\section{}
%\subsection{}
%*******
%%%%%%%%%%%%%%%%%%%%%%%%%%%%%%%%%%%%%%%%%%

%% Optional
%\sampleavailability{Samples of the compounds ... are available from the authors.}
%%%%%%%%%%%%%%%%%%%%%%%%%%%%%%%%%%%%%%%%%%
%% Only for journal Encyclopedia
%\entrylink{The Link to this entry published on the encyclopedia platform.}
%%%%%%%%%%%%%%%%%%%%%%%%%%%%%%%%%%%%%%%%%%

%%%%%%%%%%%%%%%%%%%%%%%%%%%%%%%%%%%%%%%%%%
\end{paracol}%%% nemazat
\reftitle{References}


%\externalbibliography{yes}
%\bibliography{your_external_BibTeX_file}

%=====================================
% References, variant B: internal bibliography
%=====================================

%\bibliographystyle{ieeetr}
%\bibliography{bibliography}
%\bibliographystyle{}
\bibliography{RTE_for_RS_article}

% If authors have biography, please use the format below
%\section*{Short Biography of Authors}
%\bio
%{\raisebox{-0.35cm}{\includegraphics[width=3.5cm,height=5.3cm,clip,keepaspectratio]{Definitions/author1.pdf}}}
%{\textbf{Firstname Lastname} Biography of first author}
%
%\bio
%{\raisebox{-0.35cm}{\includegraphics[width=3.5cm,height=5.3cm,clip,keepaspectratio]{Definitions/author2.jpg}}}
%{\textbf{Firstname Lastname} Biography of second author}

% The following MDPI journals use author-date citation: Arts, Econometrics, Economies, Genealogy, Humanities, IJFS, JRFM, Laws, Religions, Risks, Social Sciences. For those journals, please follow the formatting guidelines on http://www.mdpi.com/authors/references
% To cite two works by the same author: \citeauthor{ref-journal-1a} (\citeyear{ref-journal-1a}, \citeyear{ref-journal-1b}). This produces: Whittaker (1967, 1975)
% To cite two works by the same author with specific pages: \citeauthor{ref-journal-3a} (\citeyear{ref-journal-3a}, p. 328; \citeyear{ref-journal-3b}, p.475). This produces: Wong (1999, p. 328; 2000, p. 475)

%%%%%%%%%%%%%%%%%%%%%%%%%%%%%%%%%%%%%%%%%%
%% for journal Sci
%\reviewreports{\\
%Reviewer 1 comments and authors’ response\\
%Reviewer 2 comments and authors’ response\\
%Reviewer 3 comments and authors’ response
%}
%%%%%%%%%%%%%%%%%%%%%%%%%%%%%%%%%%%%%%%%%%
\end{document}